\documentclass[12pt]{article}

\usepackage{epsfig}
\usepackage{graphicx}
\usepackage{amssymb,amsmath}
\usepackage{hyperref}
\usepackage{cite}
\hypersetup{
    colorlinks=true,
    linkcolor=blue,
    citecolor=red,
}

\begin{document}

\begin{center}
\Large{\bf Lorentzian Vacuum Transitions in $f(R)$ gravity} \vspace{0.7cm}

\large  H. Garc\'{\i}a-Compe\'an$^{a,}$\footnote{e-mail address: {\tt
		hugo.compean@cinvestav.mx}}, J. Hern\'andez-Aguilar$^{b,}$\footnote{e-mail address: {\tt
jorge.hernandezag@alumno.buap.mx}}, D. Mata-Pacheco$^{c,}$\footnote{e-mail
address: {\tt daniel.mata@cinvestav.mx}}, \ \ C. Ram\'{\i}rez$^{b,}$\footnote{e-mail
address: {\tt cramirez@fcfm.buap.mx}}

\vspace{0.3cm}
{\small \em $^a$Departamento de F\'{\i}sica, Centro de
	Investigaci\'on y de Estudios Avanzados del IPN}\\
{\small\em P.O. Box 14-740, CP. 07000, Ciudad de M\'exico, M\'exico}\\

\vskip .4truecm

{\small \em $^b$Facultad de Ciencias F\'{\i}sico Matem\'aticas, \\ 
Benem\'erita Universidad Aut\'noma de Puebla\\
4 Sur 104, Puebla 72000, Puebla, M\'exico}\\
\vspace{0.3cm}

\vskip .4truecm

{\small \em $^c$New York University Abu Dhabi \\ 
Saadiyat Island, Abu Dhabi\\
PO Box 129188, United Arab Emirates}\\
\vspace{0.3cm}

\vspace*{0.3cm}
\end{center}

\begin{abstract}
 We study Lorentzian vacuum transition probabilities between two minima of a scalar field potential within the framework of $f(R)$ gravity. The analysis extends the previously considered WKB expansion of the Wheeler-DeWitt equation to modified gravity theories, up to second order. We apply the general method for homogeneous and isotropic FLRW universes, with zero and positive spatial curvature, for any $f(R)$ model. For the flat case we obtain analytic expressions for the transition probabilities for any model if we assume a constant Ricci scalar; this assumption has been considered in previous studies, in the Euclidean approach, from symmetry arguments. On the other hand, we also obtain explicit solutions without this assumption for power-law $f(R)=R^{1+n}$ models. Moreover, in the positive curvature scenario, we obtain that the assumption of a constant Ricci scalar is not consistent, but we are able to find analytical solutions in approximated regimes. In all cases we have found that the general behavior of the probabilities already found for Einstein Gravity is preserved, including the prediction of a non-singular initial state due to quantum corrections, even though the probabilities increase or decrease in a model dependent way.
 
 \vskip 1truecm

\end{abstract}

\bigskip

\newpage
\section{Introduction}
\label{S-Intro}
The phenomenon of vacuum decay represents one of the most intriguing aspects of quantum theories, with profound implications in particle physics and cosmology. The theoretical framework for understanding vacuum decay in quantum field theory in flat spacetime was established in the seminal works of Coleman and Callan \cite{Coleman:1977py,Callan:1977pt}, who developed the semiclassical instanton method to calculate decay rates of metastable vacuum states. In this formalism, the transition from a false to a true vacuum proceeds through quantum tunneling, by which bubble of true vacuum nucleate and subsequently expand, eventually filling the surrounding false vacuum. The decay rate for this process is determined by the Euclidean action of a critical bubble configuration known as the bounce solution.

Later on, Coleman and De Luccia \cite{Coleman:1980aw} incorporated gravitational effects into the vacuum decay formalism. Their work established the foundation for understanding vacuum transitions in cosmological contexts, with the inclusion of gravitational interactions, and offers a description of  spacetime geometries that shift between different vacuum configurations. Furthermore, there are remarkable properties derived in this context, such as the fact that the universe after the transition is open or the assumption of an $O(4)$ symmetry for the bounce solution. 

The development of quantum approaches to vacuum decay in gravity has provided complementary insights into the tunneling process (for a general review see \cite{Devoto:2022qen}). While the Euclidean path integral approach provides a powerful method for calculating probabilities, an alternative and equally fundamental perspective on quantum cosmology is offered by the Hamiltonian formalism. This approach is rooted in the canonical quantization of General Relativity (GR), with the central object given by the Wheeler-DeWitt (WDW) equation, which serves as the quantum counterpart to the classical constraint equations of GR \cite{Wheeler,DeWitt}. In this formalism, the universe is described by a timeless wave function, defined on a configuration space spanned by the degrees of freedom of the geometry on three-dimensional spatial slices and any matter fields present. Thus, it is possible to define a notion of transition probability by employing conditional probabilities between two solutions of different configurations. In this context a Hamiltonian formulation of gravitational tunneling was developed in \cite{FMP1,FMP2}. A key advantage of this Hamiltonian formalism is its Lorentzian nature, eliminating the need for the Wick rotation employed in the Euclidean path integral approach, which requires analytic continuation to describe post-nucleation universe evolution, thereby avoiding potential artifacts that may arise by the analytic continuation. However, the limitations of this approach allow to completely study the transitions only between two values of the cosmological constant.

Following this approach, in \cite{deAlwis:2019dkc} quantum transitions between Minkowski and de Sitter spacetimes were examined, while subsequent studies \cite{Cespedes:2020xpn} investigated whether Lorentzian vacuum transitions lead to open or closed universe configurations, being able to incorporate explicitly an scalar field with an effective approach for the wall. These Lorentzian approaches have been extended to anisotropic universes \cite{Garcia-Compean:2021syl}, alternative gravity theories such as Hořava-Lifshitz gravity \cite{Garcia-Compean:2021vcy}, and frameworks incorporating generalized uncertainty principles \cite{Garcia-Compean:2022ysy}. More recently in \cite{Garcia-Compean:2024zjr} an extension of the Lorentzian Wheeler-DeWitt formalism by including higher-order terms in the semiclassical WKB expansion was performed, obtaining analytical expressions for transition probabilities up to second-order quantum corrections in both homogeneous isotropic and anisotropic cosmological models. Interpreting the results as universe creation probabilities via vacuum decay, it is found that second-order corrections generally avoid initial singularities, though this outcome only holds for isotropic universes.

Vacuum transitions with gravitational effects have been an ongoing topic for a long time. For example, in \cite{Espinosa:2018hue,Espinosa:2018voj} the calculation of tunneling actions has been reformulated as an elementary variational problem in field space. Building upon this, in \cite{Espinosa:2021tgx, Espinosa:2022jlx} a general method to find scalar potentials with a false vacuum with exactly solvable decay at the semiclassical level was developed, including gravitational corrections and extending the methods for computing vacuum decay actions from tunneling potentials in arbitrary spacetime dimensions. These developments have been complemented by studies examining specific aspects of the tunneling process (See for example \cite{Holman:1990, Zhang:2013pna,Kristiano:2018oyv,Ghosh:2021lua, Ivo:2025fwe,Cespedes:2023jdk}), providing insights into vacuum transitions beyond the semiclassical approximation. Additionally, the studies on this topic have expanded through the influence of non-minimal couplings on false vacuum bubble nucleation \cite{Lee:2006vka}, and approaches employing the Lorentzian path integral \cite{Hayashi:2021kro,Nishimura:2023dky} represent additional directions in which the original Coleman framework has been extended and refined.

On the other hand, the inclusion of higher-order curvature terms on the gravitational action, generically emerge in effective field theories of some quantum gravity theory and could represent completions in the high energy regime of standard GR. These modified gravity theories play crucial roles in various cosmological contexts (See \cite{Ezawa:2003wh,Moreno:2023arp,Singh:2022jue,Odintsov:2025kyw,Addazi:2025qra}). In this sense, the study of vacuum decay in modified gravity theories represents a natural and necessary extension of the original Coleman-De Luccia framework. In this context, \cite{Cai:2008ht} investigates the effects of the Gauss-Bonnet term on vacuum decay, demonstrating that higher-order curvature corrections can significantly modify tunneling rates. This analysis was subsequently extended to examine vacuum transitions with the Gauss-Bonnet term in arbitrary dimensions \cite{Liu:2024aos}. Meanwhile, \cite{Gregory:2024sku} explores how higher-derivative gravity theories can be probed through tunneling phenomena. In a similar way, in \cite{Vicentini:2020lhm,Vicentini:2021qoo}, vacuum decay in quadratic gravity has been systematically investigated, examining both massless and massive cases, and new bounds on vacuum decay rates in de Sitter space were established \cite{Vicentini:2022pra}.

Among the various modified gravity theories that have been proposed over the years \cite{CANTATA:2021asi,Shankaranarayanan:2022wbx}, $f(R)$ gravity occupies a particularly prominent position due to its theoretical frame and phenomenological relevance. The $f(R)$ formalism \cite{Sotiriou:2008rp,DeFelice:2010aj,Nojiri:2017ncd} generalizes the Einstein-Hilbert action by replacing the Ricci scalar $R$ with an arbitrary function $f(R)$. This class of theories has demonstrated remarkable success in addressing outstanding problems in cosmology. For example, one of the simplest models on this class is the Starobinsky model that successfully describes a compelling mechanism for early universe inflation and matches phenomenological constraints up to date. See for example \cite{Starobinsky:1980te, Vilenkin:1985md,Linde:2025pvj,Drees:2025ngb,Planck:2018jri,Stachowski:2016zio,German:2023euc,Ketov:2025nkr,Addazi:2025qra,SidikRisdianto:2025qvk,Ivanov:2021chn}. Furthermore, in \cite{Capozziello:2014hia} $f(R)$ gravity has been employed to connect early and late universe dynamics, offering unified descriptions of inflation and late-time cosmic acceleration. Moreover, it has been shown that power-law inflation can be realized within $f(R)$ frameworks. This modification, for example, allows the theory to satisfy gravitational constraints in the solar system through a shielding mechanism, while at galactic scales the scalar field induces dynamic effects analogous to those of dark matter \cite{Yadav:2018llv,Bisabr:2010sq,Goheer:2009ss,Tsujikawa:2007gd}. Also, $f(R)$ modifications represent the simplest extension of GR that incorporates higher-order curvature effects without pathologies such as ghost instabilities. In this context, $f(R)$ theories represent an ideal testing ground for understanding how modified gravity influences quantum tunneling processes. It is well known that such theories are a particular class of Brans-Dicke theories that are equivalent, after a conformal transformation, to GR plus a scalar field dubbed the scalaron. However, it is also possible to study the coupling with an external scalar field that allows to study the cosmological consequences staying on the Jordan frame \cite{DeAngelis:2021afq,Bamonti:2021jmg,Ohta:2017trn,Figueroa:2021iwm}. Moreover, $f(R)$ theories admit consistent quantum mechanical treatments, as demonstrated by studies of the Wheeler-De Witt equation in $f(R)$ cosmology \cite{Sanyal:2001ws,Huang:2013dca,Paliathanasis:2019ega,Ramirez}. In this sense, Salehian and Firouzjahi \cite{Salehian:2018yoq} performed an analysis of vacuum decay and bubble nucleation in $f(R)$ gravity employing the Euclidean approach and a constant Ricci scalar, showing how the functional dependency on the Ricci scalar impacts decay rates and critical bubble configurations. 

The article is structured as follows: In section \ref{S-WKB} we outline the general method, as developed in \cite{Garcia-Compean:2024zjr}, for computing quantum-corrected transition probabilities in minisuperspace models, starting from a generic Hamiltonian constraint, and discuss the applicability to the theories of interest. In section \ref{S-Flat} the general formalism is applied to the case of the flat FLRW metric, obtaining general expressions for the probability amplitudes for any $f(R)$ model. In section \ref{S-ConsRS}, the scenario with constant Ricci curvature scalar is considered, and explicit expression for the transition probabilities for particular interesting models are calculated. Then, in section \ref{S-NonCR}, we avoid this simplification and compute the probabilities for some models in full generality. In section \ref{S-CFLRW}, we perform the analysis for a closed FLRW universe. We show how, in these cases, the assumption of a constant Ricci scalar is incompatible with the WDW approach, however we can obtain general expressions of the probabilities in some approximated regimes considering some particular models. Then in section \ref{S-FinalR} we present our Final Remarks. Finally in Appendix \ref{FLRW-MetricApp} we present an overview of the WDW approach to $f(R)$ theories, and in Appendix \ref{BianchiIIIApp} we present the generalization of this approach to the Bianchi III metric, showing a particular example of how this formalism can be applied in a more general scenario.
\section{Semiclassical computation of vacuum decay probabilities in the Hamiltonian formalism}
\label{S-WKB} 
This section presents the general methodology developed in \cite{Cespedes:2020xpn,Garcia-Compean:2021syl,Garcia-Compean:2024zjr}, for calculating quantum-corrected transition probabilities. The approach employs the semiclassical WKB expansion of the Wheeler-De Witt equation, incorporating quantum corrections through higher-order terms. Our analysis specifically considers transitions between false and true vacua of a scalar field potential. We adhere to the notation and conventions established in these works.

Let us consider a general form for the Hamiltonian constraint, characterized by its quadratic dependence on the momenta
        \begin{equation}\label{eq:HamConst}
             \mathcal{H}=\frac{1}{2}G^{M N}(\Phi)\pi_{M}\pi_{N}+K[\Phi] \simeq 0,
        \end{equation}
where $\Phi^M =\Phi^M(\vec{x})$ represents the 3-field configuration, with index $M$ labeling the degrees of freedom that define Wheeler's superspace, with metric $G_{MN}$, and $\pi_M$ denotes their corresponding canonically conjugate momenta. Moreover, the function $K[\Phi]$ is constituted by the scalar field potential and the remaining terms without dependence on the momenta that can appear.

The Wheeler-De Witt equation is obtained by promoting the superspace degrees of freedom and their conjugate momenta to Hermitian operators. In the coordinate representation, the Wheeler-De Witt equation is written as
        \begin{equation}
            \mathcal{H} \Psi[\Phi] = \left[ -\frac{\hbar^2}{2} G^{MN}(\Phi) \frac{\delta^2}{\delta \Phi^M \delta \Phi^N} + K[\Phi] \right] \Psi[\Phi] = 0,
        \label{eq:WDWGR}
        \end{equation}
up to ordering ambiguities. We then employ a general semiclassical WKB ansatz of the form 
        \begin{equation}
            \Psi[\Phi]=\exp \left\{\frac{i}{\hbar} S[\Phi]\right\}
        \end{equation}
with the $\hbar$-expansion
        \begin{equation}
             S[\Phi]=S_{0}[\Phi]+\hbar S_{1}[\Phi]+\hbar^2 S_{2}[\Phi]+\mathcal{O}\left(\hbar^{3}\right) ,
          \label{eq:WKBExpan}
        \end{equation}
where $S_{0}$ is the classical action and $S_{1}$ and $S_{2}$ are the first and second quantum corrections respectively. Substituting this ansatz in the WDW equation (\ref{eq:WDWGR}), we obtain an independent equation for each term in the semiclassical expansion as
        \begin{equation}
             \frac{1}{2} G^{M N} \frac{\delta S_{0}}{\delta \Phi^{M}} \frac{\delta S_{0}}{\delta \Phi^{N}}+K[\Phi]=0,
            \label{eq:WKB3O}
        \end{equation}
        \begin{equation}
             2 G^{M N} \frac{\delta S_{0}}{\delta \Phi^{M}} \frac{\delta S_{1}}{\delta \Phi^{N}}=i G^{M N} \frac{\delta^{2}}{\delta \Phi^{M} \delta \Phi^{N}} S_{0},
        \label{eq:WKB1}
        \end{equation}
        \begin{equation}
            2 G^{M N} \frac{\delta S_0}{\delta \Phi^M} \frac{\delta S_2}{\delta \Phi^N}+G^{M N} \frac{\delta S_1}{\delta \Phi^M} \frac{\delta S_1}{\delta \Phi^N}=i G^{M N} \frac{\delta^2 S_1}{\delta \Phi^M \delta \Phi^N}.
         \label{eq:WKB2}
        \end{equation}  
Under the minisuperspace approximation, the solution of Eq.~\ref{eq:WKB3O} yields the classical action $S_0$, and generates classical trajectories $\Phi_s$ in minisuperspace, with parameter $s$ given by solutions of the equation
        \begin{equation}
            C(s) \frac{d \Phi^M_s}{ds} = G^{MN} \frac{\delta S_0}{\delta \Phi^N_s},
        \label{eq:Cs}
        \end{equation}
where $C(s)$ is an auxiliary function. In this way, it can be shown \cite{Garcia-Compean:2021syl} that a general solution for $\left(\frac{d \Phi^M_s}{ds}\right)$ and $C^2(s)$ can be found, which allows to write the classical action as
        \begin{equation}
            S_{0}\left[\Phi\right]=-2 \int^{s} \frac{d s^{\prime}}{C\left(s^{\prime}\right)} \int_{X} K\left[\Phi_{s^{\prime}}\right]
            \label{eq:S0f}
        \end{equation}
and the corresponding first and second quantum corrections as
        \begin{equation}
            S_{1}=-i \operatorname{Vol}^{2}(X) \int^{s} \frac{ds'}{C^{2}(s')} \nabla^2 K ,
            \label{eq:S1}
        \end{equation}
        \begin{equation}
        	S_2=\frac{1}{2} \int^{s} ds' \frac{\operatorname{Vol}^3(X)}{C^3(s')}\nabla^{2}\left(\nabla^{2} K\right)           +\frac{1}{2} \int^{s} ds' \frac{\operatorname{Vol}^5(X)}{C^5(s')} \left[\nabla\left(\nabla^{2} K\right)\right]^{2} ,
        	\label{eq:S2}
        \end{equation}  
where we have defined
	\begin{equation}
		\nabla^{2} K=G^{M N} \frac{\partial^2 K}{\partial \Phi^{M} \partial \Phi^{N}},
	\end{equation} 
\begin{equation}
        	\quad(\nabla K)^{2}=G^{M N} \frac{\partial K}{\partial \Phi^{M}} \frac{\partial K}{\partial \Phi^{N}} ,
        \end{equation}
and $\operatorname{Vol}(X)$ stands for the volume of the spatial section. Considering a homogeneous behavior for all minisuperspace variables $\Phi^M = \Phi^M(s)$, we get
        \begin{equation}
            C^2(s) = -\frac{2 \operatorname{Vol}^2(X)}{K} (\nabla K)^2, \quad \frac{d \Phi^M_s}{ds} = \frac{K}{\operatorname{Vol}(X)} \frac{\nabla^M K}{(\nabla K)^2},
            \label{eq:Cs2}
        \end{equation}
where
        \begin{equation}
            \nabla^M K = G^{MN} \frac{\partial K}{\partial \Phi^N}.
        \end{equation}
From (\ref{eq:Cs2}), we observe that the fields satisfy
        \begin{equation}
            \frac{d\Phi^M}{d\Phi^N} = \frac{\nabla^M K}{\nabla^N K},
            \label{Relations}
        \end{equation}
for all non-vanishing $d\Phi^{M}$. These relations allow to a reduction of the transition probabilities to a single degree of freedom, allowing a coordinate transformation from $s$ to any field variable for all integrals. 

The transition probability is obtained, in this approach, from the ratio of two solutions of the WDW equation, one describing the evolution of the scalar field  from the false to the true vacuum, and another maintaining the field at the false vacuum. The tunneling transition probability for going from the false vacuum at $\phi_{A}$ to the true vacuum at $\phi_{B}$, takes the form
        \begin{equation}
            P(A \rightarrow B) = \exp[-2 \operatorname{Re}(\Gamma)],
        \label{eq:Prob}
        \end{equation}
where, up to second quantum corrections
        \begin{equation}
             \pm \Gamma=\Gamma_{0}+\Gamma_{1}+\Gamma_{2},
             \label{DefGammaGe}
        \end{equation}
where $\Gamma_{0}$ is the semiclassical contribution, while $\Gamma_1$ and $\Gamma_2$ are the first and second quantum corrections, and 
        \begin{equation}
            \begin{array}{l}
            \Gamma_0=\frac{i}{\hbar}{\big[S_0\left(\varphi_0^I, \phi_B ; \varphi_m^I, \phi_A\right)-S_0\left(\varphi_0^I, \phi_A ; \varphi_m^I, \phi_A\right)\big]}, \\
            \Gamma_1= i\big[S_1\left(\varphi_0^I, \phi_B ; \varphi_m^I, \phi_A\right)-S_1\left(\varphi_0^I, \phi_A ; \varphi_m^I, \phi_A\right)\big],\\
            \Gamma_2=i \hbar\big[S_2\left(\varphi_0^I, \phi_B ; \varphi_m^I, \phi_A\right)-S_2\left(\varphi_0^I, \phi_A ; \varphi_m^I, \phi_A\right)\big].
            \end{array}
            \end{equation}
where, $\varphi^I$ represents the degrees of freedom coming from the three metric, with conditions $\varphi^I(s=0) = \varphi_0^I$ when the scalar field is $\phi = \phi_B$, and $\varphi^I(s=s_M) = \varphi_m^I$ when $\phi = \phi_A$. Finally, the sign ambiguity appears due to the fact that the general solution of the wave functionals will be a linear superposition of exponential terms, however, we will keep only dominant terms.

In order to study a scenario, where there is a bubble of true vacuum in the background of a false vacuum, separated by a wall as in the Euclidean approach, the parameter $s$ is established such that
        \begin{equation}
            \phi(s) \approx
            \begin{cases}
            \phi_B, & 0 < s < \bar{s} - \delta s, \\
            \phi_A, & \bar{s} + \delta s < s < s_M.
            \end{cases}
            \label{eq:sparameter}
        \end{equation}
Furthermore, we will always consider the thin wall approximation in which $\delta s \to0$. AS we are in the minisuperspace approach, we cannot describe the motion of the wall. Nevertheless, we can give the solutions from the semiclassical computations on the Euclidean approach, as was shown in \cite{Cespedes:2020xpn,Garcia-Compean:2021syl}.

With this choice, the semiclassical contribution (\ref{eq:S0f}) takes the form
        \begin{equation}
         \begin{aligned}
            \Gamma_0= & -\frac{2 \operatorname{Vol}(X) i}{\hbar}\left\{\left.\int_{0}^{\bar{s}-\delta s} \frac{d s}{C(s)} K\right|_{\phi=\phi_B}-\left.\int_{0}^{s-\delta s} \frac{d s}{C(s)} K\right|_{\phi=\phi_A}\right. \\
             & \left.+\int_{\bar{s}-\delta s}^{\bar{s}+\delta s} d s\left[\frac{K}{C(s)}-\left.\frac{K}{C(s)}\right|_{\phi=\phi_A}\right]\right\}.
         \label{eq:gammazero}
        \end{aligned}
        \end{equation}
The first quantum correction (\ref{eq:S1}) is given by
        \begin{equation}
          \begin{aligned}
            \Gamma_{1}& = \operatorname{Vol}^{2}(X)\left[\int_{0}^{\bar{s}-\delta s}\right. \left.\frac{d s}{C^{2}(s)} \nabla^2 K \right|_{\phi=\phi_{B}}-\left.\int_{0}^{\bar{s}-\delta s} \frac{d s}{C^{2}(s)} \nabla^2 K\right|_{\phi=\phi_{A}} \\
             &\left. +\int_{\bar{s}-\delta s}^{\bar{s}+\delta s} ds\frac{1}{C^2(s)}\left(\nabla^2 K-\left. \nabla^2 K\right|_{\phi=\phi_{A}}\right)\right] ,
             \label{eq:gamma1}
            \end{aligned}
        \end{equation}
and the second quantum correction (\ref{eq:S2}) is written as
        \small\begin{equation}
            \begin{aligned}
            \Gamma_2= & \frac{i \hbar}{2}\left[\left.\int_{0}^{s-\delta s} ds \frac{\operatorname{Vol}^3(X)}{C^3(s)} \nabla^{2}\left(\nabla^{2} K\right)\right|_{\phi=\phi_B}+\int_{0}^{s-\delta s} d s \frac{\operatorname{Vol}^5(X)}{C^5(s)} \left[\nabla\left(\nabla^{2} K\right)\right]^{2}\right|_{\phi=\phi_B} \\ & \left.-\int_{0}^{\bar{s}-\delta s} d s \frac{\operatorname{Vol}^3(x)}{C^3(s)} \nabla^{2}\left(\nabla^{2} K\right)\right|_{\phi=\phi_A}-\left.\int_{0}^{\bar{s}-\delta s} d s \frac{\operatorname{Vol}^5(x)}{C^5(s)} \left[\nabla\left(\nabla^{2} K\right)\right]^{2} \right|_{\phi=\phi_A} \\ & +\int_{\bar{s}-\delta_s}^{\bar{s}+\delta_s} d s \frac{\operatorname{Vol}^3(X)}{C^3(s)}\left(\nabla^{2}\left(\nabla^{2} K\right)-\left.\nabla^{2}\left(\nabla^{2} K\right)\right|_{\phi=\phi_A}\right) \\ & \left.\left.+\int_{\bar{s}-\delta_s}^{\bar{s}+\delta_s} d s \frac{\operatorname{Vol}^5(x)}{C^5(s)}\bigg(\left[\nabla\left(\nabla^{2} K\right)\right]^{2}-\left[\nabla\left(\nabla^{2} K\right)\right]^{2} \right|_{\phi=\phi_A}\bigg) \right].
            \end{aligned}
            \label{eq:gamma2}
            \end{equation}\normalsize
Let us remark that this general procedure only depends on the minisuperspace homogeneous approximation and the possibility to describe a generic Hamiltonian constraint with a quadratic dependence on the momenta. As remarked in \cite{Garcia-Compean:2024zjr}, it is not restricted to a specific metric or theory of gravity. In the appendices we present an overview of how these conditions are satisfied for the homogeneous isotropic and anisotropic metrics of interest in $f(R)$ theories. Furthermore, this framework allows, in principle, computation of transition probabilities to arbitrary order. However, as demonstrated in \cite{Garcia-Compean:2024zjr}, the second-order corrections reveals significant physical effects, while remaining tractable. Thus, in the following we will restrict our analysis to this order. 

Let us remark an important point regarding the integrals on the region where the scalar field is not constant in the expressions (\ref{eq:gammazero})-(\ref{eq:gamma2}). It was shown in \cite{Cespedes:2020xpn,Garcia-Compean:2021syl} that for the semiclassical solution these terms correspond to the bubble tension term appearing in the Euclidean approach. For this reason in the subsequent studies \cite{Garcia-Compean:2021vcy,Garcia-Compean:2022ysy,Garcia-Compean:2024zjr} it was considered that if such integrals take the generic form
	\begin{equation}
		\int_{\bar{s}-\delta s}^{\bar{s}+\delta s}N(\phi)O(\varphi^{I})ds ,
        \label{eq:TenSep}
	\end{equation}
for some functions $N(\phi)$ and $O(\varphi^{I})$, then on the thin wall limit, the integral can be substituted by the evaluation of $O(\phi^{I})$ on $\bar{s}$ times an independent parameter that are generically called \textit{tension terms}. (Although only on the semiclassical result they are actual tension terms, for quantum corrections, they represent independent parameters related to scalar field fluctuations on the wall \cite{Garcia-Compean:2024zjr}). However, in $f(R)$ theories, the separability condition in \eqref{eq:TenSep} may not be satisfied. If it can be fulfilled, we will consider such terms in the same way as tension terms. On the other hand, if we can not fulfill this condition, the integral has to be performed for a particular ansatz of the potential and it will therefore be model dependent, we will try to avoid this scenario as much as possible.
\subsection{Quantum cosmology for FLRW metrics in \texorpdfstring{$f(R)$}{f(R)}}
In this work we consider  $f(R)$ modified gravity theories coupled to a canonical external scalar field with potential $V(\phi)$ that has a false and a true minimum. In natural units ($c = 1$, $8\pi G = 1$), the complete action takes the form
        \begin{equation}
            S = \frac{1}{2} \int d^4x \sqrt{-g} f(R) - \int d^4x \sqrt{-g} \left( \frac{1}{2} g^{\mu\nu} \partial_\mu \phi \partial_\nu \phi + V(\phi) \right).
        \label{eq:action}
        \end{equation}
We will focus on the study of the flat and closed FRLW metrics which are written in general as
        \begin{equation}
            ds^2 = -N^2(t)dt^2 + a^2(t)dX^{3} ,
        \end{equation}
where $N(t)$ is the lapse function and $dX^{3}$ represents the metric of the $3$-sphere in the closed case and the $3$-Euclidean plane in the flat case. In this way we are led to the Hamiltonian constraint 
        \begin{equation}
            H = N\left[ \frac{-\pi_a \pi_R}{3 f_{RR} a^2} + \frac{1}{3} \frac{f_R}{f_{RR}^2} \frac{\pi_R^2}{a^3} + \frac{\pi_\phi^2}{2  a^3} - \frac{a^3}{2}\left(f - R f_R\right) + a^3 V - 3 \kappa a f_R \right] \approx 0,
            \label{eq:FPHamilConst}
        \end{equation}
where $\kappa$ stands for the curvature parameter for flat or closed spatial geometries with values $\{0,1\}$, respectively; and the scale factor $a$ and the Ricci curvature scalar $R$ are treated as independent variables of the minisuperspace.\footnote{See Appendix \ref{FLRW-MetricApp} for a detailed derivation.}. Comparing with the general Hamiltonian constraint (\ref{eq:HamConst}), we observe its quadratic dependence on the momenta. This allows the identification of the minisuperspace defined by the coordinates $\{\Phi^M\} = \{a, R, \phi\},$ with the metric
        \begin{equation}
            (G^{MN}) = \begin{pmatrix}
            0 & -\frac{1}{3 f_{RR} a^2} & 0 \\
            -\frac{1}{3 f_{RR} a^2} & \frac{2}{3} \frac{f_R}{f_{RR}^2 a^3} & 0 \\
            0 & 0 & \frac{1}{a^3}
            \end{pmatrix}
            \label{eq:gmnpositive}
        \end{equation}
and the function
        \begin{equation}
            K = \frac{a^3}{2}\left[2 V - \left(f - R f_R\right)\right] - 3 \kappa f_R a,
        \label{eq:ppositive}
        \end{equation}
which are the starting point for the application of the formalism. 
\section{Transitions with a flat FLRW metric}
\label{S-Flat}
Having established the theoretical framework for calculating quantum transition probabilities, we now proceed to apply this methodology to the spatially flat FLRW metric, which serves as the simplest background describing the large-scale structure and evolution of the homogeneous and isotropic universe with spatially flat slices. Thus from the expression (\ref{eq:FPHamilConst}), taking the case $\kappa=0$ we obtain that for any $f(R)$, the auxiliary function takes the form
        \begin{equation}
             C^2(s)=\frac{2 \operatorname{Vol}^2(X) R\left(3 f-6 V_{A, B}-2 R f_R\right)}{3\left(f-2 V_{A, B}-R f_R\right)}.
        \end{equation}
Following the general approach we obtain
        \begin{equation}
            \frac{d R}{d s}=\frac{\left(-f+R f_R+2 V\right)\left(-3 f+6 V+R f_R\right)}{2 \operatorname{Vol}(X) f_{R R}\left(-3 R f+6 R V+2 R^2 f_R-6 V^{(1) 2}\right)},
            \label{eq:dRdsflat}
        \end{equation}
         \begin{equation}
            \frac{da}{ds}=\frac{a\left(f-2 V_{A, B}-R f_R\right)}{2 \operatorname{Vol}^2(X)\left(3 f-6 V_{A, B}-2 R f_R+6 V^{(1) 2}\right)},
            \label{eq:dadsflat}
        \end{equation}
where we can relate $a$ and $R$ in the form
        \begin{equation}
            \frac{da}{dR}=\frac{a R f_{R R}}{\left(-3 f+6 V+R f_R\right)}.
            \label{eq:dadRFlat}
        \end{equation}
Let us point out that although we have started with $a$ and $R$ as independent variables of minisuperspace, the WDW approach has led us to a relation between these two variables, which can be supplemented with the classical definition of $R$ in terms of $a$, In this way, this approach gives us more information than in the GR case, which can allow us to obtain the actual form of the scale factor. Thus, the semiclassical contribution to the transition probability in (\ref{eq:gammazero}), takes the general form 
        \small \begin{equation}
            \begin{aligned}
            \Gamma_0 = & \mp \frac{\sqrt{6} \operatorname{Vol}(X) i}{\hbar}\left\{\int_{a_0}^{\bar{a}} a^2 \sqrt{\frac{-f+2 V_B+R f_R}{R\left(-3 f+6 V_B+2 R f_R\right)}}\left(-3 f+6 V_B+2 R f_R\right) da\right. \\
            & \left.-\int_{a_0}^{\bar{a}} a^2 \sqrt{\frac{-f+2 V_A+R f_R}{R\left(-3 f+6 V_A+2 R f_R\right)}}\left(-3 f+6 V_A+2 R f_R\right) da\right\} \\
            & +\frac{\operatorname{Vol}(X)}{\hbar} a^3 T_0 +\frac{2 \operatorname{Vol}(X) i}{\hbar} \int_{\bar{s}-\delta s}^{\bar{s}+\delta s} \frac{d s}{C(s)}\left[\frac{a^3}{2}\left(f-R f_R\right)-\left.\frac{a^3}{2}\left(f-R f_R\right)\right|_{\phi=\phi_A}\right], \\
            &
            \end{aligned}
            \label{eq:g0flat}
        \end{equation} \normalsize
where by using (\ref{eq:dadsflat}) we have written the integrals over the scale factor. However, by virtue of (\ref{eq:dadRFlat}), we can see that $R$ is functionally dependent upon the scale factor $a$ and the potential $V(\phi)$; that is, $R=R(a,V)$ (that we must find by solving such differential equation). Consequently, the term associated with $f(R)$ must be retained until its explicit functional form is determined for each specific $f(R)$ function. Furthermore, based on the structure of the initial term within the integral, which involves a non-constant potential, it is possible to define a tension term $T_0$. This term is identical to the one obtained in the GR \cite{Cespedes:2020xpn,Garcia-Compean:2024zjr}. Additionally, an extra tension can appear in the third line if the separability condition is fulfilled. On the other hand, we find
        \begin{equation}
            \nabla^2 K=V^{(2)}-R+\frac{f_R\left(f_{R R}+R f_{R R R}\right)}{3 f_{R R} ^2}.
        \end{equation}
Then, using (\ref{eq:gamma1}), we obtain that the first quantum correction takes the form
         \begin{equation}
            \begin{aligned}
            \Gamma_1=& \operatorname{Vol}(X)\left\{\int_{a_0}^{\bar{a}} \frac{\left[3 f_{R R}^2\left(V_B^{(2)}- R\right)+ f_R\left(f_{R R}+R f_{R R R}\right)\right]}{a R  f_{R R}^2} d a\right. \\
            & \left.-\int_{a_0}^{\bar{a}} \frac{\left[3 f_{R R}^2\left(V_A^{(2)}- R\right)+ f_R\left(f_{R R}+R f_{R R R}\right)\right]}{a R  f_{R R}^2} d a\right\} \\
            & +\frac{\operatorname{Vol}(X)}{2 } T_1 \\ & +\operatorname{Vol}^2 (X)\int_{\bar{s}-\delta s}^{\bar{s}+\delta s} \frac{d s}{C^2(s)}\left[\frac{f_R\left(f_{R R}+R f_{R R R}\right)}{3 f_{R R}^2}-R\right.\\ & \left.-\left.\frac{f_R\left(f_{R R}+R f_{R R R}\right)}{3 f_{R R}^2}\right|_{\phi=\phi_A}+\left.R\right|_{\phi=\phi_A}\right],
             \label{eq:g1flat}
            \end{aligned}
        \end{equation}
where $V^{(n)}$ denote the $n$-th derivative of the potential with respect to the scalar field. Once again the actions shall be evaluated for particular functions. Moreover, we have obtained the same tension term associated to the GR case, $T_{1}$, and another integral that may lead to a new tension term associated with the function $f(R)$ and its derivatives. Finally, for the second correction we have
       \begin{equation}
            \begin{aligned}
            \nabla^2\left(\nabla^2 K\right)= & \frac{2 f_R}{9 a^3 f_{R R}^4}\left[-3 R f_{R R R}^2+f_{R R}\left(f_{R R R}+2 R f_{R R R R}\right)\right]+\frac{V^{(4)}}{a^3} \\
            & +\frac{2 f_R^2}{9 a^3 f_{R R}^6}\left[6 R f_{R R R}^2-2 f_{R R} f_{R R R}\left(f_{R R R}+3 R f_{R R R R}\right)\right. \\ & \left.+f_{R R}^2\left(f_{R R R R}+R f_{R R R R R}\right)\right] ,
            \end{aligned}
        \end{equation}
and
        \begin{equation}
            \begin{aligned}
            {\left[\nabla\left(\nabla^2 K\right)\right]^2=} & \frac{2 f_R}{27 a^3 f_{R R}^4}\left(-2 f_{R R}+R f_{R R R}\right)^2+\frac{V^{(3) 2}}{a^3} \\
            & +\frac{4 R f_R^2}{27 a^3 f_{R R}^6}\left(-2 f_{R R}+R f_{R R R}\right)\left(-2 f_{R R R}^2+f_{R R} f_{R R R R}\right) \\
            & +\frac{2 R^2 f_R^3}{27 a^3 f_{R R}^8}\left(-2 f_{R R R}^2+f_{R R} f_{R R R R}\right)^2.
            \end{aligned}
        \end{equation}
Thus, the expression for $\Gamma_2$ becomes        
        \begin{equation}
            \begin{aligned}
            \Gamma_2 =\frac{i \hbar \operatorname{Vol}(X)}{2}&\left\{\int_{a_0}^{\overline{a}} \frac{\left(9 f_{R R}^6 V_B^{(4)}+U\right)}{\sqrt{6} a^4 R f_{R R}^6} \sqrt{\frac{-f+2 V_B+R f_R}{R\left(-3 f+6 V_B+2 R f_R\right)}} \right. \\
            & +\int_{a_0}^{\bar{a}} \frac{\left(27 f_{R R}^8 V_B^{(3) 2}+W\right)}{2 \sqrt{6} a^4 R  f_{R R}^8}\left[\frac{-f+2 V_B+R f_R}{R\left(-3 f+6 V_B+2 R f_R\right)}\right]^{3 / 2} \\
            & -\int_{a_0}^{\bar{a}} \frac{\left(9 f_{R R}^6 V_A^{(4)}+U\right)}{\sqrt{6} a^4 R f_{R R}^6} \sqrt{\frac{-f+2 V_A+R f_R}{R\left(-3 f+6 V_A+2 R f_R\right)}} \\
            & \left.-\int_{a_0}^{\bar{a}} \frac{\left(27 f_{R R}^8 V_A^{(3) 2}+W\right)}{2 \sqrt{6} a^4 R f_{R R}^8}\left[\frac{-f+2 V_A+R f_R}{R\left(-3 f+6 V_A+2 R f_A\right)}\right]^{3 / 2}\right\} \\
            & +\frac{\hbar \operatorname{Vol}(X)}{\bar{a}^3} T_2 \\
            & +\frac{i \hbar}{2} \int_{\bar{s}-\delta s}^{\bar{s} +\delta s} \frac{\operatorname{Vol}^3(X) d s}{C^3(s)}\left(\frac{1}{9 a^3 }\right)\left(\frac{U}{f_{R R}^6}-\frac{U}{f_{R R}^6}\left.\right|_{\phi=\phi_A}\right) \\
            & +\frac{i \hbar}{2} \int_{\bar{s}-\delta s}^{\bar{s} +\delta s} \frac{\operatorname{Vol}^5(X) d s}{C^5(s)}\left(\frac{1}{27a^3 }\right)\left(\frac{W}{f_{R R}^8}-\frac{W}{f_{R R}^8}\left.\right|_{\phi=\phi_A}\right),
            \end{aligned}
            \label{eq:g2flat}
        \end{equation}
where we have defined
        \begin{equation}
            \begin{aligned}
                U &= 2 f_{R} \left\{f_{RR}^2 \left[f_{R} (f_{RRRR}+f_{RRRRR} R)-3 f_{RRR}^2 R\right]
                \right.\\& \left. -2 f_{R} f_{RR} f_{RRR} (f_{RRR}+3 f_{RRRR} R)+6 f_{R} f_{RRR}^3 R+f_{RR}^3 (f_{RRR}+2 f_{RRRR} R)\right\},
            \end{aligned}    
        \end{equation}
         \begin{equation}
            \begin{aligned}
            W & = 2  f_R f_{R R}^4\left(-2 f_{R R}+R f_{R R R}\right)^2 \\ & +4 R  f_{R}^{2} f_{R R}^2\left(-2 f_{R R}+R f_{R R R}\right)\left(-2 f_{R R R}^2+f_{R R} f_{RRRR}\right) \\
            & +2 R^2  f_{R}^{3}\left(-2 f_{R R R}^{2}+f_{R R} f_{R R R R}\right)^2,
            \end{aligned}
        \end{equation}
and a tension term $T_{2}$ which coincides with the one appearing in GR. From the general results found for $\Gamma_0$, $\Gamma_1$ and $\Gamma_2$ we can see that the tension terms obtained in GR are obtained in the same way here, but we obtain new integrals that may lead to new tension terms if the separability condition is satisfied which will depend on the modified gravity model. Furthermore, we note that in principle the object $\operatorname{Vol}(X)$ is a divergent quantity for this metric. However we can always consider a compactified spatial slice so we can forget about this term which will be a general overall constant.
\subsection{Constant Ricci Scalar Scenario}
\label{S-ConsRS}
Before having to consider specific $f(R)$ functions, let us consider a simplified scenario, We can see, that a possible solution in the expression (\ref{eq:dRdsflat}) is to consider an approach where the Ricci scalar is constant (i.e., $dR=0$). This consideration, although restrictive, allows us to simplify the probability calculations. Additionally, based on symmetry arguments of the bounce solution, this simplification was considered in the Euclidean treatment for $f(R)$ theories and closed FLRW metrics in \cite{Salehian:2018yoq}. In this way, we will be interested in the situation in which $dR=0$ holds without imposing $da=0$ in the expression (\ref{eq:dRdsflat}) and (\ref{eq:dadRFlat}), this leads us to
        \begin{equation}
            -3 f+6 V+R f_{R}=0.
        \end{equation}
On the other hand, the Hubble parameter is defined as $H=\frac{\dot{a}}{a}$, and since we have the freedom to choose the lapse function, choosing $N=1$, it follows that for the flat FLRW metric
        \begin{equation}
            R=6\left[2 H^{2}+\dot{H}\right].
        \end{equation}
However, it can be shown that a constant Ricci scalar implies a constant Hubble parameter as well. In this way, $\dot{R}=0$, and then the expression
        \begin{equation}
            6 V+R f_{R}-\left.3 f\right|_{R=12 H^{2}}=0
            \label{eq:condflat}
        \end{equation}
is satisfied. Nevertheless, we cannot take this expression as defining a differential equation for $f(R)$ to be solved because the restriction of a constant $R$ has already been made. Rather, it is an algebraic equation whose solution will give us the constant Hubble parameter given a particular form of the function $f$. Moreover, in this case the relationship between $R$ and $a$ is completely determined. Furthermore, the solution of this equation will tell us how $R$ depends on $V$ given a particular $f(R)$ function, which will be needed to find the tension terms in each case. Consequently, the semiclassical contribution (\ref{eq:g0flat}) can be written as
        \begin{equation}
        \begin{aligned}
            \Gamma_0= & \left.\mp \frac{\operatorname{Vol}(X)}{\hbar} i\left[L\left(V_B\right)-L\left(V_A\right)\right] a^3\right|_{a_0} ^{\bar{a}} +\frac{\operatorname{Vol}(X)}{\hbar} \bar{a}^3 T_0 \\ & +\frac{2 \operatorname{Vol}(X) i}{\hbar} \int_{\bar{s}-\delta s}^{\bar{s}+\delta s} \frac{d s}{C(s)}\left[\frac{a^3}{2}\left(f-R f_R\right)-\left.\frac{a^3}{2}\left(f-R f_R\right)\right|_{\phi=\phi_A}\right] ,
        \end{aligned}
        \label{eq:Lfuncaux}
        \end{equation}
with
        \begin{equation}
            L\left(V_{A, B}, f(R),f_R\right)=\sqrt{\frac{2}{3}} \sqrt{\frac{-f+ R f_R +2 V_{A, B}}{R (-3 f+2 R f_R+6 V_{A, B})}} (-3 f+2  R f_R +6 V_{A, B}).
            \label{eq:LfuncR}
        \end{equation}
This expression allows to consider as initial value $a_0=0$, so we can have  access to a possible initial singularity. However, let us remark that since we have considered the simplification of a constant Ricci scalar, the factor of $a^{3}$ in the first part will just be a number which depends on the specific $f$ function. Furthermore, the term appearing in the last integral, that is $f-Rf_{R}$ is actually just a constant in the region where the scalar field is constant. On the other hand, in the region where the scalar field varies, its value can only depend on the scalar field, not on the scale factor. Therefore, we note that this term will always fulfill the separability condition, and thus, the last integral in the above expression will only lead to a tension term with the same dependence on the scale factor as in GR, thus we can finally write 
    \begin{equation}
            \Gamma_0= \mp \frac{\operatorname{Vol}(X)}{\hbar} i\left[L\left(V_B\right)-L\left(V_A\right)\right] \bar{a}^3+\frac{\operatorname{Vol}(X)}{\hbar} \bar{a}^3 \left( T_{0}+T^{f(R)}_{0}\right). 
        \label{eq:Lfunc}
    \end{equation}
Therefore, we can expect that the general behavior of the semiclassical probability to be the same as in GR, with the only modification of being increased or decreased in the middle region on a model dependent form.

In the same way following the general expressions we define
        \begin{equation}
            M\left(V_{A, B}\right)=\frac{3 f_{R R}^2\left(V_{A, B}^{(2)}- R\right)+ f_R\left(f_{R R}+R f_{R R R}\right)}{ R f_{R R}^2} \bigg|_{\phi=\phi_A,\phi_B},
        \end{equation}
then the first quantum correction is written as
        \begin{equation}
            \begin{aligned}
                \Gamma_1=&\operatorname{Vol}(X) \left\{\left.\left[M\left(V_B\right)-M\left(V_A\right)\right] \ln(a)\right|_{a_0} ^{\bar{a}}\right\} +\frac{\operatorname{Vol}(X)}{2 } T_1 \\ & +\operatorname{Vol}^2 (X)\int_{\bar{s}-\delta s}^{\bar{s}+\delta s} \frac{d s}{C^2(s)}\left[\frac{f_R\left(f_{R R}+R f_{R R R}\right)}{3 f_{R R}^2}-R\right.\\ & \left.-\left.\frac{f_R\left(f_{R R}+R f_{R R R}\right)}{3 f_{R R}^2}\right|_{\phi=\phi_A}+\left.R\right|_{\phi=\phi_A}\right].
            \end{aligned}
        \end{equation}
In order to avoid the divergence at $a_0 = 0$, and  keep exploring this region, we impose the matching condition $M(V_A) = M(V_B)$, which imposes conditions over the second derivatives of the potential in the false and true vacuum regions. Furthermore, for the integral over the wall, we note that the argument used for $\Gamma_{0}$ follows in the same way and then we can substitute the overall integral just by a tension term. Thus, we obtain a very similar result than in GR in the form
        \begin{equation}
                \Gamma_1= \frac{\operatorname{Vol}(X)}{2 } \left(T_1+T^{f(R)}_1\right) , 
            \label{eq:G1Rcons}
        \end{equation}
with the condition 
        \footnotesize
        \begin{equation}
            \frac{3 f_{R R}^2\left(V_A^{(2)}- R\right)+ f_R\left(f_{R R}+R f_{R R R}\right)}{ R f_{R R}^2}\bigg|_{\phi=\phi_A}=\frac{3 f_{R R}^2\left(V_B^{(2)}- R\right)+ f_R\left(f_{R R}+R f_{R R R}\right)}{ R f_{R R}^2}\bigg|_{\phi=\phi_B}.
            \label{eq:flatg1constraint}
        \end{equation} \normalsize
Let us remark that this kind of condition is also needed for the same reason in GR\cite{Garcia-Compean:2024zjr}, the only thing that has changed is the specific form of the condition. Finally we define
        \begin{equation}
            Q(V_{A,B},V_{A,B}^{(n)})=\frac{\left(9 f_{R R}^6 V_{A,B}^{(4)}+U\right)}{\sqrt{6} R  f_{R R}^6} \sqrt{\frac{-f+2 V_{A,B}+R f_R}{R\left(-3 f+6 V_{A,B}+2 R f_R\right)}}\bigg|_{\phi=\phi_A,\phi_B} ,
        \end{equation}
and
        \begin{equation}
            S(V_{A,B},V_{A,B}^{(n)})=\frac{\left(27 f_{R R}^8 V_{A,B}^{(3) 2}+W\right)}{2 \sqrt{6} R f_{R R}^8}\left[\frac{-f+2 V_{A,B}+R f_R}{R\left(-3 f+6 V_{A,B}+2 R f_R\right)}\right]^{3 / 2}\bigg|_{\phi=\phi_A,\phi_B}.
        \end{equation}
Thus, the expression for the second quantum correction in (\ref{eq:g2flat}) can be written as
        \begin{equation}
            \begin{aligned}
            \Gamma_2= \pm \frac{i \hbar \operatorname{Vol}(X)}{2}\bigg\{ & {\left.\left[Q\left.\right|_{\phi=\phi_B}-Q\left.\right|_{\phi=\phi_A}\right]\left(\frac{1}{3 a^3}\right)\right|_{a_0} ^{\bar{a}}}  \\
            & \left.+\left.\left[S\left.\right|_{\phi=\phi_B}-S\left.\right|_{\phi=\phi_A} \right]\left(\frac{1}{3 a^3}\right)\right|_{a_0} ^{\bar{a}}\right\} \\
            & +\frac{\hbar \operatorname{Vol}(X)}{\bar{a}^3} T_2 \\
            & +\frac{i \hbar}{2} \int_{\bar{s}-\delta s}^{\bar{s} +\delta s} \frac{\operatorname{Vol}^3(X) d s}{C^3(s)}\left(\frac{1}{9 a^3 }\right)\left(\frac{U}{f_{R R}^6}-\frac{U}{f_{R R}^6}\bigg|_{\phi=\phi_A}\right) \\
            & +\frac{i \hbar}{2} \int_{\bar{s}-\delta s}^{\bar{s} +\delta s} \frac{\operatorname{Vol}^5(X) d s}{C^5(s)}\left(\frac{1}{27a^3 }\right)\left(\frac{W}{f_{R R}^8}-\frac{W}{f_{R R}^8}\bigg|_{\phi=\phi_A}\right).
            \end{aligned}
        \end{equation}  
Analogously to the assumptions for the first quantum correction and to the scenario in GR, to implement the condition $a_0 = 0$, we need to impose the constraints 
        \begin{equation}
            Q_A +S_A = Q_B + S_B.
            \label{eq:flatg2constraint}
        \end{equation}
Moreover, for the two last integrals we note the same scenario than in the semiclassical and first quantum correction, ans thus we can just redefine a new tension term analogously to the GR case. Thus we finally get
       \begin{equation}
            \Gamma_2= \frac{\hbar \operatorname{Vol}(X)}{\bar{a}^3} \left(T_{2}+T_2^{f(R)}\right) .
            \label{eq:G2Rconst}
        \end{equation}  
Therefore, we have obtained that the transition probabilities within the semiclassical and quantum corrected contributions behave in the same general way as GR. This arises because the overall dependence on the scale factor is the same for all $f(R)$ models, for the integrals and for the tension terms. The only difference will be given by the numerical factors on each term and the specific conditions upon the potential function and its derivatives in order to safely access the initial singularity which will be model dependent. Therefore, within this approximation, the general behavior encountered in \cite{Garcia-Compean:2024zjr} is preserved, that is the semiclassical contribution will start at 1 and decay to zero, the first quantum corrected solution will start in a lower value and fall in the same way, and the singularity will be avoided when taking into account the second quantum correction. The only difference will appear in the intermediate region where the probability may decay faster or slower than in GR depending on the specific model under consideration. We now move on to obtain specific expressions for particular proposals.
\subsubsection{\texorpdfstring{$f(R)=R^{1+n}$}{f(R)=R\textasciicircum(1+n)}}
\begin{figure}[ht]
            \centering
            \includegraphics[width=\textwidth]{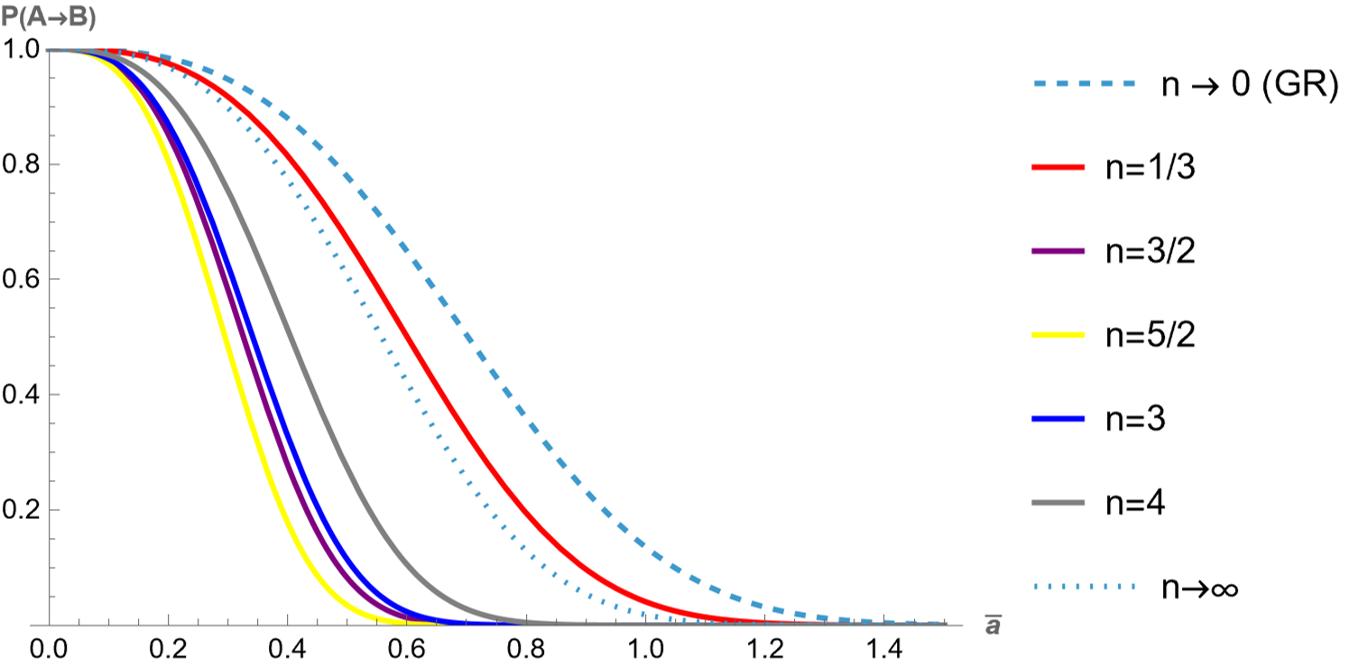}
                \caption{ Semiclassical contribution to the transition probability for $f(R)=R^{1+n}$ for the GR Result $n\to 0 $ (Dashed line), $n=1/3$ (Red line), $n=3/2$ (Purple line), $n=5/2$ (Yellow line), $n=3$ (Blue line), $n=4$ (Gray line) and $n\to \infty$ (Dotted line) . With the parameters $V_{B}=5$, $V_{A}=10$, $\operatorname{Vol}(X)=1$ and $T_{0}=1$.}
            \label{fig:RnG0PP}
        \end{figure}

As our initial study case, we examine the power-law $f(R)$ gravity model given by $f(R) = R^{1+n}$.  This model is of great interest for  phenomenological reasons (see for example \cite{Carloni:2004kp,Allemandi:2004ca,Carloni:2005ii,Leach:2006br}). Considering the expression (\ref{eq:condflat}), we obtain for the Hubble parameter
        \begin{equation}
           H=\frac{1}{2\sqrt{3}}\left(\frac{6V}{2-n}\right)^{\frac{1}{2(1+n)}}.
        \end{equation}
In this way, we note that we must exclude the value $n=2$. We then finally obtain for the semiclassical contribution          
        \begin{equation}
                \Gamma_0 = \mp \frac{\operatorname{Vol}(X)}{\hbar} i\left[L\left(V_B\right)-L\left(V_A\right)\right] \bar{a}^3 +\frac{\operatorname{Vol}(X)}{\hbar} \bar{a}^3 \left[ 1+ \left( \frac{3n}{2-n} \right) \right]T_0,
            \label{eq:g0Rdelta}
        \end{equation}
where in this case it is possible to write $T_{0}^{f(R)}$ in terms of $T_{0}$, moreover in this case
    \begin{equation}
            L(V_{A,B})=\frac{2(n+1)}{3}\left(\frac{6V}{2-n}\right)^\frac{2n+1}{2(n+1)}
        \end{equation}
We note from (\ref{eq:g0Rdelta}) that the new tension term induces a sign change in the vicinity of $n=2$. Notably, in the limit where $n \to 0$, this expression recovers the standard result for the semiclassical contribution from GR. At the opposite end, $n \to \infty$, the term accompanying the tension converges to $-2$, remaining finite. We must identify the correct combination of signs in the sign ambiguities that assures well defined probabilities depending of the values of $n$ and $V_{A,B}$ under consideration. First, given $V_{A,B}> 0$, there are two scenarios to consider based on the value of $n$.
\begin{itemize}
    \item For $0 < n < 2$, the function $L$ does not contribute to the expression. Consequently, the choice of sign in Eq. (\ref{eq:g0Rdelta}) is irrelevant.
    \item For $n > 2$, the function $L$ does contribute. In this regime, the negative sign in Eq. (\ref{eq:g0Rdelta}) must be chosen to ensure the probability is well defined.
\end{itemize}
For these reasons, in general we will keep the negative sign in this sign ambiguity. For the sign concerning (\ref{DefGammaGe}), we need to choose a positive sign. Even with the change of sign in the tension terms, causing the probability to increase, it remains bounded  even as $n$ approaches infinity. Conversely, if $V_{A,B}<0$, we need to choose a positive/negative sign in the exponential for the regions $0 < n < 2$ and $n>2$, respectively, and consider the negative sign in (\ref{eq:g0Rdelta}), for a well behaved probability for $n>0/\{2\}$. Additionally, in the limit $n \rightarrow \infty$ we have approximately
        \begin{equation}
        \Gamma_0 = \bar{a}^3 \frac{\operatorname{Vol}(X)}{\hbar} \left[\pm 4 i (V_A-V_B)-2T_{0}\right],
        \label{eq:InfR1+n}
        \end{equation}
which only have the tension term as contribution to the probability. Now, to depict the overall behavior of probability considering different values of $n$, in Figure \ref{fig:RnG0PP} we present the semiclassical contribution, specifically considering the scenario for positive potentials and choosing the conditions $V_{B}=5$, $V_{A}=10$, Vol$(X)$, $\hbar=1$ and $T_{0}=1$.
        
\begin{figure}[ht]
            \centering
            \includegraphics[width=\textwidth]{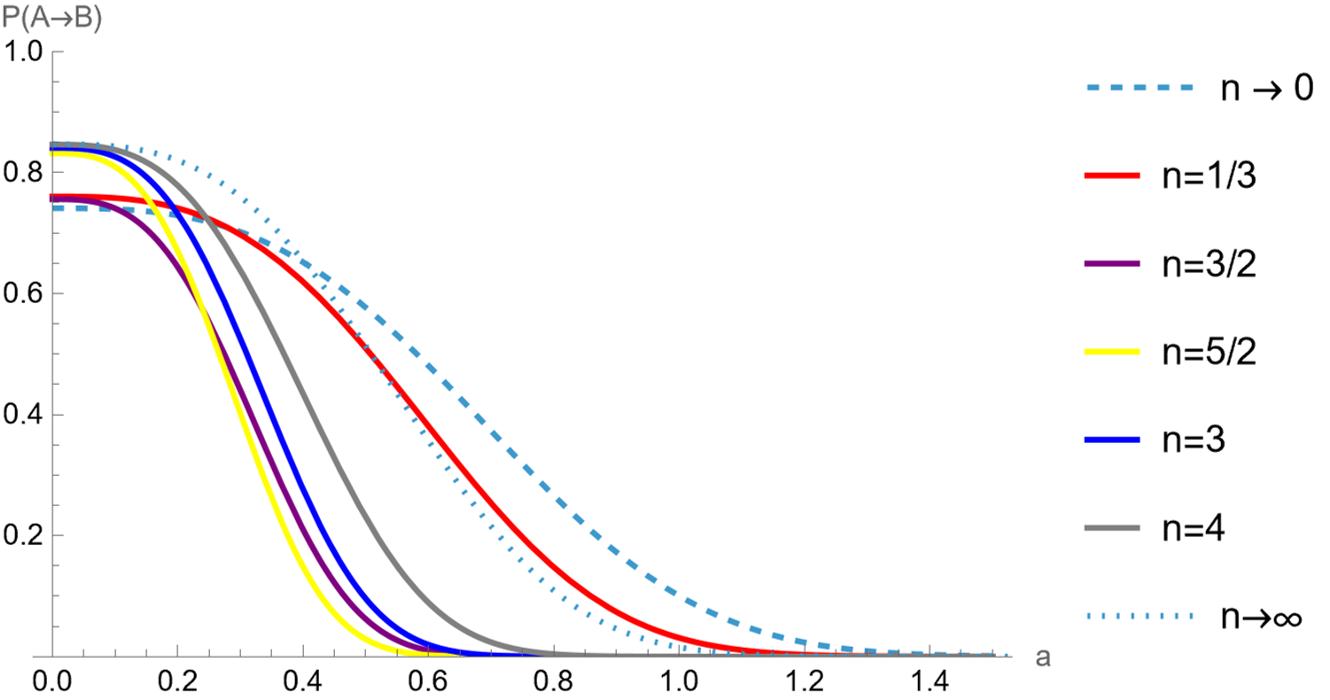}
                \caption{First Quantum Correction to the transition probability $\Gamma_0 + \Gamma_1$ for $f(R)=R^{1+n}$ using  the GR result $n \to 0 $ (Dashed line), $n=1/3$ (Red line), $n=3/2$ (Purple line), $n=5/2$ (Yellow line), $n=3$ (Blue line), $n=4$ (Gray line) and $n\to\infty$ (Dotted line). With the parameters $V_{B}=5$, $V_{A}=10$, $T_{0}=1$, $T_{1}=0.1$, and $T_{1}^{f(R)}=0.05$.}
            \label{fig:RnG1PP}
        \end{figure}

As illustrated in the figure, for $0<n<2$ the probability decreases faster as we approach $n=2$. However, for $n > 2$, where the change of sign in the tension occurs, the probability decays slower as the scale factor increases. Upon analyzing the limits of small and big values for the scale factor, it is observed that the calculated probabilities for the $f(R)=R^{1+n}$ model consistently remain within the bounds defined by the GR case. In these limit scenarios, considering positive potentials, the sole contribution arises from the tension term. Consequently, the primary distinction is found in the factor of $-2$ associated with the tension term in expression (\ref{eq:InfR1+n}); accounting for the appropriate sign convention, this contribution is double that observed in GR. Moving forward with the first quantum correction, in this case we find
         \begin{equation}
            \Gamma_1= \frac{\operatorname{Vol}(X)}{2 } T_1 + \operatorname{Vol}(X) \left(\frac{2^{2+n}}{3^n}\right)^{\frac{1}{1+n}}\left(\frac{1}{2-n}\right)^{\frac{1}{1+n}} T^{f(R)}_1,
            \label{eq:g1Rdelta}
        \end{equation}
with the restriction
            \begin{equation}
                 V_{B}^{(2)} \left(-\frac{V_B}{n -2}\right)^{-\frac{1}{n +1}}=V_{A}^{(2)} \left(-\frac{V_A}{n -2}\right)^{-\frac{1}{n +1}},
                 \label{eq:R1+ng1constraint}
            \end{equation}
in order to be able to consider $a_0=0$. It must be noted that in the limit $n \to 0$, the condition imposed upon the potential and its derivatives is precisely the one in GR. In this expression we also have that the new tension term depends on $n$ and changes sign around the critical point $n=2$. Furthermore, the expression remains finite in the limit $n \to \infty$, giving
        \begin{equation}
            \lim _{n \rightarrow \infty} \operatorname{Vol}(X) \left(\frac{2^{2+n}}{3^n}\right)^{\frac{1}{1+n}}\left(\frac{1}{2-n}\right)^{\frac{1}{1+n}} T^{f(R)}_1 = \frac{2}{3} \operatorname{Vol}(X) T^{f(R)}_1. 
        \end{equation}
In this manner, Figure \ref{fig:RnG1PP} shows the sum of the semiclassical contribution and the first quantum correction, considering positive potentials using the parameter values $V_{B}=5$, $V_{A}=10$, $T_{0}=1$, $T_{1}=0.1$, and $T_{1}^{f(R)}=0.05$. We can see that the maximum value of the probability is lower in the range $0 < n < 2$ than it is for $n > 2$, but both below $1$. This effect is attributed to a sign change in the term derived from $f(R)$.
        
\begin{figure}[ht]
            \centering
            \includegraphics[width=\textwidth]{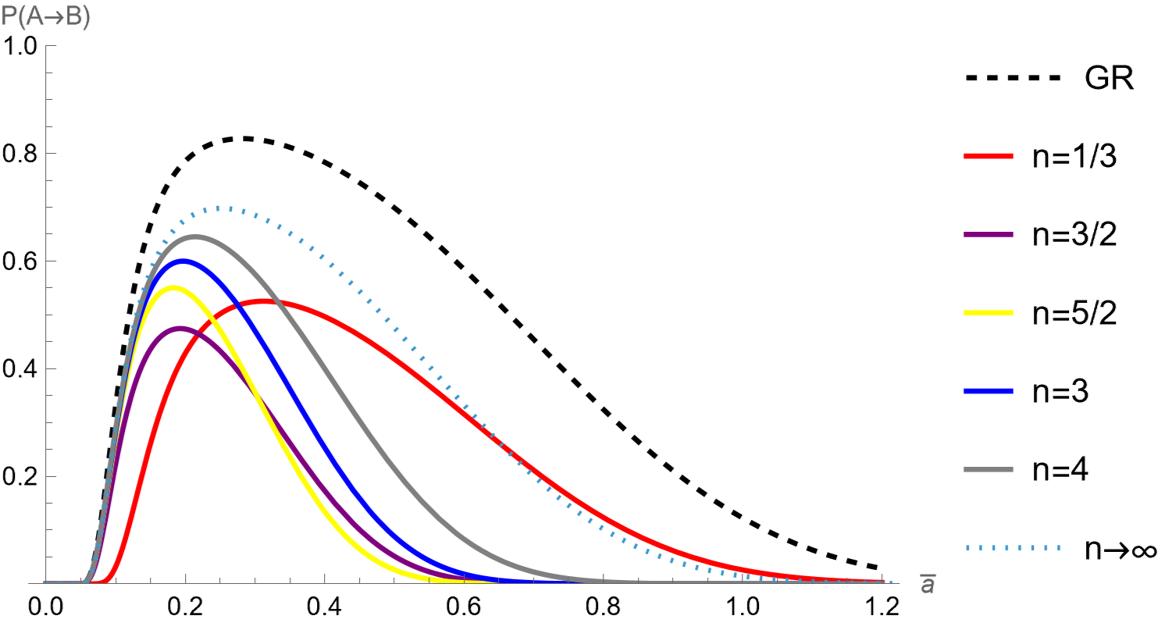}
                \caption{ Second Quantum Correction to the transition probability $\Gamma_0 + \Gamma_1+\Gamma_2$ for $f(R)=R^{1+n}$ using $n=0.05 $ (Dashed line), $n=1/3$ (Red line), $n=3/2$ (Purple line), $n=5/2$ (Yellow line), $n=3$ (Blue line),  $n=4$ (Gray line) and $n\to\infty$ (dotted line) and the GR result (Black dashed line). With the parameters $V_{B}=5$, $V_{A}=10$, $T_{0}=1$, $T_{0}^{f(R)}=1$, $T_{1}=0.1$, $T_{1}^{f(R)}=0.05$, $T_{2}=10^{-3}$, and $T_{2}^{f(R)}=10^{-5}$.}
            \label{fig:RnG2PP}
        \end{figure}

Finally, for the second quantum correction in \eqref{eq:G2Rconst}, we obtain
    \begin{equation}
        \Gamma_2 = \frac{\hbar \operatorname{Vol}(X)}{\bar{a}^3} T_2 + \frac{\hbar \operatorname{Vol}(X)}{\bar{a}^3} \frac{8\left[\frac{-6}{(n-2)}\right]^{\frac{2-n}{n+1}}}{27 n^2(n+1)} T_2^{f(R)},
        \label{eq:g2Rdelta}
    \end{equation}
where the condition (\ref{eq:flatg2constraint}) is satisfied with
        \begin{equation}
            Q\left(V_{A, B}\right)=3 V_{A, B}^{(4)} \left(\frac{2-n}{6 V_{A, B}}\right)^{\frac{3}{2 n+2}} 
        \end{equation}
and 
        \begin{equation}
            S\left(V_{A, B}\right)=  3V_{A, B}^{(3) 2}\left(\frac{2-n}{6 V_{A, B}}\right)^{\frac{5}{2(n+1)}} +\frac{8}{9 n^2(n+1)}\left(\frac{2-n}{6 V_{A, B}}\right)^{\frac{2 n+1}{2(n+1)}}.
        \end{equation}
Of particular significance is the coefficient of the new tension term $T_2^{f(R)}$, which exhibits a more difficult dependence on the power-law exponent $n$. Unlike the semiclassical and first contributions, this term does not admit a well defined $n \to 0$ limit that would recover the corresponding GR results, indicating that the second quantum correction term presents fundamentally different quantum properties of the minisuperspace model and does not fulfill the standard limiting relation valid in a classical setting. A similar behavior was encountered in \cite{Garcia-Compean:2024zjr} for the comparison between anisotropic and isotropic metrics. In Figure \ref{fig:RnG2PP} we show the transition probability incorporating the second quantum correction for the selected $f(R)$ model and GR. The calculation was performed considering the following set of physical parameters: $V_{B}=5$, $V_{A}=10$, $T_{0}=1$, $T_{1}=0.1$, $T_{1}^{f(R)}=0.05$, $T_{2}=5\times10^{-4}$, and $T_{2}^{f(R)}=10^{-4}$. The probability distribution behaves in such a way that as $n$ approaches zero, the transition probability decreases due to the divergent behavior of the $f(R)$ dependent tension term. It is observed that the maximum value of the transition probability increases as the parameter $n$ does; nevertheless, this probability remains bounded, as the tension term associated with $f(R)$ asymptotically approaches zero in the limit $n\to \infty$, and similarly, the probability distribution of the modified models remain bounded within the GR result.
We want to emphasize that the $n$ dependent coefficient multiplying the modified gravity tension term $T_2^{f(R)}$ shows distinct behavior across different intervals. Specifically, when $n$ lies in the interval $(0,2)$, this coefficient acquires real values throughout the entire range. On the other hand, for all values $n>2$, the coefficient becomes complex, and thus this tension term is in general ill defined and does not contribute, finally it vanishes as $n\to\infty$, thus giving the highest probability for this limit.
\subsubsection{Models with \texorpdfstring{$R^{2}$}{R\textasciicircum(2)} terms}
\begin{figure}[ht]
            \centering
            \includegraphics[width=\textwidth]{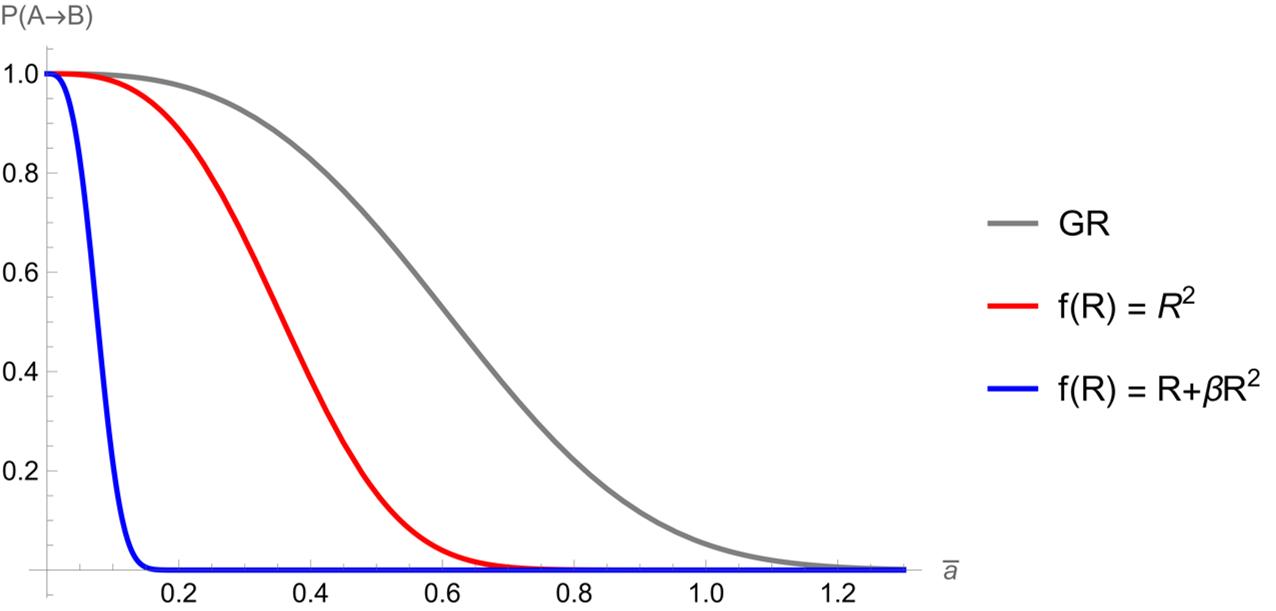}
                \caption{Semiclassical contribution for the transition probability for the models $f(R)=R^2$ (Red line), $f(R)=R+\beta R^2$ (Blue line) and GR (Gray line), with parameters $V_B = -5$, $V_A = -1$, $\beta=10^{3}$, $T_0 = 1$, and $T_0^{f(R)} = 0.5$.}
            \label{fig:StaroR2G0}
        \end{figure}

Among the $f(R)$ theories of phenomenological interest, the Starobinsky model \cite{Starobinsky:1980te} stands as one of the earliest and most compelling formulations of cosmic inflation. In this framework, the accelerated expansion of the early universe is driven by quantum corrections to the Einstein–Hilbert action, specifically through the inclusion of a term quadratic in the Ricci scalar, of the form $R + \beta R^2$ \cite{Linde:2025pvj}. This model is distinguished by its compatibility with Cosmic Microwave Background (CMB) observations, particularly regarding the measured scalar spectral index and the low tensor-to-scalar ratio \cite{Drees:2025ngb}. Furthermore, it can be conformally mapped to a scalar-tensor theory, demonstrating its equivalence to a single-field inflationary model characterized by a plateau-like potential , which is in good agreement with Planck satellite data \cite{Planck:2018jri}. Consequently, the Starobinsky model remains a benchmark in inflationary cosmology, underpinned by its theoretical motivation derived from quantum gravity effects and its robust observational support (for further discussion, see \cite{Vilenkin:1985md,Stachowski:2016zio,German:2023euc,Ketov:2025nkr,SidikRisdianto:2025qvk,Ivanov:2021chn}). Motivated by this, in this section, we examine the Starobinsky model, defined by $f(R) = R + \beta R^2$. Utilizing the results derived in the preceding section, we will compare this formulation with the $f(R)=R^{2}$ (which corresponds to the $n=1$ case of the previous results) and GR. For this, we consider the condition given in (\ref{eq:condflat}) and solve the Hubble parameter in the aforementioned models. This yields the following expressions
        \begin{equation}
           f(R)=R+\beta R^2 \Rightarrow H=\sqrt{\frac{\sqrt{1+6 V \beta}-1}{12\beta}}.
        \end{equation}
With this solution (\ref{eq:LfuncR}) is consequently written as 
        \begin{equation}
            L_{R+\beta R^{2}}(V_{A,B})= -\frac{2 \sqrt{\frac{\beta }{\sqrt{6 \beta  V_{A,B}+1}-1}} \left(-4 \beta  V_{A,B}+\sqrt{6 \beta  V_{A,B}+1}-1\right)}{\beta }.
        \end{equation}
For this model the semiclassical contribution is expressed as
        \begin{equation}
            \begin{aligned}
            R+\beta R^2: \Gamma_0= & \mp \frac{\operatorname{Vol}(X)}{\hbar} i\left[L_{R+\beta R^2}\left(V_B\right)-L_{R+\beta R^2}\left(V_A\right)\right] \bar{a}^3 +\frac{\operatorname{Vol}(X)}{\hbar} \bar{a}^3 T_0 \\ & +\frac{\operatorname{Vol}(X)}{\hbar}\frac{\bar{a}^3}{\beta} T_{0}^{f(R)}.
            \end{aligned}
        \end{equation}
To contrast this with a purely quadratic term, we set $n=1$ in the corresponding expression from the preceding section, yielding
        \begin{equation}
            R^2: \Gamma_0= \mp \frac{\operatorname{Vol}(X)}{\hbar} i\bigg[L_{R^2}\left(V_B\right)-L_{R^2}\left(V_A\right)\bigg] \bar{a}^3 +\frac{\operatorname{Vol}(X)}{\hbar} \bar{a}^3 T_0,
        \end{equation}
where $L_{R^{2}}(V_{A,B})=\frac{4\ 2^{3/4} V_{A,B}^{3/4}}{3^{1/4}}$. To highlight the contrasting behaviors of the two models, with the GR case, we present in Figure \ref{fig:StaroR2G0} the semiclassical contribution for negative potential scenarios. Specifically, we adopt the parameter values\footnote{In the context of the Starobinsky model, the coupling parameter $\beta$ behaves as $\beta\sim 1 / (6 M^2)$, where $M$ denotes the mass of the inflaton scalar field, which typically lies within the range $M \sim 1.3 \times 10^{-5} M_{Pl}$. Here, $M_{Pl}$ represents the Planck mass. Within the specific set of units established in this work, namely $\hbar=c=8 \pi G=1$, and considering the definition $M_{Pl}=\sqrt{\frac{\hbar c}{8\pi G}}$, the parameter $\beta$ is projected to be in the range $\sim 10^{9}$ \cite{Drees:2025ngb}. Consequently, large numerical values for this parameter shall be considered in the subsequent analysis.} $V_B = -5$, $V_A = -1$, $T_0 = 1$, $T_0^{f(R)} = 0.5$ and $\beta=10^{3}$, while choosing positive/negative signs for $\Gamma$ and the constant field region, respectively, to ensure a well defined probability. As we can see, in Starobinsky model, the probability decays faster compared to $R^2$. For the case of positive potentials, there is no contribution from the $L$ function in any model, so the only difference lies in the additional tension term present in Starobinsky. 
Furthermore, it is observed that both quadratic cases exhibit probability distributions that decay more rapidly than those predicted by GR thereby indicating that the transitions nucleate universes with smaller sizes in both cases. 

\begin{figure}[h!]
            \centering
            \includegraphics[width=\textwidth]{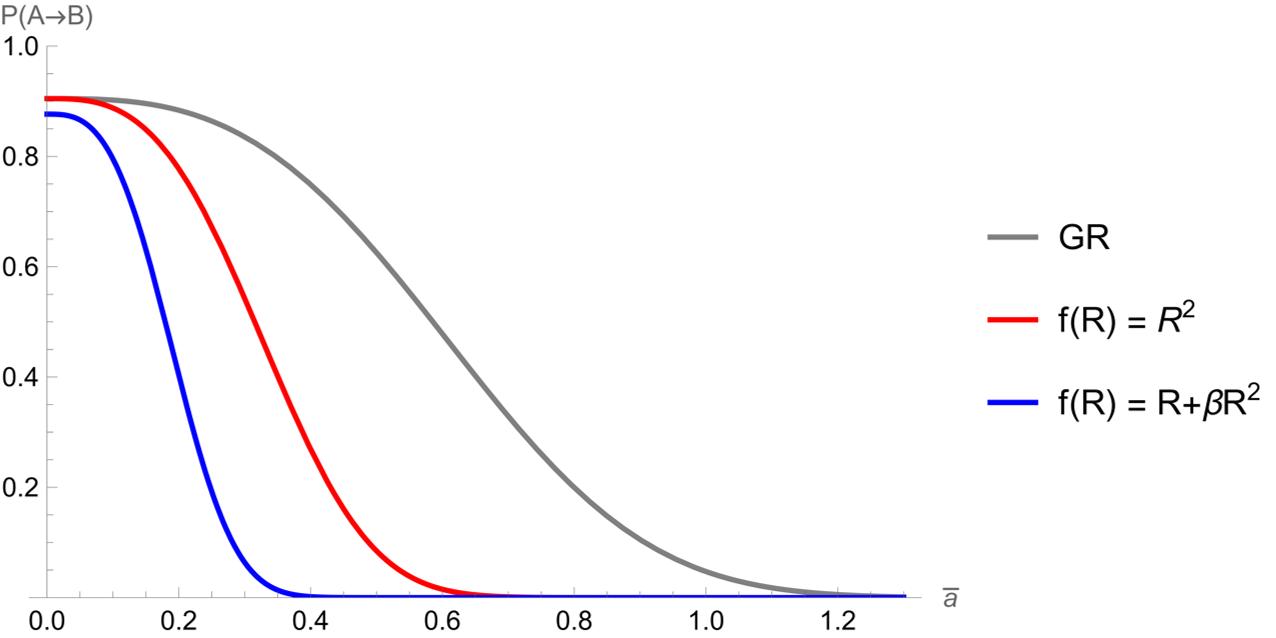}
                \caption{ First quantum correction for the transition probability $\Gamma_{0}+\Gamma_{1}$ for the models $f(R)=R^2$ (Red line), $f(R)=R+\beta R^2$ (Blue line) and GR (Gray line) with parameters$V_B = -5$, $V_A = -1$, $T_0 = 1$, $T_0^{f(R)} = 0.5$, $T_1=0.1$ and $T_1^{f(R)}=0.5$.}
            \label{fig:StaroR2G1}
        \end{figure}

Considering the first quantum correction as given in (\ref{eq:G1Rcons}) we obtain for both cases
        \begin{equation}
            R+\beta R^2 : \Gamma_1= \frac{\operatorname{Vol}(X)}{2 } T_1 +\frac{\operatorname{Vol}(X)}{\sqrt{\beta}} T_{1}^{f(R)},
        \end{equation}
and
        \begin{equation}
            R^2 : \Gamma_1= \frac{\operatorname{Vol}(X)}{2 } T_1.
        \end{equation}
Now, taking into account this first correction, we show the transition probability in Figure \ref{fig:StaroR2G1}, considering $V_B = -5$, $V_A = -1$, $T_0 = 1$, $T_0^{f(R)} = 0.5$, $T_1=0.1$, $T_1^{f(R)}=0.5$ preserving the same sign convention, we can see that the qualitative difference between models observed in the semiclassical contribution is preserved, thus the Starobinsky model continues to exhibit a more rapid probability decay compared to the $R^2$ model and leaving the additional Starobinsky tension term as a distinguishing feature between the models.
        
Following with the analysis of the second order quantum corrections, we obtain 
        \begin{equation}
            R+\beta R^2: \Gamma_2= +\frac{\hbar \operatorname{Vol}(X)}{\bar{a}^3} T_2 + +\frac{\hbar \operatorname{Vol}(X)}{\bar{a}^3}\left(\frac{3+64 \beta^4}{432 \beta^6}\right)  T_{2}^{f(R)}.
        \end{equation}
For the $R^2$ case, substituting $n=1$ into (\ref{eq:g2Rdelta}), we obtain
        \begin{equation}
            R^2: \Gamma_2= \frac{\hbar \operatorname{Vol}(X)}{\bar{a}^3} T_2.
        \end{equation}
Finally, with the inclusion of the second order quantum correction, in Figure \ref{fig:StaroR2G0}, considering the parameters $V_B = -5$, $V_A = -1$, $T_0 = 1$, $T_0^{f(R)} = 0.5$, $T_1=0.1$ and $T_1^{f(R)}=0.5$, $T_2=0.001$ and $T_{2}^{f(R)}=5\times 10^{-4}$, we represent this correction for both models and GR. The characteristic general behavior of the Starobinsky model, which exhibits a lower maximal probability, is preserved.
        \begin{figure}[h!]
            \centering
            \includegraphics[width=\textwidth]{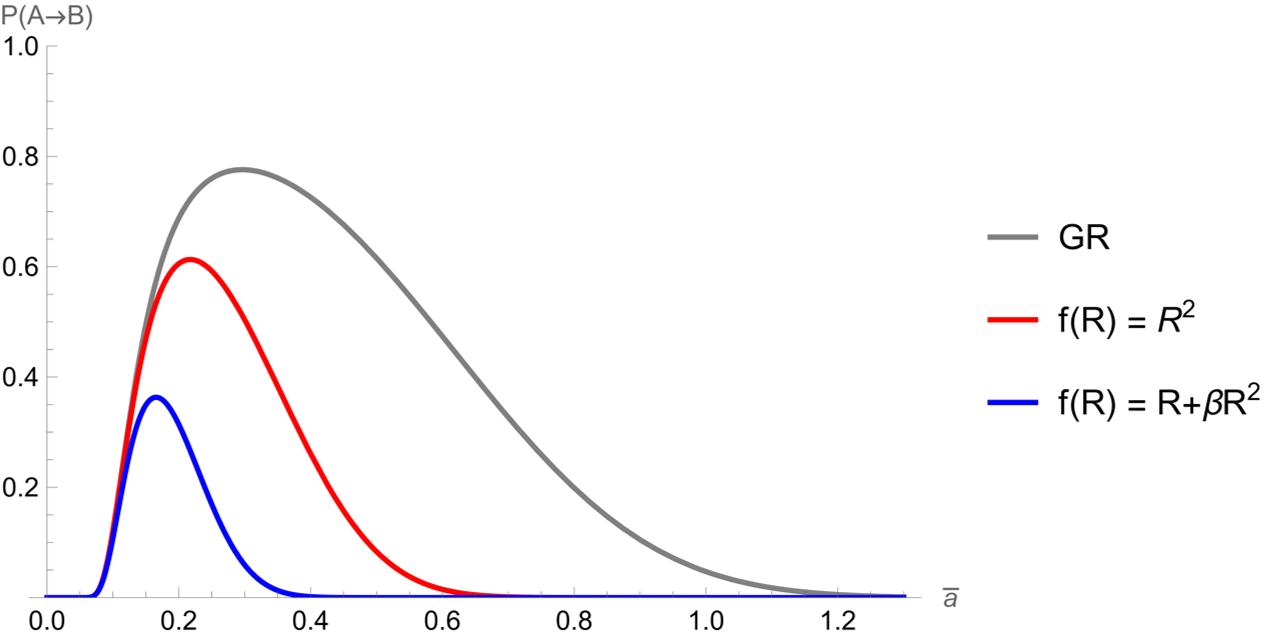}
                \caption{ Second quantum correction for the transition probability $\Gamma_{0}+\Gamma_{1}+\Gamma_{2}$ for the models $f(R)=R^2$ (Red line), $f(R)=R+\beta R^2$ (Blue line) and GR (Gray line) with parameters $V_B = -5$, $V_A = -1$, $T_0 = 1$, $T_0^{f(R)} = 0.5$, $T_1=0.1$ and $T_1^{f(R)}=0.5$, $T_2=0.001$ and $T_{2}^{f(R)}=5\times 10^{-4}$.}
            \label{fig:StaroR2G2}
        \end{figure}
\subsection{Non-constant Ricci Scalar}
\label{S-NonCR}
So far, we have shown that under the assumption of a constant Ricci scalar, the general behavior of all transition probabilities remains the same as in GR. However, within the Hamiltonian approach, this condition is a helpful simplification but it is not mandatory, thus we are allowed to abandon it to study more general scenarios. Therefore, in general there is no reason to expect that the integrals over the constant field regions exhibit the same dependence on the scale factor for all models. Instead, we need to compute the integrals and analyze the possible tension terms in the expressions (\ref{eq:g0flat}), (\ref{eq:g1flat}), and (\ref{eq:g2flat}), which may potentially yield more intricate dependencies of the scale factor. Moreover, the functional dependence of the curvature scalar $R$ on the scale factor will be dictated by the differential equation (\ref{eq:dadRFlat}). The ability to solve this equation analytically and the separability requirement of the integrals performed in the region where the scale factor varies will severely restrict the kind of models that we can safely study. Thus, in this section we study models of the general form $R^{1+n}$ that admit an analytical solution to the equation (\ref{eq:dadRFlat}), for a non-constant Ricci scalar.
\subsubsection{Case \texorpdfstring{$R^{1+n}$}{R\textasciicircum(1+n)}}
\label{NumR1+nFlat}
\begin{figure}[ht]
            \centering
            \includegraphics[width=\textwidth]{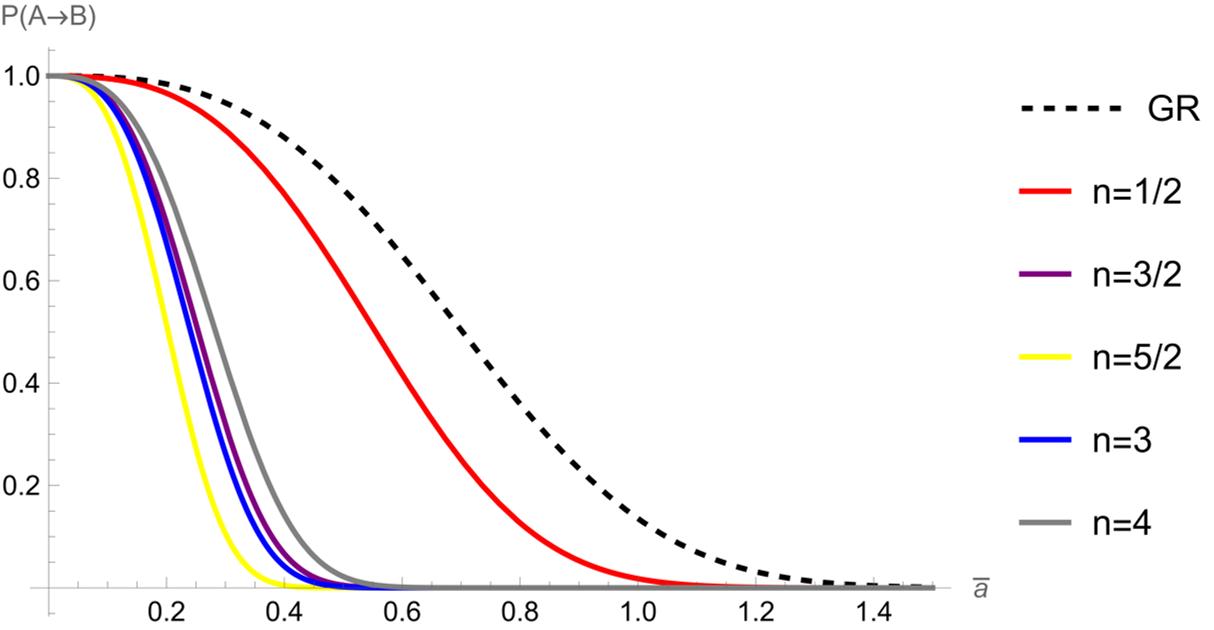}
                \caption{Semiclassical contribution for non-constant Ricci scalar in the $f(R)=R^{1+n}$ framework for $n=1/2$ (Red line), $n=3/2$ (Purple line), $n=5/2$ (Yellow line), $n=3$ (Blue line), $n=4$ (Gray line) and the GR case (Dashed Black line). With the parameters $V_{B}=5$, $V_{A}=10$, $\operatorname{Vol}(X)=1$, $\hbar=1$, $k=0.15$ and $T_{0}=1$.}
            \label{fig:NumR1+nG0}
        \end{figure}

For the $f(R)=R^{1+n}$ model, Eq. (\ref{eq:dadRFlat}) takes the form
        \begin{equation}
            \frac{da}{dR}=\frac{a n  (n +1) R^{n }}{  (n -2) R^{n +1}+6 V}.
            \label{eq:R1+ddadR}
        \end{equation}
The solution of the above equation is given by 
        \begin{equation}
            \ln[a]=\frac{n  \ln \left[  (n -2) R^{n +1}+6 V\right]}{n -2}+k',
        \end{equation}
where $k'$ is an integration constant. This relationship allows the scale factor to be expressed as a function of $R$, and conversely, so $R$ may be written in terms of the scale factor as
        \begin{equation}\label{ScaleFactorFlatPLAux} 
             a=\left[6 V+R^{1+n}(-2+n)\right]^{\frac{n}{-2+n}} k,
        \end{equation} 
        \begin{equation}
            R=\left(\frac{6 V}{n-2}\right)^{\frac{1}{n+1}}\left\{\left[\frac{k}{a(6 V)^{\frac{n}{2-n}}}\right]^{\frac{2-n}{n}}-1\right\}^{\frac{1}{n+1}},
            \label{eq:R1+dRela}
        \end{equation}
where the integration constant $k'$ is redefined as $k=e^{k'}$. Thus, by considering the general expressions, the semiclassical contribution is written as
        \begin{equation}
            \begin{aligned}
                \Gamma_0=&\frac{\mp \sqrt{6}  \operatorname{Vol}(X) i}{\hbar}\left\{\int_{a_0}^{\bar{a}} a ^{2} \frac{\sqrt{\left(2 V_B+R^{1+n} n\right)\left[\left(6 V_B+R^{1+n}(-1+2 n)\right]\right.}}{\sqrt{R}}da\right. \\
                & -\left.\int_{a_0}^{\bar{a}} a ^{2} \frac{\sqrt{\left(2 V_A+R^{1+n} n\right)\left[\left(6 V_A+R^{1+n}(-1+2 n)\right]\right.}}{\sqrt{R}}da\right\}\\
                & +\frac{\operatorname{Vol}(X)}{\hbar} \bar{a}^3 T_0+\frac{2 \operatorname{Vol}(X) i}{\hbar} \int_{\bar{s}- \delta s}^{\bar{s}+\delta s} \frac{ds}{C(s)}\left(-\frac{1}{2} a^3 R^{1+n} n+\frac{1}{2} a^3 R^{1+n} n \bigg|_{\phi_{A}}\right).
            \end{aligned}   
            \label{eq:G0RnNC}
        \end{equation}
When $R$ is expressed in terms of the scale factor, the integrals over the constant field regions do not admit analytical solutions for an arbitrary $n$. However, they remain manageable for numerical evaluation avoiding $n=2$. In order to interpret the last term, we note that since $k$ is just an integration constant we can choose $k=k_A$. Then this term can be rewritten as
        \begin{equation}
             \frac{2 \operatorname{Vol}(X) i}{\hbar} \int_{\bar{s}-\delta s}^{\bar{s}+\delta s} \frac{d s}{C(s)}\left(\frac{3n}{n-2}\right)a^3\left(V-V_{A}\right).
        \end{equation}
Thus we will be able to interpret this term also as a tension term, and actually it will be the same as the one appearing for the case with the constant Ricci scalar in (\ref{eq:g0Rdelta}). Now, to depict the behavior of the semiclassical contribution, the transition probability is presented in Figure \ref{fig:NumR1+nG0}. For this, the parameters are assigned as $V_A=10$, $V_B=5$, $\operatorname{Vol}(X)=1$, and $\hbar=1$, with the tension set to $T_0=1$. Regarding the constant $k$, a value of $k=0.15$ is selected to ensure a well-behaved probability. It is important to emphasize that, while the limit $n \to 0$ should ideally recover the GR framework, such limit is not consistent with the solution found in Eq. (\ref{ScaleFactorFlatPLAux}), thus in this scenario we will not be able to recover GR by just taking an appropriate limit in contrast to the simplified scenario presented earlier. In the Figure we present plots starting at $n = 1/2$. Similarly, the asymptotic limit $n \to \infty$ is identified with the results previously derived in (\ref{eq:InfR1+n}). Under the appropriate selection of a negative sign in (\ref{DefGammaGe}), this limit recovers a probability profile identical to that of $n = 1/2$, a behavior dictated by the dominance of the tension term. Finally, we also add the GR result for comparison. We can see from this Figure that the general behavior of the probability remains the same as in GR, the modification serves once again  to reduce or decrease the decay depending upon the specific choice of the parameter $n$, in the same way as the results found for the constant Ricci scalar scenario. Notably, the GR result corresponds to the profile that decays in the slowest form and thus has the greatest probability.

\begin{figure}[ht]
            \centering
             \includegraphics[width=\textwidth]{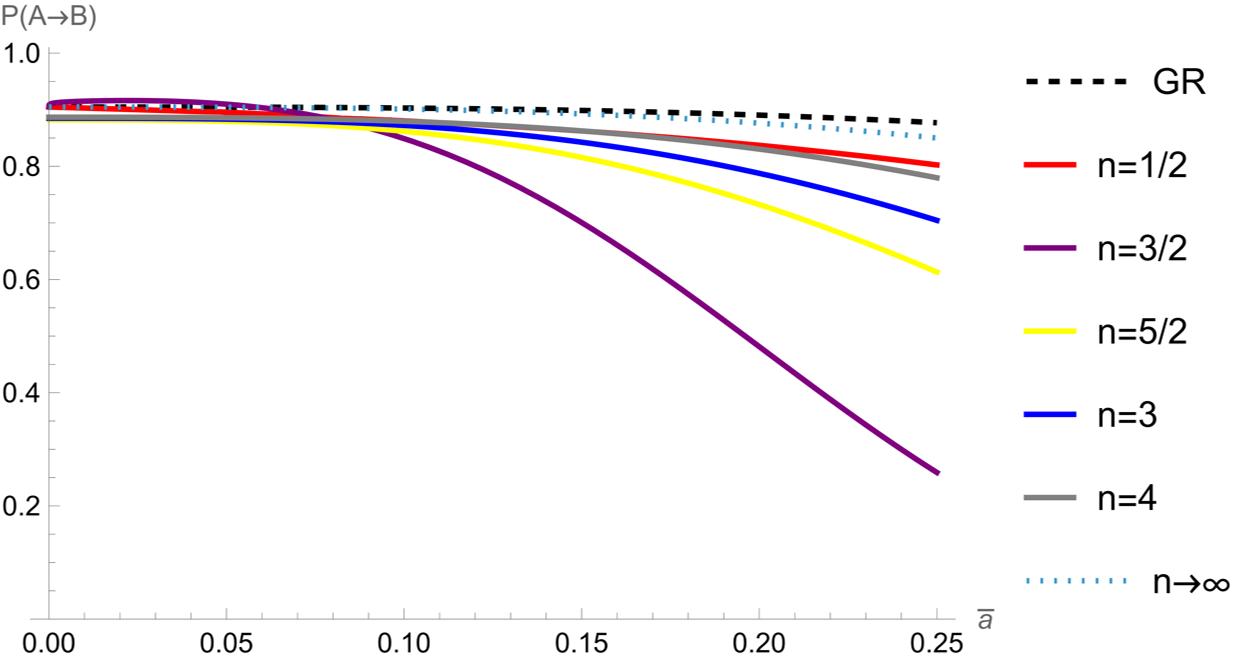}
            \caption{First quantum correction to the transition probability for General Relativity (black dashed line) and within the $f(R)=R^{1+n}$ framework for various values of $n$: $n=1/2$ (red), $n=3/2$ (purple), $n=5/2$ (yellow), $n=3$ (blue), and $n=4$ (gray). The numerical evaluation is performed with the parameters $V_B=5$, $V_A=10$, $\operatorname{Vol}(X)=1$, $\hbar=1$, $k=0.15$, $T_{0}=1$, $T_{1}=0.1$ and $T_{1}^{f(R)}=0.015$.}
            \label{fig:NumR1+nG1}
        \end{figure}

Proceeding with the analysis of the first quantum correction and utilizing expression (\ref{eq:g1flat}), we obtain
        \begin{equation}
            \begin{aligned}
            \Gamma_1=&\operatorname {Vol}(X)\left\{\int_{a_0}^{\bar{a}}\left(\frac{-2 R+3 V_B^{(2)}}{a R}\right) da -\int_{a_0}^{\bar{a}}\left(\frac{-2 R+3 V_A^{(2)}}{a R}\right) d a\right\} \\
             & +\frac{\operatorname{Vol}(X)}{2} T_1 \\
             & +\operatorname{Vol}(X) \int_{\bar{s}-\delta s}^{\bar{s}+\delta s} \frac{d s}{C^2(s)}\left(-\frac{2}{3} R+\left.\frac{2}{3} R\right|_{\phi_A}\right).
            \end{aligned}
            \label{eq:G1RnNC}
        \end{equation}
Taking into account (\ref{eq:R1+dRela}) and computing the integrals in the regions of constant field, we obtain
        \begin{equation}
            \begin{aligned}
            \Gamma_1=&\operatorname{Vol}(X)\left\{3 V_B^{(2)} \int_{a_0}^{\bar{a}} \frac{d a}{\left.aR\right|_B}-3 V_A^{(2)} \int_{a_0}^{\bar{a}} \frac{d a}{\left.aR\right|_A} \right\}\\
            & +\frac{\operatorname{Vol}(X) T_1}{2}-\frac{2}{3} \operatorname{Vol}^2(X)\int_{\bar{s}-\delta s} ^{\bar{s}+\delta s} \frac{ds}{C^2(s)}\left(R-\left.R\right|_A\right),
            \end{aligned}
        \end{equation}
where, substituting the expression for $R$ given in (\ref{eq:R1+dRela}) we get
        \begin{equation}
            \begin{aligned}
            \int \frac{d a}{\left.aR\right|_{A,B}} = & \quad \frac{(1+n)}{6(n-2)}\left(\frac{V_{A,B}}{n-2}\right)^{-\frac{1}{1+n}}\left[\frac{\left(\frac{k}{a}\right)^{-1+\frac{2}{n}}}{V_{A,B}}-6\right]^{\frac{n}{1+n}} \\
            & \times { }_2 F_1\left[1, \frac{n}{1+n}, 2-\frac{1}{1+n},\frac{1-\left(\frac{k}{a}\right)^{-1+\frac{2}{n}}}{6 V_{A,B}}\right],
            \end{aligned}
            \label{eq:2F1Integral}
        \end{equation}
where ${ }_2 F_1$ denotes the hypergeometric function. Furthermore, given the functional form of $R$, in general the separability condition in (\ref{eq:TenSep}) is not satisfied. However, let us introduce the variable
        \begin{equation}
            x = \left[ \frac{k}{a(6V)^{\frac{n}{2-n}}} \right]^{\frac{2-n}{n}}.
        \end{equation}
In the regime where $x \gg 1$ (which implies $a \ll 1$ for $0<n < 2$), the binomial expansion yields
        \begin{equation}
            \begin{aligned}
            (x-1)^\alpha & = x^\alpha \left( 1 - \frac{1}{x} \right)^\alpha \\
            & \approx x^\alpha \left[ 1 - \alpha x^{-1} + \frac{\alpha(\alpha-1)}{2!} x^{-2} - \frac{\alpha(\alpha-1)(\alpha-2)}{3!} x^{-3} + O(x^{-4}) \right],
            \end{aligned}
            \label{eq:binomexp}
        \end{equation}
where $\alpha = \frac{1}{n+1}$. Consequently, by considering the leading order, we obtain for the Ricci scalar in this approximated regime
        \begin{equation}
            R \approx \left( \frac{k}{a} \right)^{\frac{2-n}{n(n+1)}} \left( \frac{1}{n-2} \right)^{\frac{1}{1+n}} - \frac{6V}{1+n} \left( \frac{1}{n-2} \right)^{\frac{1}{1+n}} \left[ \left( \frac{k}{a} \right)^{\frac{2-n}{n}} \right]^{-1+\frac{1}{1+n}}.
        \end{equation}
Conversely, for the case $n > 2$, we consider the limit where $x \ll 1$. Applying the generalized binomial expansion in this regime results in
        \begin{equation}
            (x-1)^{\frac{1}{1+n}} \approx (-1)^{\frac{1}{1+n}} \left[ 1 - \frac{x}{1+n} - \frac{nx^2}{2(1+n)^2} + \mathcal{O}(x^3) \right],
        \end{equation}
from which it follows that
        \begin{equation}
            R \approx (-6)^{\frac{1}{n+1}} \left( \frac{V}{n-2} \right)^{\frac{1}{n+1}} \left( 1 - \frac{\left( \frac{k}{a} \right)^{\frac{n-2}{n}}}{6V(n+1)} \right).
        \end{equation}
These results ensure that the separability condition is satisfied within the regime of small scale factor values. Also, under this approximation, the integral in (\ref{eq:2F1Integral}) can be approximated yielding a divergent term at $a_{0}=0$ in the same way as in the constant Ricci scenario. Therefore, in order to  ensure a well behaved probability distribution we impose the same condition (\ref{eq:R1+ng1constraint}) and obtain simply contributions on the form of tension terms. In this way, the first quantum correction is expressed as
        \begin{equation}
            \Gamma_1=\operatorname{Vol}(X) T_1+ \begin{cases}\operatorname{Vol}(X)\frac{4}{1+n}\left(\frac{1}{n-2}\right)^{\frac{1}{1+n}}\left(\frac{k}{\bar{a}}\right)^{\frac{n-2}{n+1}} T_{1}^{f(R)}, &  0<n<2 \\ \operatorname{Vol}(X)\frac{2}{3}\left(\frac{6}{2-n}\right)^{\frac{1}{n+1}} T_1^{f(R)}, &  n>2.\end{cases}
        \end{equation}
With this result, the effect of the first quantum correction on the transition probability is illustrated in Figure \ref{fig:NumR1+nG1}, using $V_B=5$, $V_A=10$, $\operatorname{Vol}(X)=1$, $\hbar=1$, $k=0.15$, $T_{0}=1$, $T_{1}=0.1$ and $T_{1}^{f(R)}=0.015$. It should be noted that, due to the nature of the approximations employed, these results are strictly valid for small values of the scale factor and do not encompass the standard operational range. However, within this approach, the general behavior remains consistent with the one encountered in GR and in the previous cases. Furthermore, within each interval of $n$, the previously identified behavior is preserved, characterized by the probabilities for $n > 2$ initiating at higher values than those for $n < 2$. It is also noteworthy that the tension terms exhibit distinct differences due to their respective functional dependencies. Upon the inclusion of the first correction, the limit $n \to \infty$ is different from the $n = 1/2$ case, a direct consequence of the presence of the additional tension term.

\begin{figure}[ht]
            \centering
             \includegraphics[width=\textwidth]{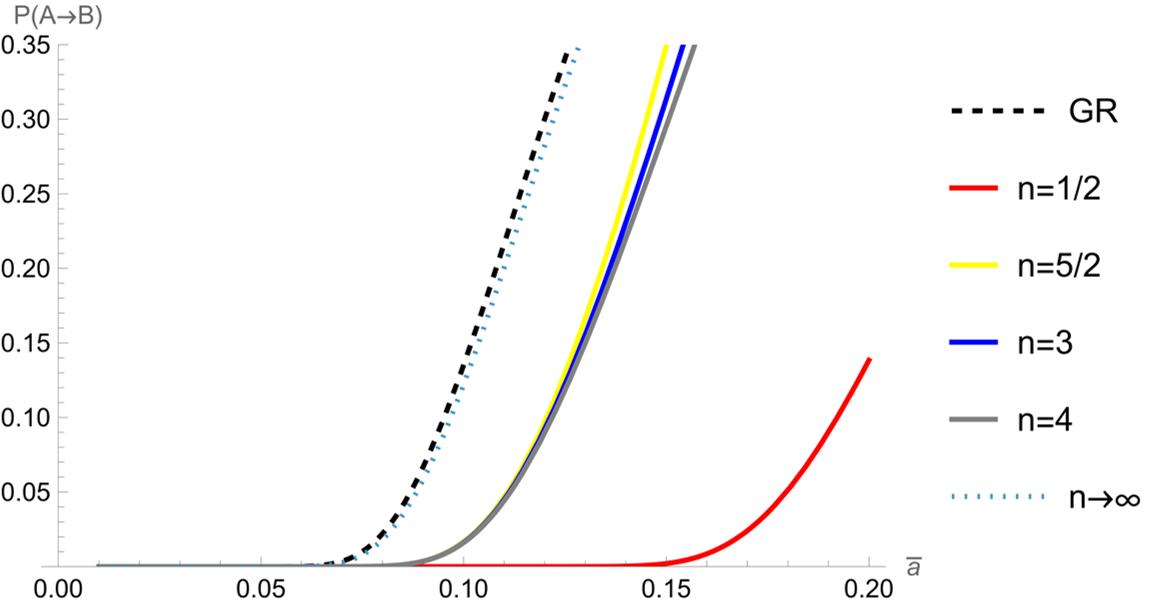}
            \caption{Second quantum correction to the transition probability for General Relativity (black dashed line) and within the $f(R)=R^{1+n}$ framework for various values of $n$: $n=1/2$ (red), $n=5/2$ (yellow), $n=3$ (blue), and $n=4$ (gray). The evaluation is performed with the parameters $V_B=5$, $V_A=10$, $\operatorname{Vol}(X)=1$, $\hbar=1$, $k=0.15$, $T_{0}=1$, $T_{1}=0.1$,  $T_{1}^{f(R)}=0.015$, $T_{2}=0.001$ and $T_{2}^{f(R)}=0.0005$.}
            \label{fig:NumR1+nG2}
        \end{figure}

Regarding the second quantum correction, we consider the general expression provided in (\ref{eq:g2flat}), which yields in this case
        \small \begin{equation}
            \begin{aligned}
            \Gamma_2 & =\frac{3 i \hbar \operatorname{Vol}(X)}{2 \sqrt{2}} V_B^{(4)} \int_{a_0}^{\bar{a}} \frac{d a}{a^4 R} \sqrt{\frac{3 R^{1+n} n+6 V_B}{R\left[6 V_B+R^{1+n}(2 n-1)\right]}} \\
            & +\frac{i \hbar \operatorname{Vol}(X)}{4 \sqrt{6}} \int_{a_0}^{\bar{a}} \frac{d a}{R^{1+n}}\left[\frac{2 V_A+R^{1+n} n}{6 R V+R^{2+n}(2 n-1)}\right]^{\frac{3}{2}} \frac{\left[8 R^2+27 R^n(1+n) n^2 V_B^{(3) 2}\right]}{a^4 n^2(1+n)} 
            \\ & -\frac{3 i \hbar \operatorname{Vol}(X)}{2 \sqrt{2}} V_A^{(4)} \int_{a_0}^{\bar{a}} \frac{d a}{a^4 R} \sqrt{\frac{3 R^{1+n} n+6 V_A}{R\left[6 V_A+R^{1+n}(2 n-1)\right]}} \\
            & -\frac{i \hbar \operatorname{Vol}(X)}{4 \sqrt{6}} \int_{a_0}^{\bar{a}} \frac{d a}{R^{1+n}}\left[\frac{2 V_A+R^{1+n} n}{6 R V_A+R^{2+n}(2 n-1)}\right]^{\frac{3}{2}} \frac{\left[8 R^2+27 R^n(1+n) n^2 V_B^{(3) 2}\right]}{a^4 n^2(1+n)}
            \\
            & +\frac{\hbar \operatorname{Vol}(X)}{\bar{a}^{3}} T_2 \\
            & +\frac{i \hbar}{2} \int_{\overline{s}-\delta s}^{\overline{s}+\delta s}\frac{ \operatorname{Vol}(X)^5}{C^5(s)}ds\left[\frac{8}{27 n^2(1+n)}\right]\left(\frac{1}{a^3}\right)\left(R^{2-n}-R^{2-n}\bigg|_{\phi_A}\right).
            \end{aligned}
        \end{equation} \normalsize
Applying the expansion from Eq. (\ref{eq:binomexp}) once more, we approximate the term $R^{2-n}$ as
        \begin{equation}
            \begin{aligned}
            R^{2-n} & \approx \left(\frac{1}{n-2}\right)^{\frac{(2-n)^2}{n+1}}\left(\frac{k}{a}\right)^{\frac{(2-n)^2}{n(n+1)}} \\
            & -\left(\frac{1}{n-2}\right)^{\frac{2-n}{n+1}}\left(\frac{2-n}{n+1}\right)\left(\frac{k}{a}\right)^{\frac{(2-n)(2-n)}{(n+1) n}} 6 V, 
            \end{aligned}
        \end{equation}
for $0<n<2$ and $a<<1$. And for $n>2$ we have
        \begin{equation}
            R^{2-n}\approx(-6)^{\frac{(2-n)}{n+1}} \left(\frac{V}{n-2}\right)^{\frac{2-n}{n+1}} \left(1-\frac{(2-n) \left(\frac{k}{a}\right)^{\frac{2-n}{n}}}{6 (n+1) V}\right).
        \end{equation}
Allowing us to fulfill the separability condition, and consequently define an extra tension term, so we finally get
        \small \begin{equation}
            \begin{aligned}
            \Gamma_2 & =\frac{3 i \hbar \operatorname{Vol}(X)}{2 \sqrt{2}} V_B^{(4)} \int_{a_0}^{\bar{a}} \frac{d a}{a^4 R} \sqrt{\frac{3 R^{1+n} n+6 V_B}{R\left[6 V_B+R^{1+n}(2 n-1)\right]}} \\
            & +\frac{i \hbar \operatorname{Vol}(X)}{4 \sqrt{6}} \int_{a_0}^{\bar{a}} \frac{d a}{R^{1+n}}\left[\frac{2 V_B+R^{1+n} n}{6 R V_B+R^{2+n}(2 n-1)}\right]^{\frac{3}{2}} \frac{\left[8 R^2+27 R^n(1+n) n^2 V_B^{(3) 2}\right]}{a^4 n^2(1+n)} 
            \\ & -\frac{3 i \hbar \operatorname{Vol}(X)}{2 \sqrt{2}} V_A^{(4)} \int_{a_0}^{\bar{a}} \frac{d a}{a^4 R} \sqrt{\frac{3 R^{1+n} n+6 V_A}{R\left[6 V_A+R^{1+n}(2 n-1)\right]}} \\
            & -\frac{i \hbar \operatorname{Vol}(X)}{4 \sqrt{6}} \int_{a_0}^{\bar{a}} \frac{d a}{R^{1+n}}\left[\frac{2 V_A+R^{1+n} n}{6 R V_A+R^{2+n}(2 n-1)}\right]^{\frac{3}{2}} \frac{\left[8 R^2+27 R^n(1+n) n^2 V_A^{(3) 2}\right]}{a^4 n^2(1+n)}
            \\
            & +\frac{\hbar \operatorname{Vol}(X)}{\bar{a}^{3}} T_2 \\
            & + \begin{cases}\hbar\operatorname{Vol}(X) \left(\frac{1}{n-2}\right)^{\frac{2-n}{n+1}}\left(\frac{2-n}{n+1}\right) \left(k\right)^{\frac{(2-n)(2-n)}{(n+1) n}}  \left(\frac{1}{\bar{a}}\right)^{3+\frac{(2-n)(2-n)}{(n+1) n}} T_2, &  0<n<2 \\ \hbar\operatorname{Vol}(X) \left[\frac{8}{27 n^2(1+n)}\right]\left(\frac{1}{2-n}\right)^{\frac{2-n}{n+1}}\left(\frac{1}{\bar{a}^3}\right)T_2^{f(R)},& n>2.\end{cases}
            \end{aligned}
        \end{equation} \normalsize
In Figure \ref{fig:NumR1+nG2} we present the contribution of these terms to the transition probability considering the parameters $V_B=5$, $V_A=10$, $\operatorname{Vol}(X)=1$, $\hbar=1$, $k=0.15$, $T_{0}=1$, $T_{1}=0.1$, $T_{1}^{f(R)}=0.015$, $T_{2}=0.001$ and $T_{2}^{f(R)}=0.0005$. It is observed that the characteristic behavior of GR is preserved once again and the probability only increases or decreases depending on the value of $n$.
\section{\textbf{Transitions with a Closed FLRW metric}}\label{S-CFLRW}
Now that we have presented a full analysis of the flat FLRW metric for different proposals and considerations. Let us study the closed FLRW metric which is of great interest because its spatial slice is a non-divergent constant\footnote{For this metric, the volume of the spatial slide $X$ is given by  $\operatorname{Vol}(X) = \int_{0}^{2 \pi} \int_{0}^\pi \int_{0}^\pi \sin^2 r \sin \theta \, \mathrm{d}r \, \mathrm{d}\theta \, \mathrm{d}\phi = 2 \pi^2.$ }, which is the reason why the Euclidean approach has only been pursued for this kind of metrics. Following the general treatment we find in this case that the semiclassical contribution is given by 
    \footnotesize \begin{equation}
            \begin{aligned}
            \Gamma_0= & \pm 2 \sqrt{\frac{3}{2}} \frac{\operatorname{Vol}(X) i}{\hbar}\left\{\int_{a_0}^{\bar{a}} a^2\left(-3 f + 6 V_B+2 R f_R\right) \sqrt{\frac{-a^2 f+2 a^2 V_B+\left(-6+a^2 R\right) f_R}{\left(-6+a^2 R\right)\left(-3 f+6 V_B+2 R f_R\right)}} d a\right. \\
            & \left.-\int_{a_0}^{\bar{a}} a^2\left(-3 f + 6 V_A+2 R f_R\right) \sqrt{\frac{-a^2 f+2 a^2 V_A+\left(-6+ a^2 R\right) f_R}{\left(-6+a^2 R\right)\left(-3 f+6 V_A+2 R f_R\right)}} d a\right\} \\
            & +\frac{\operatorname{Vol}(X)}{\hbar} \bar{a}^3 T_0
            \\
            & +\frac{2 \operatorname{Vol}(X) i}{\hbar} \int_{\bar{s}-\delta s}^{\bar{s}+\delta s} \frac{d s}{C(s)}\left[\frac{a^3}{2}\left(f-R f_R\right)+3 f_R a -\left.\frac{a^3}{2}\left(f-R f_R\right)\right|_{\phi=\phi_A} - 3 f_R a \left.\right|_{\phi=\phi_A} \right].
            \end{aligned}
            \label{ClosedSemiclassical}
        \end{equation}
        \normalsize
On the other hand, in this case we have 
        \begin{equation}
            \nabla^2 K=\frac{2}{a^{2}}+V^{(2)}-R+\frac{f_R\left[a^{2}f_{R R}+(a^{2}R-6) f_{R R R}\right]}{3 a^{2}f_{R R} ^2},
        \end{equation}
then the first quantum correction takes the form  \footnotesize
        \begin{equation}
            \begin{aligned}
            \Gamma_1=\operatorname{Vol}(X) & {\left\{\int_{a_0}^{\bar{a}} \frac{3 f_{R R}^2\left[\left(2-a^2 R\right) +a^2 V_B^{(2)}\right]+ f_R\left[a^2 f_{R R}+\left(-6+a^2 R\right) f_{R R R}\right]}{a\left(-6+a^2 R\right) f_{R R}^2} d a\right.} \\
            & \left.-\int_{a_0}^{\bar{a}} \frac{3 f_{R R}^2\left[\left(2-a^2 R\right) +a^2 V_A^{(2)}\right]+ f_R\left[a^2 f_{R R}+\left(-6+a^2 R\right) f_{R R R}\right]}{a\left(-6+a^2 R\right) f_{R R}^2} d a\right\} \\
            & +\frac{\operatorname{Vol}(X)}{2 } T_1
            \\ & +\operatorname{Vol}^2 (X)\int_{\bar{s}-\delta s}^{\bar{s}+\delta s} \frac{d s}{C^2(s)}\left[\frac{f_R\left(f_{R R}+R f_{R R R}\right)}{3 f_{R R}^2}-R -\frac{2 f_R f_{R R R}}{a^2 f_{R R}^2} \right.
            \\ & \left.-\left.\frac{f_R\left(f_{R R}+R f_{R R R}\right)}{3 f_{R R}^2}\right|_{\phi=\phi_A}+\left.R\right|_{\phi=\phi_A} \left. -\frac{2 f_R f_{R R R}}{a^2 f_{R R}^2}\right|_{\phi=\phi_A} \right]
            \end{aligned}
        \end{equation} \normalsize
Finally, we also obtain 
        \begin{equation}
            \begin{aligned}
                \nabla^2\left(\nabla^2 K\right) =&  \left.\frac{1}{9 a^5 f_{R R}^6}\left\{-24 f_{R R}^4 f_{R R R}+2 f_R f_{R R}^2\left[-3\left(-14+a^2 R\right) f_{R R R}^3\right.\right. \right.
                \\ & \left.+f_{R R}\left(a^2 f_{R R R}+2\left(-12+a^2 R\right) f_{R R R R}\right)\right] 
                \\ &  +9 a^2 f_{R R}^6 V^{(4)}+2 f_R^2\left[6\left(-6+a^2 R\right) f_{R R R}^3-\right. 
                \\ & -2 f_{R R} f_{R R R}\left(a^2 f_{R R R}+3\left(-6+a^2 R\right) f_{R R R R}\right) 
                \\ &  \left.+f_{R R}^2\left(a^2 f_{R R R R}+\left(-6+a^2 R\right) f_{R R R R R}\right)\right]\left.\right\} ,
            \end{aligned}
        \end{equation}
and \small
        \begin{equation}
            \begin{aligned}
            {\left[\nabla\left(\nabla^2 K\right)\right]^2=} & \frac{1}{27 a^7 f_{R R}^8}\left\{3 f_{R R}^6\left[-16 a^2 f_{R R}+8\left(-6+a^2 R\right) f_{R R R}+9 a^4 f_{R R}^2 V^{(3) 2}\right]\right. \\
            & +4\left(-6+a^2 R\right) f_R^2 f_{R R}^2\left[-2 a^2 f_{R R}+\left(-12+a^2 R\right) f_{R R R}\right] \\
            & \times {\left[-2 f_{R R R}^2+f_{R R} f_{R R R R}\right]} \\
            & +2\left(-6+a^2 R\right)^2 f_R^3\left(-2 f_{R R R}^2+f_{R R} f_{R R R R}\right)^2 \\
            & +2 f_R f_{R R}^4\left[4 a^4 f_{R R}^2+\left(-42+a^2 R\right)\left(-6+a^2 R\right) f_{R R R}^2+\right. \\
            & \left.\left.+4 f_{R R}\left(a^2\left(12-a^2 R\right) f_{R R R}+3\left(-6+a^2 R\right) f_{R R R R}\right)\right]\right\}.
            \end{aligned}
        \end{equation} \normalsize
That leads to the second quantum correction term in the form 
        \begin{equation}
            \begin{aligned}
            \Gamma_2= &\pm \frac{i \hbar \operatorname{Vol}(X)}{2}\left\{\int_{a_0}^{\bar{a}} \frac{\left(U+9 a^2 f_{R R}^6 V_B^{(4)}\right)}{\sqrt{6} a^4\left(-6+a^2 R\right) f_{R R}^6} \sqrt{\frac{-a^2 f+2 a^2 V_B+\left(-6+a^2 R\right) f_R}{\left(-6+a^2 R\right)\left(-3 f+6 V_B+2 R f_R\right)}} d a\right. \\
            & +\int_{a_0}^{\bar{a}} \frac{\left(W+27 a^4 f_{R R}^2 V_B^{(3) 2}\right)}{2 \sqrt{6} a^6\left(-6+a^2 R\right) f_{R R}^8}\left[\frac{-a^2 f+2 a^2 V_B+\left(-6+a^2 R\right) f_R}{\left(-6+a^2 R\right)\left(-3 f+6 V_B+2 R f_R\right)}\right]^{\frac{3}{2}} d a \\
            & -\int_{a_0}^{\bar{a}} \frac{\left(U+9 a^2 f_{R R}^6 V_A^{(4)}\right)}{\sqrt{6} a^4\left(-6+a^2 R\right) f_{R R}^6} \sqrt{\frac{-a^2 f+2 a^2 V_A+\left(-6+a^2 R\right) f_R}{\left(-6+a^2 R\right)\left(-3 f+6 V_A+2 R f_R\right)}} d a \\
            & \left.-\int_{a_0}^{\bar{a}} \frac{\left(W+27 a^4 f_{R R}^2 V_A^{(3) 2}\right)}{2 \sqrt{6} a^6\left(-6+a^2 R\right) f_{R R}^8}\left[\frac{-a^2 f+2 a^2 V_A+\left(-6+a^2 R\right) f_R}{\left(-6+a^2 R\right)\left(-3 f+6 V_A+2 R f_R\right)}\right]^{\frac{3}{2}} d a\right\} \\
            & +\frac{i \hbar}{2} \int_{s_0}^{s- \delta s} \frac{\operatorname{Vol}^3(X) d s}{C^3(s)}\left(\frac{1}{a^3 }\right)\left(V^{(4)}-V_A^{(4)}\right) \\
            & +\frac{i \hbar}{2} \int_{s_0}^{\bar{s}-\delta_s} \frac{\operatorname{Vol}^5(X) d s}{C^5(s)}\left(\frac{1}{a^3 }\right)\left(V^{(3) 2}-V_A^{(3) 2}\right) \\
            & +\frac{i \hbar}{2} \int_{\bar{s}-\delta s}^{\bar{s} +\delta s} \frac{\operatorname{Vol}^3(X) d s}{C^3(s)}\left(\frac{1}{9 a^5 }\right)\left(\frac{U}{f_{R R}^6}-\frac{U}{f_{R R}^6}\bigg|_{\phi=\phi_A}\right) \\
            & +\frac{i \hbar}{2} \int_{\bar{s}-\delta s}^{\bar{s} +\delta s} \frac{\operatorname{Vol}^5(X) d s}{C^5(s)}\left(\frac{1}{27a^7 }\right)\left(\frac{W}{f_{R R}^8}-\frac{W}{f_{R R}^8}\bigg|_{\phi=\phi_A}\right),
            \end{aligned}
        \end{equation}
where the auxiliary functions are defined as
        \begin{equation}
            \begin{aligned}
            U= & -24 f_{R R}^4 f_{R R R}+2 f_R f_{R R}^2\left\{-3\left(-14+a^2 R\right) f_{R R R}^2\right.
            \\& \left.+f_{R R}\left[a^2 f_{R R R}+2\left(-12+a^2 R\right) f_{R R R R}\right]\right\}
            \\& +2 f_R^2\left\{\left.6\left(-6+a^2 R\right) f_{R R R}^3 -2 f_{R R} f_{R R R}\left[a^2 f_{R R R}+3\left(-6+a^2 R\right) f_{R R R R}\right]\right.\right.
            \\ & \left.+f_{R R}^2\left[a^2 f_{R R R R}+\left(-6+a^2 R\right) f_{R R R R R}\right]\right\} ,
            \end{aligned}
        \end{equation}
and
    \small \begin{equation}
        \begin{aligned}
        W = & +3f_{R R}^6\left[-16 a^2 f_{R R}+8\left(-6+a^2 R\right) f_{R R R}\right] \\
        & -4\left(-6+a^2 R\right) f_R^2 f_{RR}^2\left[2 a^2 f_{R R}+\left(12-a^2 R\right) f_{R R R}\right]\left(-2 f_{R R R}^2+f_{R R} f_{R R R R}\right) \\
        & +2\left(-6+a^2 R\right)^2 f_R^3\left(-2 f_{R R R}^2+f_{R R} f_{R R R R}\right)^2+2 f_R f_{R R}^4\left\{4 a^4 f_{R R}^2\right. \\
        & \left.+\left(252-48 a^2 R+a^4 R^2\right) f_{R R R}^2 \right.\\&\left. -4 f_{R R}\left[a^2\left(-12+a^2 R\right) f_{R R R}-3\left(-6+a^2 R\right) f_{R R R R}\right]\right\}.
        \end{aligned}
    \end{equation} \normalsize
In this approach, we also obtain
     \begin{equation}
            \frac{da}{ds}=\frac{\left(a^2 R-6\right) \left[a^2 (f-2 V)+\left(6-a^2 R\right) f_{R}\right]}{2 a \operatorname{Vol}(X) \left[\left(a^2 R-6\right) \left(3 f-2 R f_{R}-6 V\right)+6 a^2 V^{(1)2}\right]}
            \label{eq:dadsClosed}
        \end{equation}
and
        \begin{equation}
            \frac{dR}{ds}=\frac{-3 a^4 (f-2 V){}^2+4 a^2 \left(a^2 R-3\right) f_{R} (f-2 V)+\left(36-a^4 R^2\right) f_{R}{}^2}{2 a^2 \operatorname{Vol}(X) f_{RR} \left[\left(a^2 R-6\right) \left(3 f-2 R f_{R}-6 V\right)+6 a^2 V^{(1)2}\right]} ,
            \label{eq:dRdsClosed}
        \end{equation}
which leads to the general relation
    \begin{equation}
            \frac{da}{dR}=\frac{a \left(a^2 R-6\right) f_{RR}}{\left(a^2 R+6\right) f_{R}-3 a^2 (f-2 V)}.
            \label{RelationClosed}
        \end{equation}     
We can see that the general expression for  the semiclassical result (\ref{ClosedSemiclassical}) coincides with the result obtained using the Euclidean approach in \cite{Salehian:2018yoq}, albeit with the integrals written in a different form. Moreover, as we showed in subsection \ref{S-ConsRS} the simplification of considering a constant Ricci scalar made the analysis of the transition probabilities possible for any $f$ function and it was well motivated by symmetry arguments in \cite{Salehian:2018yoq}, so let us pursue this simplification. From (\ref{eq:dadsClosed}) and (\ref{eq:dRdsClosed}) we note that in order to have a constant Ricci scalar but an evolving scale factor we need 
    \begin{equation}
            (6 V+R f_{R}-3 f) + \frac{6}{a^2}=0,
    \end{equation}
which is not solvable for any function since $a$ is not a constant. Therefore, we note that the Hamiltonian approach described in the WDW equation does not admit the simplification of a constant Ricci scalar for this metric. This signals an important deviation between the Lorentzian and Euclidean formalisms. Furthermore, we have seen in the previous section that the non-constant Ricci scalar case is very difficult to manage if we want a full description of the transition probabilities since we need expressions that allow us to solve the differential equation (\ref{RelationClosed}) at the same time that fulfills the separability condition for the tension terms. Moreover, in this case there is an extra term that needs to satisfy the separability conditions. In addition, the differential equation to be solved relating $R$ with $a$ is more difficult as well. Therefore, this metric represents a more challenging scenario to compute the transition probabilities. However, we note that we can write Eq. (\ref{RelationClosed}) as 
    \begin{equation}
        \frac{da}{dR}=\frac{\left(R-\frac{6}{a^2}\right)af_{RR}}{\left(R+\frac{6}{a^2}\right)f_{R}-3f_{R}+6V}, 
        \label{eq:approxdadrClosed}
    \end{equation}
which motivate us to look for solutions of this equation on approximated regimes of the scale factor.
\subsection{Non-divergent \texorpdfstring{$R$}{R}}
First of all, let us assume that $R$ does not diverge when $a \to 0$, in that case for big values of the scale factor we can ignore the $1/a^2$ term compared with $R$ in the latter expression, this leads us to
    \begin{equation}
        \frac{da}{dR}\simeq\frac{Raf_{RR}}{Rf_{R}-3f_{R}+6V} . 
    \end{equation}
We note that this differential equation is the same as the one obtained in the flat FLRW case in (\ref{eq:dadRFlat}). Therefore, in this case the relation between $R$ and $a$ is of the same form as before, thus we know that we can solve the equations for the models of the form $f(R)=R^{n+1}$ as 
        \begin{equation*}
            R=\left(\frac{6 V}{n-2}\right)^{\frac{1}{n+1}}\left\{\left[\frac{k}{a(6 V)^{\frac{n}{2-n}}}\right]^{\frac{2-n}{n}}-1\right\}^{\frac{1}{n+1}}.
        \end{equation*}
Then, in general we can obtain the transition probabilities up to second quantum corrections in the same way as before, employing the same approximations, in this case we obtain 
        \begin{equation}
            \begin{aligned}
            \Gamma_0 \simeq & \mp \frac{\mp \sqrt{6} \operatorname{Vol}(X) i}{\hbar} \\ & \times \left\{\int_{a_0}^{\bar{a}} a ^{2} \frac{\sqrt{\left\{2 V_B+R^n\left[\left(R-\frac{6}{a^2}\right) n-\frac{6}{a^2}\right]\right\}\left[6 V_B+R^{1+n}(-1+2 n)\right]}}{\sqrt{R-\frac{6}{a^2}}}da\right. \\
                & -\left.\int_{a_0}^{\bar{a}} a ^{2} \frac{\sqrt{\left\{2 V_B+R^n\left[\left(R-\frac{6}{a^2}\right) n-\frac{6}{a^2}\right]\right\}\left[6 V_A+R^{1+n}(-1+2 n)\right]}}{\sqrt{R-\frac{6}{a^2}}}da\right\}\\ \\ & 
            +\left\{\begin{array}{l}
            \frac{\operatorname{Vol}(X)}{\hbar}\left\{\bar{a}^{3}\left[1+\left(\frac{3 n}{2-n}\right)\right] T_0 + (1+n)\left(\frac{6}{2-n}\right)^{\frac{n}{1+n}} \bar{a}T_{0}^{f(R)}\right\}, \quad 0<n<2 \\
            \frac{\operatorname{Vol}(X)}{\hbar}\left\{\bar{a}^{3}\left[1+\left(\frac{3 n}{2-n}\right)\right]+\frac{3 n k^{\frac{n-2}{n^2+n}}}{(n-2)^{\frac{n}{n+1} }} \bar{a}^{\frac{2+n^2}{n+n^2}}\right\} T_0, \quad n>2,
            \end{array}\right.
            \end{aligned}
        \end{equation}
        
        \begin{equation}
            \begin{aligned}
                \Gamma_1 \simeq &  \operatorname{Vol}(X)\left[\int_{a_0}^{\bar{a}} \frac{\left(2 R-3 V_B^{(2)}-\frac{6}{a^2 n}\right) d a}{a\left(\frac{6}{a^2}-R\right)}-\int_{a_0}^{\bar{a}} \frac{\left(2 R-3 V_A^{(2)}-\frac{6}{a^2 n}\right) d a}{a\left(\frac{6}{a^2}-R\right)}\right] \\ & +\operatorname{Vol}(X) T_1+ \begin{cases} \operatorname{Vol}(X)\frac{2}{3}\left(\frac{6}{2-n}\right)^{\frac{1}{n+1}} T_1^{f(R)}, &  0<n<2 \\  \operatorname{Vol}(X)\frac{4}{1+n}\left(\frac{1}{n-2}\right)^{\frac{1}{1+n}}\left(\frac{k}{\bar{a}}\right)^{\frac{n-2}{n+1}} T_{1}^{f(R)}, &  n>2\end{cases}
            \end{aligned}
        \end{equation}
and\small
         \begin{equation}
            \begin{aligned}
            \Gamma_2 \simeq & \pm \frac{3 i \hbar \operatorname{Vol}(X)}{2 \sqrt{2}} V_{B}^{(4)} \int_{a_0}^{\bar{a}} \frac{d a}{a^4\left(R-\frac{6}{a^2}\right)^{\frac{3}{2}}} \sqrt{\frac{3 R^{n} \left[ \left( R-\frac{6}{a^2} \right)n -\frac{6}{a^2}\right] +6 V_B}{\left[6 V_B+R^{1+n}(2 n-1)\right]}} 
            \\
            & \pm \frac{i \hbar \operatorname{Vol}(X)}{4 \sqrt{6}} \int_{a_0}^{\bar{a}} \frac{d a}{R^{n}} \left\{\frac{2 V_B+R^n\left[n\left(R-\frac{6}{a^2}\right)-\frac{6}{a^2}\right]}{\left(R-\frac{6}{a^2}\right)\left[6 V_B+R^{1+n}(2 n-1)\right]}\right\}^{3 / 2} \\ & \times \frac{\left[8 R\left(R-\frac{6}{a^2}\right)+27 R^n n^2(1+n) V_B^{(3) 2}\right]}{ a^4 n^2(1+n)\left(R-\frac{6}{a^2}\right)} 
            \\ & \mp  \frac{3 i \hbar \operatorname{Vol}(X)}{2 \sqrt{2}} V_{A}^{(4)} \int_{a_0}^{\bar{a}} \frac{d a}{a^4\left(R-\frac{6}{a^2}\right)^{\frac{3}{2}}} \sqrt{\frac{3 R^{n} \left[ \left( R-\frac{6}{a^2} \right)n -\frac{6}{a^2}\right] +6 V_A}{\left[6 V_A+R^{1+n}(2 n-1)\right]}} \\
            & \mp \frac{i \hbar \operatorname{Vol}(X)}{4 \sqrt{6}} \int_{a_0}^{\bar{a}} \frac{d a}{R^{n}} \left\{\frac{2 V_A+R^n\left[n\left(R-\frac{6}{a^2}\right)-\frac{6}{a^2}\right]}{\left(R-\frac{6}{a^2}\right)\left[6 V_A+R^{1+n}(2 n-1)\right]}\right\}^{3 / 2} \\ & \times \frac{\left[8 R\left(R-\frac{6}{a^2}\right)+27 R^n n^2(1+n) V_A^{(3) 2}\right]}{ a^4 n^2(1+n)\left(R-\frac{6}{a^2}\right)} 
            \\
            & +\frac{\hbar \operatorname{Vol}(X)}{\bar{a}^{3}} T_2 \\
            & + \begin{cases} \hbar\operatorname{Vol}(X) \left[\frac{8}{27 n^2(1+n)}\right]\left(\frac{6}{2-n}\right)^{\frac{2-n}{n+1}}\left(\frac{1}{\bar{a}^3}\right)T_2^{f(R)}, &  0<n<2, \\ \hbar\operatorname{Vol}(X) \left(\frac{1}{n-2}\right)^{\frac{2-n}{n+1}}\left(\frac{2-n}{n+1}\right) \left(k\right)^{\frac{(2-n)(2-n)}{(n+1) n}}  \left(\frac{1}{\bar{a}}\right)^{\frac{4 n^2-n+4}{n^2+n}} T_2, & n>2,\end{cases}
            \\
            & + \begin{cases} \hbar\operatorname{Vol}(X) \left[\frac{48}{27 n^2(n+1)}\right]\left(\frac{6}{2-n}\right)^{\frac{1-n}{n+1}}\left(\frac{1}{\bar{a}^5}\right) T_2^{(f(R), 1)}, &  0<n<2, \\ \hbar\operatorname{Vol}(X) \frac{48(n-1)}{27 n^2(n+1)^2}\left(\frac{1}{n-2}\right)^{\frac{1-n}{n+1}}(k)^{\frac{(2-n)(1-n)}{n(n+1)}}\left(\frac{1}{\bar{a}}\right)^{\frac{6 n^2+2 n+2}{n^2+n}} T_2, &  n>2,\end{cases}
            \end{aligned}
        \end{equation}
        \normalsize
where $T_{2}^{(f(R),1)}$ is a new tension term. From the preceding expressions, we observe that in the integrals over constant field regions, if one further assumes the limit $1/a^2 \to 0$ (as well as with the differential equation (\ref{eq:approxdadrClosed})), the expressions obtained for the flat FLRW case in the previous section will be identically recovered. Consequently, the distinction between these scenarios will be mainly dominated by the tension terms, which appear due to the additional term $-3 f_R a$ in the function $K$. 

Moreover, we note that in the semiclassical contribution $\Gamma_0$, an additional tension term $T_{0}^{f(R)}$ is defined exclusively for the range $0 < n < 2$. On the other hand, for $n>2$, the term $T_0$ is maintained but instead of being modified just by a numerical factor, in this case the contribution has a dependence on different powers of $\bar{a}$. However since such powers are positive, the new term will lead to the same decay than in the standard GR case. Thus, once again the overall behavior for the probability will be the same as the GR result. In fact the behavior of this scenario exhibits the behavior depicted in Figure \ref{fig:NumR1+nG0} for the flat case, where the transition probabilities for $n > 2$ decay more rapidly as the value of $n$ approaches two. For the interval $0 < n < 2$, the probability decays at larger values of the scale factor, asymptotically approaching the GR limit. Notably, the GR case remains as the upper bound for all the  probability distributions.

On the other hand, for the first quantum correction $\Gamma_1$, we can again focus on the tension terms given the approximations for big scale factors; here, an additional constant contribution arises for $0 < n < 2$. Conversely, for $n > 2$, an additional term dependent on the scale factor $\bar{a}$ is present. It is important to note, however, that as $\bar{a} \to 0$, this term vanishes, leaving only the constant contribution. Thus, despite the presence of different functional dependence with the scale factor, the general behavior found in GR is once again preserved. Therefore, once again the only difference is to increase or decrease the probability depending on the particular choice of $n$. Consequently, in this scenario, given that the interval $0 < n < 2$ involves two constant tension terms and the integrals within the constant field regions vanish as $\bar{a} \to 0$, the transition probability originates from a lower initial value compared to the $n > 2$ cases and it follows a subsequent decay. This behavior is the same as the one encountered for the constant and non-constant Ricci scalar in the flat result illustrated in Figures \ref{fig:RnG1PP} and \ref{fig:NumR1+nG1}, wherein the decay is more rapid for values over $n=2$; however, as the parameter $n$ increases, the decay of the probability shifted toward larger values of the scale factor, where the approximation used here relies.

Finally, for the second quantum correction, substantial different contributions emerge within the tension terms. In the interval $0 < n < 2$, it is necessary to define two additional tension terms, namely $T_{2}^{f(R)}$ and $T_{2}^{(f(R),1)}$, whereas for $n > 2$, additional contributions are present, yet the tension term remains uniquely $T_2$. Consequently, we obtain terms that decay for big values of the scale factor, consistent with the behavior observed previously. Thus once again, the general behavior observed for GR is preserved and the probability is decreased or increased depending on the choice of $n$ following the same behavior as  it was found in Figures \ref{fig:RnG2PP} and \ref{fig:NumR1+nG2} for the flat case, that is  for values within the interval $0 < n < 2$, the transition probability reach its maximum at a non-zero scale factor, exhibiting a decay that is less abrupt than that observed for $n > 2$. Nevertheless, these probabilities consistently remain below the values of the GR result.

\begin{figure}[ht]
            \centering
            \includegraphics[width=\textwidth]{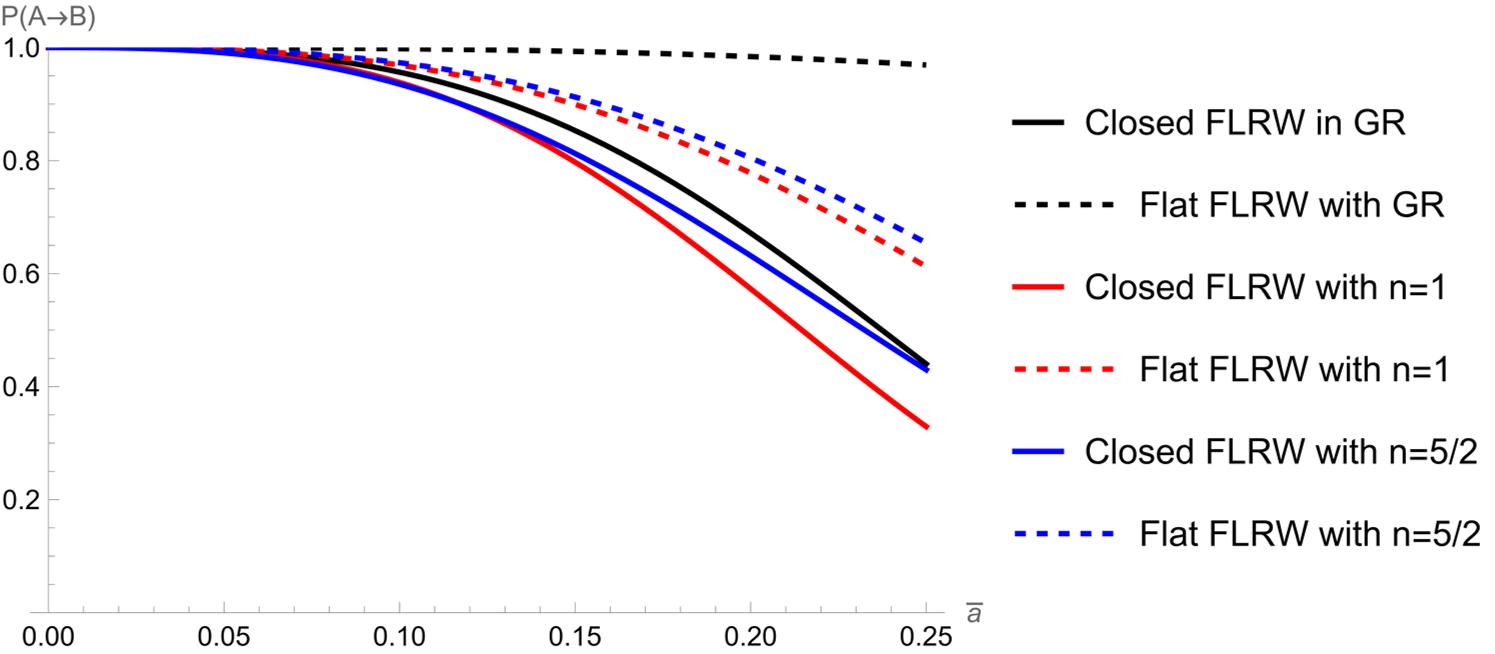}
                \caption{Transition probability considering the semiclassical contribution for $f(R)=R^{1+n}$, in Closed (solid) and Flat (Dashed) FLRW frameworks. For $n=1$ (Red lines), $n=5/2$ (Blue lines) and for GR (Black lines). With the parameters $V_{B}=5$, $V_{A}=10$, $\operatorname{Vol}(X)=2\pi^{2}$ for closed case, $\operatorname{Vol}(X)=1$ for flat case, $c_{0}=0.5$, $\hbar=1$ and $T_{0}=1$.
                }
            \label{fig:ClosedGRVSR1+nG0}
        \end{figure}

On the other hand, in order to consider the opposite limit for he transition probability constituted of small scale factors, we note that since the idea is not to depart too drastically from GR, if $R$ does not diverge with $a$, then $f(R)$ can not diverge with $a \to 0$ as well. Therefore, the relation (\ref{RelationClosed}) leads in this limit to
    \begin{equation}
        \frac{da}{dR}\simeq-\frac{6af_{RR}}{Rf_{R}} .
    \end{equation}
Considering in particular the model that we have studied ($f(R)=R^{n+1}$) we find the solution
    \begin{equation}
        R=\frac{6n}{c_{0}+\ln a},
        \label{eq:smallaRrel}
    \end{equation}
which is well behaved for $c_{0}\neq0$ and it is in agreement with the approximation used. Using this ansatz we can compute the transition probabilities and its two quantum corrections leading to 
        \begin{equation}
            \begin{aligned}
            \Gamma_0 \simeq & \mp \frac{\mp \sqrt{6} \operatorname{Vol}(X) i}{\hbar} \\ & \times \left\{\int_{a_0}^{\bar{a}} a ^{2} \frac{\sqrt{\left\{2 V_B+R^n\left[\left(R-\frac{6}{a^2}\right) n-\frac{6}{a^2}\right]\right\}\left[6 V_B+R^{1+n}(-1+2 n)\right]}}{\sqrt{R-\frac{6}{a^2}}}da\right. \\
                & -\left.\int_{a_0}^{\bar{a}} a ^{2} \frac{\sqrt{\left\{2 V_B+R^n\left[\left(R-\frac{6}{a^2}\right) n-\frac{6}{a^2}\right]\right\}\left[6 V_A+R^{1+n}(-1+2 n)\right]}}{\sqrt{R-\frac{6}{a^2}}}da\right\} \\ & 
             +\frac{\operatorname{Vol}(X)}{\hbar}\bar{a}^3T_{0} ,
            \end{aligned}
        \end{equation}
        
        \begin{equation}
            \begin{aligned}
                \Gamma_1 & \simeq  \operatorname{Vol}(X)\left[\int_{a_0}^{\bar{a}} \frac{\left(2 R-3 V_B^{(2)}-\frac{6}{a^2 n}\right) d a}{a\left(\frac{6}{a^2}-R\right)}-\int_{a_0}^{\bar{a}} \frac{\left(2 R-3 V_A^{(2)}-\frac{6}{a^2 n}\right) d a}{a\left(\frac{6}{a^2}-R\right)}\right] \\ & +\operatorname{Vol}(X) T_1 ,
            \end{aligned}
            \label{eq:G1ClosedSmalla}
        \end{equation}
and
         \begin{equation}
            \begin{aligned}
            \Gamma_2 \simeq & \pm \frac{3 i \hbar \operatorname{Vol}(X)}{2 \sqrt{2}} V_{B}^{(4)} \int_{a_0}^{\bar{a}} \frac{d a}{a^4\left(R-\frac{6}{a^2}\right)^{\frac{3}{2}}} \sqrt{\frac{3 R^{n} \left[ \left( R-\frac{6}{a^2} \right)n -\frac{6}{a^2}\right] +6 V_B}{\left[6 V_B+R^{1+n}(2 n-1)\right]}} 
            \\
            & \pm \frac{i \hbar \operatorname{Vol}(X)}{4 \sqrt{6}} \int_{a_0}^{\bar{a}} \frac{d a}{R^{n}} \left\{\frac{2 V_B+R^n\left[n\left(R-\frac{6}{a^2}\right)-\frac{6}{a^2}\right]}{\left(R-\frac{6}{a^2}\right)\left[6 V_B+R^{1+n}(2 n-1)\right]}\right\}^{3 / 2} \\ & \times \frac{\left[8 R\left(R-\frac{6}{a^2}\right)+27 R^n n^2(1+n) V_B^{(3) 2}\right]}{ a^4 n^2(1+n)\left(R-\frac{6}{a^2}\right)} 
            \\ & \mp  \frac{3 i \hbar \operatorname{Vol}(X)}{2 \sqrt{2}} V_{A}^{(4)} \int_{a_0}^{\bar{a}} \frac{d a}{a^4\left(R-\frac{6}{a^2}\right)^{\frac{3}{2}}} \sqrt{\frac{3 R^{n} \left[ \left( R-\frac{6}{a^2} \right)n -\frac{6}{a^2}\right] +6 V_A}{\left[6 V_A+R^{1+n}(2 n-1)\right]}} \\
            & \mp \frac{i \hbar \operatorname{Vol}(X)}{4 \sqrt{6}} \int_{a_0}^{\bar{a}} \frac{d a}{R^{n}} \left\{\frac{2 V_A+R^n\left[n\left(R-\frac{6}{a^2}\right)-\frac{6}{a^2}\right]}{\left(R-\frac{6}{a^2}\right)\left[6 V_A+R^{1+n}(2 n-1)\right]}\right\}^{3 / 2} \\ & \times \frac{\left[8 R\left(R-\frac{6}{a^2}\right)+27 R^n n^2(1+n) V_A^{(3) 2}\right]}{ a^4 n^2(1+n)\left(R-\frac{6}{a^2}\right)} 
            \\
            & +\frac{\hbar \operatorname{Vol}(X)}{\bar{a}^{3}} T_2.
            \end{aligned}
            \label{eq:G2ClosedSmalla}
        \end{equation}
It is observed that, since the solution for $R$ in Eq. (\ref{eq:smallaRrel}) exhibits no functional dependence on the potential $V$, the additional terms arising from $f(R)$ that could potentially lead to new tension terms actually vanish  and we are led with the usual GR tension terms $T_0$, $T_1$, and $T_2$. Thus, we expect that the general behavior is also preserved in this case, we will only obtain a less or more rapidly decaying probability depending on the choice of $n$. Furthermore, in the non-constant Ricci scalar scenario in flat FLRW we also obtained the results as approximations for small values of the scale factor (see section \ref{NumR1+nFlat}), thus we directly compare the results with both spatial curvatures. Thus in Figure \ref{fig:ClosedGRVSR1+nG0}, we compare the behavior of the semiclassical contribution to the transition probability for scale factor values from $0$ to $0.25$ with the parameters $V_{B}=5$ and $V_{A}=10$, where the spatial volume is assigned as $\operatorname{Vol}(X)=2\pi^{2}$ for the closed case and $\operatorname{Vol}(X)=1$ for the flat case. Furthermore, the constant in Eq. (\ref{eq:smallaRrel}) is set to $c_{0}=0.5$, while $\hbar=1$. The tension term is set as $T_{0}=1$ and we also added the GR result for comparison. It is observed that the probability for the flat universe decays slower than for the closed universe case and both results consistently remain below the probabilities predicted by the GR result.

Following the results for the first quantum correction (\ref{eq:G1ClosedSmalla}) and the second one (\ref{eq:G2ClosedSmalla} we obtain consistent behaviors with previous findings by taking different values of the $n$ parameter (see Figures \ref{fig:NumR1+nG0}, \ref{fig:NumR1+nG1}, and \ref{fig:NumR1+nG2}). Furthermore, making the direct comparison with the flat results of the last section, we find consistency with Figure \ref{fig:ClosedGRVSR1+nG0}, that is the flat result has in all cases a bigger probability but both results withing the $f(R)$ framework decay faster than the standard GR result. 
        
\subsection{Divergent \texorpdfstring{$R$}{R}}
So far we have assumed that $R$ does not diverge with $a$ in the limit $a\to0$. However, from the classical definition we can see that a divergence with $a$ is possible and natural. Then, let us consider the case in which $R$ diverges. First of all let us consider the ansatz
    \begin{equation}
        R=\frac{\alpha}{a^2} ,
    \end{equation}
that is, the special case in which $Ra^2$ is a constant. In this case, we  obtain in general 
    \begin{equation}
        \frac{da}{dR}=\frac{a(\alpha-6)f_{RR}}{(\alpha+6)f_{R}+a^2(6V-3f)}.
    \end{equation}
Let us consider the particular ansatz of a power law for the $f$ function. Then, for small scale factors we obtain
    \begin{equation}
        \frac{da}{dR}=\frac{\alpha^{n-1}(\alpha-6)n(n+1)}{(\alpha+6)(n+1)\alpha^n-3\alpha^{n+1}}a^3 ,
    \end{equation}
that is consistent with the ansatz proposed and it is once again independent of the potential. On the other hand, for big values of the scale factor we find
    \begin{equation}
        \frac{da}{dR}=\frac{\alpha^{n-1}(\alpha-6)n(n+1)}{6V}a ,
    \end{equation}
which is inconsistent with the ansatz. Therefore, we see that this behavior can only be encountered for small scale factors, at least for this ansatz. Finally, let us assume that $R$ diverges with $a\to0$ faster than $1/a^2$. In this case, for small values of the scale factor we encounter the same form that the one obtained in the flat FLRW result. On the other hand, for big values of the scale factor we are led to 
    \begin{equation}
        -\frac{a}{6}\frac{da}{dR}=\frac{f_{RR}}{6V-3f} ,
    \end{equation} 
which leads to the solution
    \begin{equation}
        R^n=\frac{6V}{n+1}\left[c_{o}-\frac{a^2}{12}\right] ,
    \end{equation}
which contradicts the assumption. Then, this behavior is incompatible with the solutions of the WDW equation. Therefore, the only ansatz that can successfully lead to approximated solutions for both small and large scale factors is when the Ricci scalar $R$ does not diverge with $a$.

\section{Final Remarks}
\label{S-FinalR} 
In this work we have studied vacuum transition probabilities between two minima of a scalar field potential including quantum corrections  in $f(R)$ theories. This analysis used a Lorentzian Hamiltonian formalism as developed in \cite{Garcia-Compean:2024zjr} to include up to second order quantum corrections encoded in the semiclassical WKB expansion, thus extending its applicability to modified theories of gravity, in particular to $f(R)$ theories. 

Firstly we considered a homogeneous, isotropic and spatially flat ($k=0$) FLRW universe. Starting from the general $f(R)$ action (\ref{eq:action}), coupled to a scalar field, we treated the Ricci scalar $R$ as an independent field enforced by a Lagrange multiplier (we presented the details in Appendix \ref{FLRW-MetricApp}). This procedure successfully recasts the theory into the required Hamiltonian constraint form, which is quadratic in the momenta. We obtained explicit expressions for the semiclassical, first order and second order quantum correction for an arbitrary $f(R)$ model. We found that these expressions preserve the standard tension terms known from GR, but introduce new model dependent terms that may led to tension terms if a separability condition is satisfied. Then, we introduced the simplification of a constant Ricci scalar. This assumption is motivated by arguments made over maximally symmetric spacetimes in the Euclidean treatment of this kind of transitions studied in previous works. Using this assumption we obtained that all the transition probabilities, being the semiclassical results or including the quantum corrections behave in the general way in exactly the same form as in GR \cite{Garcia-Compean:2024zjr} independently of the model under consideration. In particular, once again the initial singularity is avoided when the second quantum correction is taken into account. The only difference that we can expect is a slower or faster decaying probability and the conditions to have access to the initial singularity to take different forms in a model dependent way. 

Then we considered specific models to obtain explicit analytical results.  We analyze the power-law model $f(R)=R^{1+n}$ within the constant Ricci scalar approximation. The transition probability for the semiclassical contribution was presented in Figure \ref{fig:RnG0PP}, the first quantum corrected result in Figure \ref{fig:RnG1PP} and the second quantum corrected one in \ref{fig:RnG2PP}. We found that for $n<2$ the probability is reduced (and moved towards smaller values for the second quantum correction), whereas for $n>2$ the effect is the contrary. However, for the semiclassical result the GR probability is the less decaying curve in all cases, whereas for the first quantum correction results the GR result is in the middle of both behaviors and in the second quantum corrected result the GR curve has the peak on the a bigger value the scale factor in general. Thus, in general taking into account the second quantum correction under this approximation, these models predict an avoidance of the initial singularity with a peak for finite values of the scale factor, and as was also pointed out in Figure \ref{fig:RnG2PP}, the $f(R)$ framework leads to probabilities below those predicted by GR. Notably, the $f(R)$-dependent tension term $T_2^{f(R)}$ appearing on the second quantum corrected result does not have a well defined $n \to 0$ GR limit, thus at this order we encountered a fundamental aspect of the minisuperspace that does not simplify with a limit valid on the classical theory, this happened as well with the anisotropic limit of the Bianchi III metric in \cite{Garcia-Compean:2024zjr}.

Furthermore, we also studied the Starobinsky $f(R)=R+\beta R^2$ model and compared the results with the quadratic $f(R)=R^2$ model and GR. The semiclassical, first quantum corrected and second quantum corrected results were shown in Figures \ref{fig:StaroR2G0}, \ref{fig:StaroR2G1} and \ref{fig:StaroR2G2} respectively. We found in all cases that the biggest probability is encountered with GR and the lowest probability is always the Starobinsky curve. In particular, for the second quantum correction result we see that the Starobinky model leads to a peak of its probability for smaller scale factors. 

We then noted that within this approach, the constant Ricci scalar was not a mandatory condition. Therefore we also studied the transition probabilities without such assumption. The separability condition and the ability to solve differential equations analytically restricted us to study only the $f(R)=R^{1+n}$ models. We presented the probability for the semiclassical contribution in Figure \ref{fig:NumR1+nG0}, we found that the general behavior of GR is preserved in this scenario as well, and the curves present a similar behavior as with the constant Ricci scenario. Furthermore, we were able to obtain the first quantum corrected result for small values of the scale factor whose probabilities were presented in Figure  \ref{fig:NumR1+nG1}, were we also noted that the general behavior of GR was preserved. Finally, in Figure \ref{fig:NumR1+nG2} we also presented the result including the second quantum corrections for small scale factor as well. We also found that the behavior of GR is once again encountered, that is, the singularity gets avoided and the behavior regarding $n$ follows the same trend as was shown for constant Ricci in the intervals $0<n<2$ and $n>2$. The fact that  the general behavior of GR is encountered even if we do not make the assumption over the Ricci scalar is remarkable, suggesting that the avoidance of the initial singularity is robust against considering modified theories of gravity.

Finally, we considered a homogeneous, isotropic and spatially positive ($k=1$) FLRW universe. Surprisingly, we encountered that the assumptions of a constant Ricci scalar is inconsistent within this approach. This signals a remarkable departure between the Euclidean and Hamiltonian approach. However, we were able to study the probabilities in the more general non-constant Ricci scalar scenario on approximated regimes of the scale factor for $f(R)=R^{1+n}$ models. In order to accomplish this we investigated different behaviors of the Ricci scalar and found that the only consistent ansatz that allows to solve for small and big scale factors is when $R$ remains finite as $a \to 0$. In this case we presented the transition probability considering only the semiclassical contribution in Figure \ref{fig:ClosedGRVSR1+nG0} showing that the behavior in this case remains the same. Remarkably, even if the concrete expressions for the transition probabilities are far more difficult than in the flat case with different dependence on the scale factor, we found that the general behavior of GR is preserved for all the cases. Thus, in particular once again the second quantum correction leads to an avoidance of the initial singularity even in this case. In comparison with the flat metric results, it is further observed that the transition probability predicted for the closed universe scenario consistently remains below of the flat configuration.

We would like to remark that in all cases where we were able to study explicitly the transition probabilities, we encountered in general the same behavior as the results for GR. In particular, we showed that in all models we can obtain an avoidance of the initial singularity by incorporating the second quantum corrections, thus these features seem to be a robust feature beyond GR. Moreover, in all cases studied without approximations on the Ricci scalar, the GR result stayed as the curve which exhibits the greatest probability in all scenarios for the semiclassical and second quantum corrected result, whereas this feature is preserved for the first quantum corrected result for big enough values of the scale factor. It would be interesting to study if such features can still hold on other kinds of modifies theories of gravity. Furthermore, the inconsistency of the constant Ricci scalar approximation in closed universes with the formalism employed, highlights a new departure between the Euclidean and Lorentzian formalism that should be studied further.
\newpage
\vspace{1cm}
\centerline{\bf Acknowledgments} \vspace{.5cm} We would like to thank Fernando Quevedo and Luis Zapata for useful discussions. J. H. Aguilar (No. 933342) would like to thank SECIHTI for a grant. D. Mata-Pacheco acknowledges support of Tamkeen under the NYU Abu Dhabi Research Award (ADHPG–AD3547).
\appendix
\section{FLRW Metrics}
\label{FLRW-MetricApp} 
To construct the quantum cosmological framework for $f(R)$ gravity coupled to a scalar field, we begin with the total action comprising both gravitational and matter sectors as
        \begin{equation}
            S=\frac{1}{2} \int d^4 x f(R) \sqrt{-g}-\int d^4 x \sqrt{-g}\left(\frac{1}{2} g^{\mu \nu} \partial_\mu \phi \partial_\nu \phi+V\right).
            \label{eq:action}
        \end{equation}
This action generalizes the Einstein-Hilbert formulation by replacing the Ricci scalar $R$ with an arbitrary function $f(R)$, while maintaining the standard kinetic and potential terms for the scalar field $\phi$. The coupling between geometry and matter through the metric determinant $\sqrt{-g}$ ensures general covariance of the theory.

Let us consider first the spatially flat FLRW metric in Cartesian coordinates,
        \begin{equation}
            ds^2 = -N^2(t)dt^2 + a^2(t)(dx^2 + dy^2 + dz^2).
        \end{equation}
The Ricci scalar for this homogeneous and isotropic spacetime takes the form
        \begin{equation}
            R = 6\left(\frac{\dot{a}^2}{a^2N^2} + \frac{\ddot{a}}{aN^2} - \frac{\dot{a}\dot{N}}{aN^3}\right),
            \label{eq:RicciFlatDD}
        \end{equation}
where the three terms respectively represent, the Hubble expansion contribution ($\dot{a}^2/a^2$), the cosmic acceleration ($\ddot{a}/a$) and the gauge-dependent term involving the lapse function ($\dot{N}$). The proper volume element is $\sqrt{-g} = Na^3$.

To appropriately address the $f(R)$ term within the action, the methodology established in \cite{Vilenkin:1985md,Huang:2013dca} is adopted. This approach suggests that, in the general scenario, the elimination of the second derivatives of the field $a$ through integration by parts cannot be carried out. Instead, the standard procedure for canonical quantization in these circumstances requires to treat the scalar curvature $R$ as an auxiliary field. This is equivalent to the reformulation of the original fourth-order differential equation into a coupled system of two second-order differential equations. Consequently, we can consider (\ref{eq:RicciFlatDD}) as a constraint, and by employing the method of Lagrange multipliers, the action can be expressed as
        \begin{equation}
            \frac{1}{2}\int d^4x \sqrt{-g}f(R) = \frac{1}{2}\int d^4x \left\{Na^3f(R) - \lambda\left[R - 6\left(\frac{\dot{a}^2}{a^2N^2} + \frac{\ddot{a}}{aN^2} - \frac{\dot{a}\dot{N}}{aN^3}\right)\right]\right\}.
        \end{equation}
Here, $\lambda$ serves as a Lagrange multiplier. The variation of the action with respect to the Ricci scalar $R$ yields
        \begin{equation}
            \frac{\delta S}{\delta R} = Na^3f_R(R) - \lambda = 0,
        \end{equation}
from which we immediately identify the multiplier as
        \begin{equation}
            \lambda = Na^3f_R(R).
        \end{equation}
Substituting this back into our constrained action and performing integration by parts on the terms involving time derivatives, we obtain after discarding boundary terms
        \begin{equation}
            \frac{1}{2}\int d^4x \sqrt{-g}f(R) = \int d^4x \left\{\frac{Na^3}{2}(f - Rf_R) - 3\dot{R}f_{RR}\frac{a^2\dot{a}}{N} - 3f_R\frac{a\dot{a}^2}{N}\right\},
        \end{equation}
where we note that the terms on the right-hand side represent: the potential-like term $(f - Rf_R)$ and the standard kinetic term modified by $f_R$. For the scalar field sector, assuming homogeneity (\(\phi = \phi(t)\)), we obtain:
        \begin{equation}
            \int d^4x \sqrt{-g}\left(\frac{1}{2}g^{\mu\nu}\partial_\mu\phi\partial_\nu\phi + V(\phi)\right) = \int d^4x \left(-\frac{a^3}{2N}\dot{\phi}^2 + Na^3V(\phi)\right).
        \end{equation}
Combining both contributions, we arrive at the complete Lagrangian
        \begin{equation}
            \mathcal{L} = \frac{Na^3}{2}(f - Rf_R) - 3\dot{R}f_{RR}\frac{a^2\dot{a}}{N} - 3f_R\frac{a\dot{a}^2}{N} - \frac{a^3}{2N}\dot{\phi}^2 + Na^3V(\phi).
        \end{equation}
This Lagrangian contains all the dynamical information of our system, including both gravitational and matter degrees of freedom. In order to obtain the Hamiltonian constraint, the canonical momenta, defined as \(\pi_M = \frac{\partial \mathcal{L}}{\partial \dot{\Phi}^M}\) where \(\Phi^M = \{N, a, R, \phi\}\), are computed to be
\begin{equation}
    \begin{cases}
        \pi_N = 0, \\
        \pi_a = -3\dot{R}f_{RR}\frac{a^2}{N} - 6f_R\frac{a\dot{a}}{N}, \\
        \pi_R = -3f_{RR}\frac{a^2\dot{a}}{N}, \\
        \pi_\phi = \frac{a^3}{N}\dot{\phi}.
    \end{cases}
\end{equation}
The vanishing of \(\pi_N\) reflects the fact that the lapse function \(N\) is a Lagrange multiplier enforcing the Hamiltonian constraint, rather than a dynamical degree of freedom as usual.

Performing the Legendre transformation, we obtain the Hamiltonian constraint
        \begin{equation}
            H = N\left[-\frac{\pi_a\pi_R}{3f_{RR}a^2} + \frac{1}{3}\frac{f_R}{f_{RR}^2}\frac{\pi_R^2}{a^3} + \frac{\pi_\phi^2}{2a^3} - \frac{a^3}{2}(f - Rf_R) + a^3V(\phi)\right] \approx 0.
            \label{eq:HamConstraintFlat}
        \end{equation}
The resulting expression is clearly quadratic in the momenta, consistent with the general restriction specified by the formalism in (\ref{eq:HamConst}). On the other hand, for the case of positive curvature FLRW metric
        \begin{equation}
            ds^2 = -N^2(t)dt^2 + a^2(t)\left(dr^2 + \sin^2 r\,d\Omega_2^2\right),
        \end{equation}
where \(r \in [0, \pi]\) and \(d\Omega_2^2\) is the metric on the unit 2-sphere, we get
        \begin{equation}
            \begin{cases}
                R = 6\left(\frac{1}{a^2} + \frac{\dot{a}^2}{a^2N^2} + \frac{\ddot{a}}{aN^2} - \frac{\dot{a}\dot{N}}{aN^3}\right), \\
                \sqrt{-g} = Na^3\sin^2 r\sin\theta.
            \end{cases}
        \end{equation}
The additional \(1/a^2\) term in the Ricci scalar represents the intrinsic curvature contribution from the spatial sections. Following the same procedure, the Hamiltonian constraint in this case takes the form
        \begin{equation}
            H = N\left[-\frac{\pi_a\pi_R}{3f_{RR}a^2} + \frac{1}{3}\frac{f_R}{f_{RR}^2}\frac{\pi_R^2}{a^3} + \frac{\pi_\phi^2}{2a^3} - \frac{a^3}{2}(f - Rf_R) + a^3V(\phi) - 3f_Ra\right] \approx 0.
            \label{eq:HamConstraintPFLRW}
        \end{equation}
As we can see, the final expressions for Hamiltonian constraints, (\ref{eq:HamConstraintPFLRW}) and (\ref{eq:HamConstraintFlat}), have the general form (\ref{eq:HamConst}) and only differs on the final term $-3f_Ra$, in this way we can propose a Hamiltonian constraint that enclose both scenarios as
        \begin{equation}
            H = N\left[ \frac{-\pi_a \pi_R}{3 f_{RR} a^2} + \frac{1}{3} \frac{f_R}{f_{RR}^2} \frac{\pi_R^2}{a^3} + \frac{\pi_\phi^2}{2  a^3} - \frac{a^3}{2}\left(f - R f_R\right) + a^3 V - 3 \kappa a f_R \right] \approx 0,
        \end{equation}
where $\kappa$ stands for the curvature parameter for flat or closed spatial geometries with values $\{0,1\}$ respectively. This is the Hamiltonian constraint (\ref{eq:FPHamilConst}) used in the main text.
\section{Bianchi III metric}
\label{BianchiIIIApp} 
In the main text we studied transitions employing only the FLRW metrics. However, we would like to point out that the general mechanism is not restricted to the isotropic case. Therefore, let us consider a homogeneous and anisotropic universe described by the Bianchi III metric.  In a local set of coordinates the metric is written as
        \begin{equation}
            d s^{2} = -N^{2}(t) d t^{2}+A^{2}(t) d x^{2}+B^{2}(t)       e^{-2 \alpha x} d y^{2}+C^{2}(t) d z^{2},
	\label{eq:BIII}
        \end{equation}
where $\alpha \neq 0$ is a constant measuring the amount of anisotropy. The Ricci scalar is given by
        \begin{equation}
            \begin{aligned}
            R=&\frac{2}{A^2 B C N^3}\bigg\{A\bigg[A N \dot{B} \dot{C}+C(N \dot{A} \dot{B}-A \dot{B} \dot{N}+A N \ddot{B})\bigg] \\
            &+B\bigg[-C\left(\alpha^2 N^3+A \dot{A} \dot{N}-A N \ddot{A}\right) \\
            &+A(N \dot{A}\dot{C}-A \dot{C} \dot{N}+A N \ddot{C})\bigg]\bigg\}.
            \end{aligned}
        \end{equation}
Then, considering a homogeneous scalar field canonically coupled to $f(R)$ gravity with the above metric, and following the same procedure as in appendix \ref{FLRW-MetricApp}, the Lagrangian takes the form
        \begin{equation}
        \begin{aligned}
            L& =\frac{N A B C}{2}\left(f-f_R R\right)-\alpha^2 f_R \frac{B C N}{A}-N A B C V(\phi)+\frac{A B C}{2 N} \dot{\phi}^2 \\
            & -\frac{f _{R R}}{N} \dot{R}(\dot{A}BC+A \dot{B} C+A B \dot{C})-\frac{f_R}{N}(\dot{A} B \dot{C}+\dot{A} \dot{B} C+A \dot{B} \dot{C}),
            \end{aligned}
        \end{equation}
which allows to obtain the canonical momenta as
        \begin{equation}
            \begin{aligned}
            & \pi_N=\frac{\partial L}{\partial \dot{N}}=0, \\
            & \pi_A=\frac{\partial L}{\partial \dot{A}}=-\frac{f_{R R}}{N} \dot{R} B C-\frac{f_R}{N}(B \dot{C}+\dot{B C}), \\
            & \pi_B=\frac{\partial L}{\partial \dot{B}}=-\frac{f_{R R}}{N} \dot{R} A C-\frac{f_R}{N}(\dot{A} C+A \dot{C}), \\
            & \pi_C=\frac{\partial L}{\partial \dot{C}}=-\frac{f_{R R}}{N} \dot{R} A B-\frac{f_R}{N}    (\dot{A}B+A \dot{B}),  \\
            & \pi_R=\frac{\partial L}{\partial \dot{R}}=-\frac{f_{R R}}{N}(\dot{A}B C+A \dot{B} C+A B \dot{C}), \\
            & \pi_\phi=\frac{\partial L}{\partial \dot{\phi}}=\frac{AB C}{N} \dot{\phi} \Rightarrow \dot{\phi}=\frac{\Pi_\phi N}{A B C}.
            \end{aligned}
        \end{equation}
Based on the above, the resulting Hamiltonian is
        \begin{equation}
            \begin{aligned}
            H=N & {\left[\frac{1}{3 A B C} \frac{1}{f_R}\left(A^2 \pi_A^2+B^2 \pi_B^2+C^2 \pi_C^2\right)\right.}+\frac{1}{3 A B C} \frac{f_R}{f_{R R}^2} \pi_R^2+\frac{1}{2 A B C} \pi_\phi^2 \\
            & -\frac{1}{3 C} \frac{1}{f_R} \pi_A \pi_B-\frac{1}{3 B} \frac{1}{f_R} \pi_A \pi_C-\frac{1}{3 A} \frac{1}{f_R} \pi_B \pi_C \\
            & -\frac{1}{3 B C} \frac{1}{f_{R R}} \pi_A \pi_R-\frac{1}{3 A C} \frac{1}{f_{R R}} \pi_B \pi_R-\frac{1}{3 A B} \frac{1}{f_{R R}} \pi_C \pi_R \\
            & \left.+\frac{1}{2} A B C\left(Rf_{R}-f\right)+A B C V(\phi)+\alpha^2 \frac{B C}{A} f_R\right] \approx 0,
            \end{aligned}
        \end{equation}
that satisfies the form of the general constraint
        \begin{equation}
            H=N\left(\frac{1}{2} G^{M N} \pi_M \pi_N+P\right).
        \end{equation}
From this equation $G^{MN}$ can be read off
        \begin{equation}
            G^{M N}=\left(\begin{array}{ccccc}
            \frac{2 A}{3 B C f_R} & -\frac{1}{3 C f_R} & -\frac{1}{3 B f_R} & -\frac{1}{3 B C f_{R R}} & 0 \\
            -\frac{1}{3 C f_R} & \frac{2 B}{3 A C f_R} & -\frac{1}{3 A f_R} & -\frac{1}{3 A C f_{R R}} & 0 \\
            -\frac{1}{3 B f_R} & -\frac{1}{3 A f_R} & \frac{2 C}{3 A B f_R} & -\frac{1}{3 A B f_{R R}} & 0 \\
            -\frac{1}{3 B C f_{R R}} & -\frac{1}{3 A C f_{R R}} & -\frac{1}{3 A B f_{R R}} & \frac{2 f_R}{3 A B C f_{R R}^2} & 0 \\
            0 & 0 & 0 & 0 & \frac{1}{A B C}
            \end{array}\right)
        \end{equation}
and
        \begin{equation}
            K=\frac{1}{2} A B C (f_R R - f)+\frac{\alpha^2 B C f_R}{A}+A B C V(\phi).
        \end{equation}
At this point, we have everything we need to apply the general formalism outlined in section \ref{S-WKB}. The solution of the general system of equations is
        \begin{equation}
            C^2=\frac{2 \operatorname{Vol}(X)\left[3 A^2(f-2 V)\left(A^2 R+2 \alpha^2\right)-2\left(A^4 R^2+2 A^2 R \alpha^2-8 \alpha^4\right) f_R\right]}{3 A^4(f-2 V)-3\left(A^4 R+2 A^2 \alpha^2\right) f_R},
        \end{equation}
        \begin{equation}\label{eq:BianchiSys1}
            \frac{d A}{d s}=-\frac{A\left(A^2 R+10 \alpha^2\right)\left[-A^2(f-2 V)+\left(A^2 R+2 \alpha^2\right) f_R\right]}{2 \operatorname{Vol}(X)\left[3 A^2(f-2 V)\left(A^2 R+2 \alpha^2\right)-2\left(A^4 R^2+2 A^2 R \alpha^2-8 \alpha^4\right) f_R\right]},
        \end{equation}
the relationships between variables of the metric and the extra degree of freedom $R$ are
        \begin{equation}
            \begin{aligned}
            & \frac{d A}{d R}=\frac{A\left(A^2 R+10 \alpha^2\right) f_{R R}}{-3 A^2 f+6 A^2 V+\left(A^2 R-2 \alpha^2\right) f_R}, \\
            & \frac{d B}{d R}=\frac{B\left(A^2 R-2 \alpha^2\right) f_{R R}}{-3 A^2 f+6 A^2 V+\left(A^2 R-2 \alpha^2\right) f_R}, \\
            & \frac{d C}{d R}=\frac{C\left(A^2 R-2 \alpha^2\right) f_{R R}}{-3 A^2 f+6 A^2 V+\left(A^2 R-2 \alpha^2\right) f_R}.
            \end{aligned}
        \end{equation}
Therefore, using (\ref{eq:BianchiSys1}) we can relate the metric degrees of freedom as
        \begin{equation}
            \begin{aligned}\label{eq:BianchiSys2}
            & \frac{d A}{d B}=\frac{A^3 R+10 A \alpha^2}{A^2 B R-2 B \alpha^2}, \\
            & \frac{d A}{d C}=\frac{A^3 R+10 A \alpha^2}{A^2 C R-2 C \alpha^2}, \\
            & \frac{d B}{d C}=\frac{B}{C}.
            \end{aligned}
        \end{equation}
We then obtain for the semiclassical contribution
        \begin{equation}
            \begin{aligned}
            \Gamma_0= & -\frac{2 \operatorname{Vol}(X) i}{\hbar} \left\{ \left.\int_0^{\bar{s}} \frac{d s}{C(s)} K\right|_{\phi=\phi_B}-\left.\int_0^{s} \frac{d s}{C(s)} K\right|_{\phi=\phi_A} \right\} \\
            & -\frac{2 \operatorname{Vol}(X) i}{\hbar} \int_{\bar{s}-\delta s}^{\bar{s}+\delta s} \frac{d s}{C(s)} A B C\left(V-V_A\right) \\
            & -\frac{2 \operatorname{Vol}(X) i}{\hbar} \int_{\bar{s}-\delta s}^{\bar{s}+\delta s} \frac{d s}{C(s)} \frac{A B C}{2}\left[\left(f_R R-f\right)-\left.\left(f_R R-f\right)\right|_{\phi_A}\right] \\
            & -\frac{2 \operatorname{Vol}(X) i}{\hbar} \int_{s-\delta s}^{\bar{s}+\delta s} \frac{d s}{C(s)} \frac{B C}{A} \alpha^2\left(f_R-f_R \mid \phi_A\right),
            \end{aligned}
        \end{equation}
where
\footnotesize
     \begin{equation}
            \begin{aligned}
             & \frac{1}{C(s)}K = - dA \frac{BC}{\sqrt{2} \left(A^4 R+10 \alpha ^2 A^2\right)}\{3 \left(A^4 R+2 \alpha ^2 A^2\right) f
             \\ & -2 \left[-3 A^4 V'^2+\left(-8 \alpha ^4+A^4 R^2+2 \alpha ^2 A^2 R\right) f_R+3 V \left(A^4 R+2 \alpha ^2 A^2\right)\right]\} \times \\ & \times \sqrt{\frac{-3 A^4 f+6 A^4 V+3 A^2 \left(2 \alpha ^2+A^2 R\right) f_R}{2 \left[-3 A^4 V'^2+\left(-8 \alpha ^4+A^4 R^2+2 \alpha ^2 A^2 r\right) f_R+3 V \left(A^4 R+2 \alpha ^2 A^2\right)\right]-3 \left(A^4 R+2 \alpha ^2 A^2\right) f}} .
            \end{aligned}
        \end{equation}
\normalsize
The first quantum correction takes the form 
        \footnotesize
        \begin{equation}
            \begin{aligned}
            \Gamma_1 & =\operatorname{Vol}(X)\left[\left.\int_0^{\bar{s}-\delta s} \frac{d s}{C^2(s)} \nabla^2 K\right|_{\phi=\phi_B}-\left.\int_0^{\bar{s}-\delta s} \frac{d s}{C^2(s)} \nabla^2 K\right|_{\phi=\phi_A}\right]
            \\ & +\operatorname{Vol}^2(X) \int_{s-\delta s}^{\bar{s}+\delta s} \frac{d s}{C^2(s)}\left(V^{(2)}-V_A^{(2)}\right) 
            \\ & + \operatorname{Vol}^2(X) \int_{\bar{s}-\delta s}^{\bar{s}+\delta s} \frac{d s}{C^2(s)}\left[\left(-2 R+\frac{f}{f_R}-\frac{2 V}{f_R}+\frac{f_R}{3 f_{R R}}+\frac{R f_R f_{R R R}}{3 f_{R R}^2}+\frac{2 \alpha^2}{3} \frac{f_R f_{R R R}}{A^2 f_{R R}^2}\right)\right. 
            \\ & \left.-\left.\left(-2 R+\frac{f}{f_R}-\frac{2 V}{f_R}+\frac{f_R}{3 f_{R R}}+\frac{R f_R f_{R R R}}{3 f_{R R}^2}+\frac{2 \alpha^2}{3} \frac{f_R f_{R R R}}{A^2 f_{R R}^2}\right)\right|_{\phi_A}\right],
            \end{aligned}
        \end{equation}
\normalsize        
where
            \begin{equation}
        \begin{aligned}
            \frac{1}{C^2(s)}\nabla^2 K= & \frac{dA}{A\left(A^2 R+10 \alpha^2\right) f_R f_{R R}^2} \biggl\{  3 A^2(f-2 V) f_{R R}^2 
            \\ & +f_R f_{R R}^2\left(-6 A^2 R+4 \alpha^2+3 A^2 V^{(2)}\right) 
            \\ & +f_R^2\left[A^2 f_{R R}+\left(A^2 R+2 \alpha^2\right) f_{R R R}\right] \biggr\}.
        \end{aligned}
        \end{equation}
Moreover the second quantum correction can be written as
        \small
        \begin{equation}
            \begin{aligned}
            \Gamma_2= & \frac{i \hbar \operatorname{Vol}(X)}{2}\left[\left.\int_{0}^{s-\delta s} \frac{ds}{C^3(s)} \nabla^{2}\left(\nabla^{2} K\right)\right|_{\phi=\phi_B}+\int_{0}^{s-\delta s} d s \frac{\operatorname{Vol}^5(X)}{C^5(s)} \left[\nabla\left(\nabla^{2} K\right)\right]^{2}\right|_{\phi=\phi_B} \\ & \left.-\int_{0}^{\bar{s}-\delta s} d s \frac{\operatorname{Vol}^3(x)}{C^3(s)} \nabla^{2}\left(\nabla^{2} K\right)\right|_{\phi=\phi_A}-\left.\int_{0}^{\bar{s}-\delta s} d s \frac{\operatorname{Vol}^5(x)}{C^5(s)} \left[\nabla\left(\nabla^{2} K\right)\right]^{2} \right|_{\phi=\phi_A} \\ & +\int_{\bar{s}-\delta_s}^{\bar{s}+\delta_s} ds \frac{\operatorname{Vol}^3(X)}{C^3(s)}\left(U_{BIII}-\left.U_{BIII}\right|_{\phi=\phi_A}\right) \\ & \left.\left.+\int_{\bar{s}-\delta_s}^{\bar{s}+\delta_s} d s \frac{\operatorname{Vol}^5(x)}{C^5(s)}\left(W_{BIII}-W_{BIII} \right|_{\phi=\phi_A}\right) \right].
            \end{aligned}
            \end{equation}
\normalsize
where
\small
        \begin{equation}
            \begin{aligned}
                & \frac{1}{C^3(s)} \nabla^{2}\left(\nabla^{2} K\right) = \frac{dA}{\sqrt{6} A^2 B C\left(A^2 R+10 \alpha^2\right) f_R^2 f_{R R}^6} \\ & \times \sqrt{\frac{A^2\left(A^2 R+2 \alpha^2\right) f_R-A^4(f-2 V)}{6\left(A^4 R+2 A^2 \alpha^2\right) V-3\left(A^4 R+2 A^2 \alpha^2\right) f+2\left(A^4 R^2+2 A^2 R \alpha^2-8 \alpha^4\right) f_R}} \times \\ & \begin{gathered}
                \times \left\{ 12 A^2(f-2 V) f_{R R}^6+6 f_R f_{R R}^4\left[f_{R R}^2\left(8 \alpha^2-3 A^2 V^{(2)}\right)-A^2(f-2 V) f_{R R R}\right] \right.\\
                +2 f_R^3 f_{R R}^2\left\{f_{R R}\left[A^2 f_{R R R}+2\left(A^2 R+4 \alpha^2\right) f_{R R R R}\right]-\left(3 A^2 R+14 \alpha^2\right) f_{R R R}^2\right\}\\
                +f_R^2 f_{R R}^4\left[32 \alpha^2 f_{R R R}+3 A^2 f_{R R}\left(3 f_{R R} V^{(4)}-2\right)\right] \\
                +2 f_R^4\left\{6\left(A^2 R+2 \alpha^2\right) f_{R R R}^3-2 f_{R R} f_{R R R}\left[A^2 f_{R R R}+3\left(A^2 R+2 \alpha^2\right) f_{R R R R}\right]\right. \\ \left.
                \left.+f_{R R}^2\left[A^2 f_{R R R R}+\left(A^2 R+2 \alpha^2\right) f_{R R R R R}\right]\right\} \right\} ,
                \end{gathered}
            \end{aligned}
        \end{equation}
\normalsize
and 
\footnotesize
       $$
               \frac{1}{C^5(s)} \left[\nabla\left(\nabla^{2} K\right)\right]^{2} =   dA \frac{\left[\left(A^2 R+2 \alpha^2\right) f_R-A^2(f-2 V)\right]}{2 \sqrt{6} A^2 B C\left(A^2 R+10 \alpha^2\right) f_R^3\left[-A^4(f-2 V)+A^2\left(A^2 R+2 \alpha^2\right) f_R\right] f_{R R}^8}
        $$
        $$
        \times \left(\frac{A^2\left(A^2 R+2 \alpha^2\right) f_R-A^4(f-2 V)}{6\left(A^4 R+2 A^2 \alpha^2\right) V-3\left(A^4 R+2 A^2 \alpha^2\right) f+2\left(A^4 R^2+2 A^2 R \alpha^2-8 \alpha^4\right) f_R}\right)^{\frac{3}{2}} 
        $$
        $$
        \times  \bigg\{ 18 A^4(f-2 V)^2 f_{R R}^8+12 A^2 f_R\left[9 A^2 V^{(1) 2}-4 \alpha^2(f-2 V)\right] f_{R R}^8 
        $$
        $$ +  4 f_R^2 f_{R R}^6\big\{3 A^2 f\left[2 A^2 f_{R R}-\left(A^2 R+4 \alpha^2\right) f_{R R R}\right]+6 A^2 V\left[f_{R R R}\left(A^2 R+4  \alpha^2\right)-2 A^2 f_{R R}\right]
        $$
        $$
        + f_{R R}^2\left(32 \alpha^4-27 A^4 V^{(1)} V^{(3)}\right)\big\}
        $$
        $$
         +  4\left(A^2 R+2 \alpha^2\right) f_R^5 f_{R R}^2\left[f_{R R R}\left(A^2 R+4 \alpha^2\right)-2 A^2 f_{R R}\right]\left(f_{R R} f_{R R R R}-2 f_{R R R}^2\right)
         $$
         $$
        +  2\left(A^2 R+2 \alpha^2\right) f_R^6\left(f_{R R} f_{R R R R}-2 f_{R R R}^2\right)^2 
        $$
        $$
        +  f_R^3 f_{R R}^4\big\{-32 A^2 \alpha^2 f_{R R}^3+16 \alpha^2\left(A^2 R+10 \alpha^2\right) f_{R R}^2 f_{R R R}+24 A^2\left(A^2 R+2 \alpha^2\right)(f-2 V) f_{R R R}^2 
        $$
        $$
        +  27 A^4 f_{R R}^4 V^{(3) 2}-12 A^2\left(A^2 R+2 \alpha^2\right)(f-2 V) f_{R R} f_{R R R R}\big\} 
        $$
        $$
        +  2 f_R^4 f_{R R}^4\left\{4 A^4 f_{R R}^2+\left(A^4 R^2-8 A^2 R \alpha^2-4 \alpha^4\right) f_{R R R}^2+4 f_{R R}\big[-f_{R R R}\big(A^4 R+4 A^2 \alpha^2\right)
        $$
        \begin{equation}
        + 2 \alpha^2\left(A^2 R+2 \alpha^2\right) f_{R R R R}\big]\big\}\bigg\} ,
        \end{equation}
\normalsize
with
        \begin{equation}
            \begin{aligned}
            & U_{BIII}= \frac{1}{9 A^3 B C f_R^2 f_{R R}^6} \\  \times & \left\{ \right. 12 A^2(f-2 V) f_{R R}^6+6 f_R\left[f_{R R}^6\left(8 \alpha^2-3 A^2 V^{(2)}\right)-A^2(f-2 V) f_{R R}^4 f_{R R R}\right] \\
            + & 2 f_R^3 f_{R R}^2\left\{f_{R R}\left[A^2 f_{R R R}+2\left(A^2 R+4 \alpha^2\right) f_{R R R R}\right]-\left(3 A^2 R+14 \alpha^2\right) f_{R R R}^2\right\} \\
            + & f_R^2 f_{R R}^4\left(-6 A^2 f_{R R}+32 \alpha^2 f_{R R R}+9 A^2 f_{R R}^2 V^{(4)}\right) \\
            + & 2 f_R^4\left\{6\left(A^2 R+2 \alpha^2\right) f_{R R R}^3-2 f_{R R} f_{R R R}\left[A^2 f_{R R R}+3\left(A^2 R+2 \alpha^2\right) f_{R R R R}\right]\right. \\
            + & \left.f_{R R}^2\left[A^2 f_{R R R R}+\left(A^2 R+2 \alpha^2\right) f_{R R R R R}\right]\right\} \left.\right\} ,
            \end{aligned}
        \end{equation}
and
\small
        \begin{equation}
            \begin{aligned}
             & W_{BIII} = \frac{1}{27 A^5 B C f_R^3 f_{R R}^8} \\ \times & 18 A^4(f-2 V)^2 f_{R R}^8+12 A^2 f_R\left(-4 \alpha^2 f+8 \alpha^2 V+9 A^2 V^{(1) 2}\right) f_{R R}^8 \\
            + & 4 f_R^2 f_{R R}^6\left\{3 f\left[2 A^4 f_{R R}-A^2\left(A^2 R+4 \alpha^2\right) f_{R R R}\right]\right. \\
            + & \left. 6 V\left[-2 A^4 f_{R R}+A^2\left(A^2 R+4 \alpha^2\right) f_{R R R}\right]+f_{R R}^2\left(32 \alpha^4-27 A^4 V^{(1)} V^{(3)}\right)\right\} \\
            + & 4\left(A^2 R+2 \alpha^2\right) f_R^5 f_{R R}^2\left[-2 A^2 f_{R R}+\left(A^2 R+4 \alpha^2\right) f_{R R R}\right]\left(-2 f_{R R R}^2+f_{R R} f_{R R R R}\right) \\
            + & 2\left(A^2 R+2 \alpha^2\right)^2 f_R^6\left(-2 f_{R R R}^2+f_{R R} f_{R R R R}\right)^2 \\
            + & f_R^3 f_{R R}^4\left[-32 A^2 \alpha^2 f_{R R}^3+16 \alpha^2\left(A^2 R+10 \alpha^2\right) f_{R R}^2 f_{R R R}\right. \\
            + & 24 A^2\left(A^2 R+2 \alpha^2\right)(f-2 V) f_{R R R}^2+27 A^4 f_{R R}^4 V^{(3) 2} \\
            - & \left.12 A^2\left(A^2 R+2 \alpha^2\right)(f-2 V) f_{R R} f_{R R R R}\right] \\
            + & 2 f_R^4 f_{R R}^4\left\{4 A^4 f_{R R}^2+\left(A^4 R^2-8 A^2 R \alpha^2-4 \alpha^4\right) f_{R R R}^2\right. \\
            - & \left.4 f_{R R}\left[\left(A^4 R+4 A^2 \alpha^2\right) f_{R R R}-2 \alpha^2\left(A^2 R+2 \alpha^2\right) f_{R R R R}\right]\right\}.
            \end{aligned}
        \end{equation}
\normalsize
Therefore we have shown that the the general formulation to compute transition probabilities up to second order quantum corrections is applicable in this anisotropic scenario as well, obtaining general expressions for an arbitrary $f(R)$ function. However, in order to obtain consistent solutions for a particular model, we will need to solve the system of equations (\ref{eq:BianchiSys1}) and (\ref{eq:BianchiSys2}) and choose one of the degrees of freedom to translate the integrals on terms of $ds$ to such variable. Furthermore, we will need to pursue a separability condition for all the integrals performed on the wall, those tasks are  significantly non-trivial even for simple models of $f(R)$. Consequently, the analysis of this scenario will not be not extended further.



\begin{thebibliography}{99}
         \bibitem{Coleman:1977py}
	S.~R.~Coleman,
	``The Fate of the False Vacuum. 1. Semiclassical Theory,''
	Phys. Rev. D \textbf{15} (1977), 2929-2936
	[erratum: Phys. Rev. D \textbf{16} (1977), 1248]
	doi:10.1103/PhysRevD.16.1248
	
	\bibitem{Callan:1977pt}
	C.~G.~Callan, Jr. and S.~R.~Coleman,
	``The Fate of the False Vacuum. 2. First Quantum Corrections,''
	Phys. Rev. D \textbf{16} (1977), 1762-1768 
	doi:10.1103/PhysRevD.16.1762
	
	\bibitem{Coleman:1980aw}
	S.~R.~Coleman and F.~De Luccia,
	``Gravitational Effects on and of Vacuum Decay,''
	Phys. Rev. D \textbf{21} (1980), 3305
	doi:10.1103/PhysRevD.21.3305

        \bibitem{Devoto:2022qen}
        F.~Devoto, S.~Devoto, L.~Di Luzio and G.~Ridolfi,
        ``False vacuum decay: an introductory review,''
        J. Phys. G \textbf{49} (2022) no.10, 103001
        doi:10.1088/1361-6471/ac7f24
        [arXiv:2205.03140 [hep-ph]].

        \bibitem{Wheeler} 
        J.A. Wheeler, 
        ``Superspace and the nature of quantum geometrodynamics,''
	pp 615-724 of {\it Topics in Nonlinear Physics}, (ed) N.J. Zabusky,
	Springer-Verlag NY, Inc. (1969).
	
	\bibitem{DeWitt} 
        B.S. DeWitt,
        ``Quantum theory of gravity I, The canonical
	theory,'' Phys. Rev. {\bf 160} (5) (1967) 1113.

        \bibitem{FMP1}
	W.~Fischler, D.~Morgan and J.~Polchinski,
	``Quantum Nucleation of
	False Vacuum Bubbles,'' Phys. Rev. D \textbf{41} (1990), 2638 
	doi:10.1103/PhysRevD.41.2638
	
	\bibitem{FMP2}
	W.~Fischler, D.~Morgan and J.~Polchinski,
	``Quantization of False
	Vacuum Bubbles: A Hamiltonian Treatment of Gravitational
	Tunneling,'' Phys. Rev. D \textbf{42} (1990), 4042-4055 
	doi:10.1103/PhysRevD.42.4042

        \bibitem{deAlwis:2019dkc}
	S.~P.~De Alwis, F.~Muia, V.~Pasquarella and F.~Quevedo, ``Quantum
	Transitions Between Minkowski and de Sitter Spacetimes,'' Fortsch.
	Phys. \textbf{68} (2020) no.9 , 2000069  doi:10.1002/prop.202000069
	[arXiv:1909.01975 [hep-th]].

        \bibitem{Cespedes:2020xpn}
	S.~Cespedes, S.~P.~de Alwis, F.~Muia and F.~Quevedo,
	``Lorentzian vacuum transitions: Open or closed universes?,''
	Phys. Rev. D \textbf{104} (2021) no.2, 026013 
	doi:10.1103/PhysRevD.104.026013 [arXiv:2011.13936 [hep-th]].

        \bibitem{Garcia-Compean:2021syl}
	H.~Garc\'\i{}a-Compe\'an and D.~Mata-Pacheco,
	``Lorentzian vacuum transitions for anisotropic universes,''
	Phys. Rev. D \textbf{104} (2021) no.10, 106014
	doi:10.1103/PhysRevD.104.106014
	[arXiv:2107.07035 [hep-th]].
	
	\bibitem{Garcia-Compean:2021vcy}
	H.~Garc\'\i{}a-Compe\'an and D.~Mata-Pacheco,
	``Lorentzian Vacuum Transitions in Ho\v{r}ava\textendash{}Lifshitz Gravity,''
	Universe \textbf{8} (2022) no.4, 237
	doi:10.3390/universe8040237
	[arXiv:2111.11571 [gr-qc]].
	
	\bibitem{Garcia-Compean:2022ysy}
	H.~Garc\'{\i}a-Compe\'an and D.~Mata-Pacheco,
	``Lorentzian vacuum transitions with a generalized uncertainty principle,''
	Class. Quant. Grav. \textbf{39} (2022) no.23, 235011
	doi:10.1088/1361-6382/ac9efc
	[arXiv:2206.06534 [gr-qc]].

        \bibitem{Garcia-Compean:2024zjr}
        H.~Garc\'\i{}a-Compe\'an, J.~Hern\'andez-Aguilar, D.~Mata-Pacheco and C.~Ram\'\i{}rez,
        ``Effects of quantum corrections to Lorentzian vacuum transitions in the presence of gravity,''
        Class. Quant. Grav. \textbf{42}, no.2, 025018 (2025)
        doi:10.1088/1361-6382/ad9fcc
        [arXiv:2406.13845 [gr-qc]].

        \bibitem{Espinosa:2018hue}
        J.~R.~Espinosa,
        ``A Fresh Look at the Calculation of Tunneling Actions,''
        JCAP \textbf{07} (2018), 036
        doi:10.1088/1475-7516/2018/07/036
        [arXiv:1805.03680 [hep-th]].

         \bibitem{Espinosa:2018voj}
	J.~R.~Espinosa,
	``Fresh look at the calculation of tunneling actions including gravitational effects,''
	Phys. Rev. D \textbf{100} (2019) no.10 , 104007 
	doi:10.1103/PhysRevD.100.104007
	[arXiv:1808.00420 [hep-th]].

        \bibitem{Espinosa:2021tgx}  
	J.~R.~Espinosa, J.~F.~Fortin and J.~Huertas,
	``Exactly solvable vacuum decays with gravity,''
	Phys. Rev. D \textbf{104} (2021) no.6, 065007 
	doi:10.1103/PhysRevD.104.065007
	[arXiv:2106.15505 [hep-th]].

        \bibitem{Espinosa:2022jlx}
	J.~R.~Espinosa and J.~F.~Fortin,
	``Vacuum decay actions from tunneling potentials for general spacetime dimension,''
	JCAP \textbf{02} (2023), 023 
	doi:10.1088/1475-7516/2023/02/023
	[arXiv:2211.13667 [hep-th]].

        \bibitem{Holman:1990}
        R.~Holman, E.~W.~Kolb, S.~L.~Vadas and Y.~Wang,
        ``False-vacuum decay in generalized extended inflation,''
        Phys. Lett. B \textbf{250} (1990), 24-28
        doi:10.1016/0370-2693(90)91148-5.

         \bibitem{Zhang:2013pna}
        Y.~l.~Zhang, R.~Saito, D.~h.~Yeom and M.~Sasaki,
        ``Coleman-de Luccia instanton in dRGT massive gravity,''
        JCAP \textbf{02} (2014), 022
        doi:10.1088/1475-7516/2014/02/022
        [arXiv:1312.0709 [hep-th]].
        
        \bibitem{Kristiano:2018oyv}
        J.~Kristiano, R.~D.~Lambaga and H.~S.~Ramadhan,
        ``Coleman-de Luccia Tunneling Wave Function,''
        Phys. Lett. B \textbf{796} (2019), 225-229
        doi:10.1016/j.physletb.2019.07.040
        [arXiv:1808.10110 [gr-qc]].
        
        \bibitem{Ghosh:2021lua}
        J.~K.~Ghosh, E.~Kiritsis, F.~Nitti and L.~T.~Witkowski,
        ``Revisiting Coleman-de Luccia transitions in the AdS regime using holography,''
        JHEP \textbf{09} (2021), 065
        doi:10.1007/JHEP09(2021)065
        [arXiv:2102.11881 [hep-th]].

        \bibitem{Ivo:2025fwe}
        V.~Ivo,
        ``One loop aspects of Coleman de Luccia instantons at small backreaction,''
        [arXiv:2509.18651 [hep-th]].

        \bibitem{Cespedes:2023jdk}
        S.~Cespedes, S.~de Alwis, F.~Muia and F.~Quevedo,
        ``Quantum transitions, detailed balance, black holes, and nothingness,''
        Phys. Rev. D \textbf{109} (2024) no.10, 105027
        doi:10.1103/PhysRevD.109.105027
        [arXiv:2307.13614 [hep-th]].

        \bibitem{Lee:2006vka}
        W.~Lee, B.~H.~Lee, C.~H.~Lee and C.~Park,
        ``The False vacuum bubble nucleation due to a nonminimally coupled scalar field,''
        Phys. Rev. D \textbf{74} (2006), 123520
        doi:10.1103/PhysRevD.74.123520
        [arXiv:hep-th/0604064 [hep-th]].

        \bibitem{Hayashi:2021kro}
        T.~Hayashi, K.~Kamada, N.~Oshita and J.~Yokoyama,
        ``Vacuum decay in the Lorentzian path integral,''
        JCAP \textbf{05} (2022) no.05, 041
        doi:10.1088/1475-7516/2022/05/041
        [arXiv:2112.09284 [hep-th]].

        \bibitem{Nishimura:2023dky}
        J.~Nishimura, K.~Sakai and A.~Yosprakob,
        ``A new picture of quantum tunneling in the real-time path integral from Lefschetz thimble calculations,''
        JHEP \textbf{09} (2023), 110
        doi:10.1007/JHEP09(2023)110
        [arXiv:2307.11199 [hep-th]].
  
        \bibitem{Ezawa:2003wh}
        Y.~Ezawa, H.~Iwasaki, M.~Ohmori, S.~Ueda, N.~Yamada and T.~Yano,
        ``Cosmology in a higher curvature gravity,''
        Class. Quant. Grav. \textbf{20} (2003), 4933-4942
        doi:10.1088/0264-9381/20/22/016
        [arXiv:gr-qc/0306065 [gr-qc]].

        \bibitem{Moreno:2023arp}
        J.~Moreno and A.~J.~Murcia,
        ``Cosmological higher-curvature gravities,''
        Class. Quant. Grav. \textbf{41} (2024) no.13, 135017
        doi:10.1088/1361-6382/ad51c5
        [arXiv:2311.12104 [gr-qc]].

        \bibitem{Singh:2022jue}
        J.~K.~Singh, H.~Balhara, K.~Bamba and J.~Jena,
        ``Bouncing cosmology in modified gravity with higher-order curvature terms,''
        JHEP \textbf{03} (2023), 191
        [erratum: JHEP \textbf{04} (2023), 049]
        doi:10.1007/JHEP03(2023)191
        [arXiv:2206.12423 [gr-qc]].

        \bibitem{Odintsov:2025kyw}
        S.~D.~Odintsov, V.~K.~Oikonomou and G.~S.~Sharov,
        ``Einstein-Gauss-Bonnet cosmology confronted with observations,''
        JHEAp \textbf{47} (2025), 100398
        doi:10.1016/j.jheap.2025.100398
        [arXiv:2503.17946 [gr-qc]].

        \bibitem{Addazi:2025qra}
        A.~Addazi, Y.~Aldabergenov and S.~V.~Ketov,
        ``Curvature corrections to Starobinsky inflation can explain the ACT results,''
        Phys. Lett. B \textbf{869} (2025), 139883
        doi:10.1016/j.physletb.2025.139883
        [arXiv:2505.10305 [gr-qc]].

        \bibitem{Cai:2008ht}
	R.~G.~Cai, B.~Hu and S.~Koh,
	``Gauss-Bonnet Term on Vacuum Decay,''
	Phys. Lett. B \textbf{671} (2009), 181-186
	doi:10.1016/j.physletb.2008.11.053
	[arXiv:0806.2508 [hep-th]].

        \bibitem{Liu:2024aos}
        Y.~Liu,
        ``Vacuum transitions with the Gauss-Bonnet term in $D$ dimensions,''
        Phys. Rev. D \textbf{111} (2025) no.4, 044012
        doi:10.1103/PhysRevD.111.044012
        [arXiv:2406.19451 [hep-th]].

        \bibitem{Gregory:2024sku}
        R.~Gregory and S.~Q.~Hu,
        ``Testing Higher Derivative Gravity through Tunnelling,''
        Particles \textbf{7} (2024) no.1, 144-160
        doi:10.3390/particles7010008
        [arXiv:2402.10620 [hep-th]].

        \bibitem{Vicentini:2020lhm}
        S.~Vicentini and M.~Rinaldi,
        ``Vacuum decay in quadratic gravity,''
        Eur. Phys. J. Plus \textbf{137}, no.3, 332 (2022)
        doi:10.1140/epjp/s13360-022-02529-6
        [arXiv:2009.04435 [gr-qc]].

        \bibitem{Vicentini:2021qoo}
        S.~Vicentini and M.~Rinaldi,
        ``Vacuum decay and quadratic gravity: the massive case,''
        Gen. Rel. Grav. \textbf{54} (2022) no.2, 22
        doi:10.1007/s10714-022-02907-6
        [arXiv:2101.03520 [gr-qc]].

        \bibitem{Vicentini:2022pra}
        S.~Vicentini,
        ``New bounds on vacuum decay in de Sitter space,''
        [arXiv:2205.11036 [gr-qc]].

        \bibitem{CANTATA:2021asi}
        E.~N.~Saridakis \textit{et al.} [CANTATA],
        ``Modified Gravity and Cosmology. An Update by the CANTATA Network,''
        Springer, 2021,
        ISBN 978-3-030-83714-3, 978-3-030-83717-4, 978-3-030-83715-0
        doi:10.1007/978-3-030-83715-0
        [arXiv:2105.12582 [gr-qc]].

         \bibitem{Shankaranarayanan:2022wbx}
        S.~Shankaranarayanan and J.~P.~Johnson,
        ``Modified theories of gravity: Why, how and what?,''
        Gen. Rel. Grav. \textbf{54} (2022) no.5, 44
        doi:10.1007/s10714-022-02927-2
        [arXiv:2204.06533 [gr-qc]].

        \bibitem{Sotiriou:2008rp}
        T.~P.~Sotiriou and V.~Faraoni,
        ``$f(R)$ Theories Of Gravity,''
        Rev. Mod. Phys. \textbf{82} (2010), 451-497
        doi:10.1103/RevModPhys.82.451
        [arXiv:0805.1726 [gr-qc]].

        \bibitem{DeFelice:2010aj}
        A.~De Felice and S.~Tsujikawa,
        ``$f(R)$ theories,''
        Living Rev. Rel. \textbf{13} (2010), 3
        doi:10.12942/lrr-2010-3
        [arXiv:1002.4928 [gr-qc]].

        \bibitem{Nojiri:2017ncd}
        S.~Nojiri, S.~D.~Odintsov and V.~K.~Oikonomou,
        ``Modified Gravity Theories on a Nutshell: Inflation, Bounce and Late-time Evolution,''
        Phys. Rept. \textbf{692} (2017), 1-104
        doi:10.1016/j.physrep.2017.06.001
        [arXiv:1705.11098 [gr-qc]].

        \bibitem{Starobinsky:1980te}
        A.~A.~Starobinsky,
        ``A New Type of Isotropic Cosmological Models Without Singularity,''
        Phys. Lett. B \textbf{91} (1980), 99-102
        doi:10.1016/0370-2693(80)90670-X
        
        \bibitem{Vilenkin:1985md}
        A.~Vilenkin, 
        ``Classical and Quantum Cosmology of the Starobinsky Inflationary Model,'' Phys. Rev. D \textbf{32} (1985), 2511 doi:10.1103/PhysRevD.32.2511

        \bibitem{Linde:2025pvj}
        A.~Linde,
        ``Alexei Starobinsky and Modern Cosmology,''
        [arXiv:2509.01675 [hep-th]].

        \bibitem{Planck:2018jri}
        Y.~Akrami \textit{et al.} [Planck],
        ``Planck 2018 results. X. Constraints on inflation,''
        Astron. Astrophys. \textbf{641} (2020), A10
        doi:10.1051/0004-6361/201833887
        [arXiv:1807.06211 [astro-ph.CO]].

        \bibitem{Stachowski:2016zio}
        A.~Stachowski, M.~Szyd{\l}owski and A.~Borowiec,
        ``Starobinsky cosmological model in Palatini formalism,''
        Eur. Phys. J. C \textbf{77} (2017) no.6, 406
        doi:10.1140/epjc/s10052-017-4981-8
        [arXiv:1608.03196 [gr-qc]].

        \bibitem{German:2023euc}
        G.~German, J.~C.~Hidalgo and L.~E.~Padilla,
        ``Solution for cosmological observables in the Starobinsky model of inflation,''
        Eur. Phys. J. Plus \textbf{139} (2024) no.3, 295
        doi:10.1140/epjp/s13360-024-05065-7
        [arXiv:2307.08257 [astro-ph.CO]].

        \bibitem{Ketov:2025nkr}
        S.~V.~Ketov,
        ``On Legacy of Starobinsky Inflation,''
        [arXiv:2501.06451 [gr-qc]].

        \bibitem{Drees:2025ngb}
M.~Drees and Y.~Xu,
``Refined predictions for Starobinsky inflation and post-inflationary constraints in light of ACT,''
Phys. Lett. B \textbf{867} (2025), 139612
doi:10.1016/j.physletb.2025.139612
[arXiv:2504.20757 [astro-ph.CO]].

        \bibitem{SidikRisdianto:2025qvk}
N.~Sidik Risdianto, R.~H.~S.~Budhi, N.~Shobcha, A.~Salim Adam and M.~A.~Syakura,
``The Preheating Stage on The Starobinsky Inflation after ACT,''
[arXiv:2507.12868 [gr-qc]].

        \bibitem{Ivanov:2021chn}
        V.~R.~Ivanov, S.~V.~Ketov, E.~O.~Pozdeeva and S.~Y.~Vernov,
        ``Analytic extensions of Starobinsky model of inflation,''
        JCAP \textbf{03} (2022) no.03, 058
        doi:10.1088/1475-7516/2022/03/058
        [arXiv:2111.09058 [gr-qc]].


        \bibitem{Capozziello:2014hia}
        S.~Capozziello, M.~De Laurentis and O.~Luongo,
        ``Connecting early and late universe by $f(R)$ gravity,''
        Int. J. Mod. Phys. D \textbf{24} (2014) no.04, 1541002
        doi:10.1142/S0218271815410023
        [arXiv:1411.2822 [gr-qc]].

        \bibitem{Yadav:2018llv}
        B.~K.~Yadav and M.~M.~Verma,
        ``Dark matter as scalaron in $f(R)$ gravity models,''
        JCAP \textbf{10} (2019), 052
        doi:10.1088/1475-7516/2019/10/052
        [arXiv:1811.03964 [gr-qc]].

        \bibitem{Bisabr:2010sq}
        Y.~Bisabr,
        ``Local Gravity Constraints and Power Law f(R) Theories,''
        Grav. Cosmol. \textbf{16} (2010), 239-244
        doi:10.1134/S0202289310030084
        [arXiv:1005.5670 [gr-qc]].

        \bibitem{Goheer:2009ss}
        N.~Goheer, J.~Larena and P.~K.~S.~Dunsby,
        ``Power-law cosmic expansion in f(R) gravity models,''
        Phys. Rev. D \textbf{80} (2009), 061301
        doi:10.1103/PhysRevD.80.061301
        [arXiv:0906.3860 [gr-qc]].

        \bibitem{Tsujikawa:2007gd}
        S.~Tsujikawa,
        ``Matter density perturbations and effective gravitational constant in modified gravity models of dark energy,''
        Phys. Rev. D \textbf{76} (2007), 023514
        doi:10.1103/PhysRevD.76.023514
        [arXiv:0705.1032 [astro-ph]].
        
        \bibitem{DeAngelis:2021afq}
        M.~De Angelis, L.~Figurato and G.~Montani,
        ``Quantum dynamics of the isotropic universe in metric f(R) gravity,''
        Phys. Rev. D \textbf{104} (2021) no.2, 024054
        doi:10.1103/PhysRevD.104.024054
        [arXiv:2105.02934 [gr-qc]].

        \bibitem{Bamonti:2021jmg}
        N.~Bamonti, A.~Costantini and G.~Montani,
        ``Features of the primordial Universe in f(R)-gravity as viewed in the Jordan frame,''
        Class. Quant. Grav. \textbf{39} (2022) no.17, 175011
        doi:10.1088/1361-6382/ac7694
        [arXiv:2103.17063 [gr-qc]].

        \bibitem{Ohta:2017trn}
        N.~Ohta,
        ``Quantum equivalence of $f(R)$ gravity and scalar{\textendash}tensor theories in the Jordan and Einstein frames,''
        PTEP \textbf{2018} (2018) no.3, 033B02
        doi:10.1093/ptep/pty008
        [arXiv:1712.05175 [hep-th]].

        \bibitem{Figueroa:2021iwm}
        D.~G.~Figueroa, A.~Florio, T.~Opferkuch and B.~A.~Stefanek,
        ``Lattice simulations of non-minimally coupled scalar fields in the Jordan frame,''
        SciPost Phys. \textbf{15} (2023) no.3, 077
        doi:10.21468/SciPostPhys.15.3.077
        [arXiv:2112.08388 [astro-ph.CO]].


        \bibitem{Sanyal:2001ws}
        A.~K.~Sanyal and B.~Modak,
        ``Quantum cosmology with $R + R^2$ gravity,''
        Class. Quant. Grav. \textbf{19}, 515-526 (2002)
        doi:10.1088/0264-9381/19/3/307
        [arXiv:gr-qc/0107070 [gr-qc]].

        \bibitem{Huang:2013dca}
        R.~N.~Huang,``The Wheeler-DeWitt equation of $f(R,L_m)$ gravity in minisuperspace,'' [arXiv:1304.5309 [gr-qc]].

        \bibitem{Paliathanasis:2019ega}
        A.~Paliathanasis,
        ``Similarity solutions for the Wheeler\textendash{}DeWitt equation in $f\left( R\right) $-cosmology,''
        Eur. Phys. J. C \textbf{79}, no.12, 1031 (2019)
        doi:10.1140/epjc/s10052-019-7553-2
        [arXiv:1910.03288 [gr-qc]].

        \bibitem{Ramirez}
        C.~Ram\'{\i}rez and V. V\'azquez-B\'aez,
        ``Time in Quantum Cosmology of FRW f (R) Theories,''
        Galaxies \textbf{6}, no.1, 1031 (2018)
        https://doi.org/10.3390/galaxies6010012.

        \bibitem{Salehian:2018yoq}
	    B.~Salehian and H.~Firouzjahi,
    	``Vacuum decay and bubble nucleation in $f(R)$ gravity,''
    	Phys. Rev. D \textbf{99} (2019) no.2, 025002
    	doi:10.1103/PhysRevD.99.025002
    	[arXiv:1810.01391 [hep-th]]. 
        

        \bibitem{Carloni:2004kp}
        S.~Carloni, P.~K.~S.~Dunsby, S.~Capozziello and A.~Troisi,
        ``Cosmological dynamics of $R^n$ gravity,''
        Class. Quant. Grav. \textbf{22} (2005), 4839-4868
        doi:10.1088/0264-9381/22/22/011
        [arXiv:gr-qc/0410046 [gr-qc]].

         \bibitem{Allemandi:2004ca}
        G.~Allemandi, A.~Borowiec and M.~Francaviglia,
        ``Accelerated cosmological models in first order nonlinear gravity,''
        Phys. Rev. D \textbf{70} (2004), 043524
        doi:10.1103/PhysRevD.70.043524
        [arXiv:hep-th/0403264 [hep-th]].

        \bibitem{Carloni:2005ii}
        S.~Carloni, P.~K.~S.~Dunsby and D.~M.~Solomons,
        ``Bounce conditions in f(R) cosmologies,''
        Class. Quant. Grav. \textbf{23} (2006), 1913-1922
        doi:10.1088/0264-9381/23/6/006
        [arXiv:gr-qc/0510130 [gr-qc]].

        \bibitem{Leach:2006br}
        J.~A.~Leach, S.~Carloni and P.~K.~S.~Dunsby,
        ``Shear dynamics in Bianchi I cosmologies with Rn-gravity,''
        Class. Quant. Grav. \textbf{23} (2006), 4915-4937
        doi:10.1088/0264-9381/23/15/011
        [arXiv:gr-qc/0603012 [gr-qc]].

\end{thebibliography}
\end{document}